\RequirePackage{etex}
\documentclass[a4paper,11pt]{article}
\usepackage{jcappub}
\pdfoutput=1
\usepackage{tikz} 
\usetikzlibrary{positioning,arrows.meta,calc}
\usepackage[T1]{fontenc}
\usepackage{paralist}
\usepackage{graphicx}
\usepackage{amsfonts}
\usepackage{amssymb}
\usepackage{xcolor}
\usepackage{mathrsfs}
\usepackage[export]{adjustbox}
\usepackage{amsmath}
\usepackage{physics}
\usepackage{slashed}
\usepackage{dcolumn}
\usepackage{verbatim}
\usepackage{float}
\usepackage{multirow}
\usepackage{soul}
\usepackage{xspace}
\usepackage[normalem]{ulem}
\usepackage{hyperref}
\usepackage{tabularx}
\usepackage{setspace}
\usepackage{booktabs}
\usepackage{mathtools}
\usepackage{subcaption}

\renewcommand{\eqref}[1]{Eq.~(\ref{#1})}

\renewcommand{\paragraph}[1]{~\\ \noindent{\bf \emph{#1} --}}

\renewcommand{\vec}{\mathbf}

\newcommand{\xhat}{\hat{\mathbf x}}
\newcommand{\yhat}{\hat{\mathbf y}}
\newcommand{\zhat}{\hat{\mathbf z}}

\newcommand{\Qrot}{\ensuremath{\mathcal T}}

\allowdisplaybreaks

\begin{document}

\title{The Role of Symmetries in Dark Matter Detector Design}

\author[a,1]{Benjamin Lillard \note{ORCID:~\href{https://orcid.org/0000-0001-8496-4808}{0000-0001-8496-4808}}}
\emailAdd{blillard@psu.edu}
\affiliation[a]{Institute for Gravitation and the Cosmos, The Pennsylvania State University, University Park, PA 16802, USA}

\author[b,2]{Jack D. Shergold \note{ORCID:~\href{https://orcid.org/0000-0001-5685-9007}{0000-0001-5685-9007}}}
\emailAdd{j.d.shergold@liverpool.ac.uk}

\author[b,3]{Juri Smirnov \note{ORCID:~\href{https://orcid.org/0000-0002-3082-0929}{0000-0002-3082-0929}}}
\emailAdd{juri.smirnov@liverpool.ac.uk}

\affiliation[b]{Department of Mathematical Sciences, University of Liverpool,
Liverpool, L69 7ZL, United Kingdom}

\abstract{
Anisotropic materials have emerged as promising candidates for the next generation of sub-GeV dark matter direct detection experiments, because their intrinsic directionality gives rise to a daily modulation signal as the detector rotates with respect to a dark matter wind.
Predicting the shape of the modulation signal requires knowledge of the electronic excited states: however, we show that the amplitude of the modulation can be estimated using only the symmetries of the material.
By decomposing the finite momentum dark matter--electron scattering form factor into spherical harmonics, we show that the 230 crystallographic space groups collapse to just 5 classes, distinguished by their suppression of the quadrupole modes of the squared form factor. 
We apply our symmetry-projection framework to the special case of molecular crystals, and derive an accurate group-theoretic estimator for the loss of daily modulation signal due to crystallisation, which depends only on the symmetries and relative orientations of the molecules within the crystal. 
Finally, we demonstrate that these estimates are linearly proportional to the absolute magnitude of the modulation signal, allowing us to rank molecular crystals without the need for expensive electronic structure calculations. 
Together, these results provide a fast and interpretable route to large-scale screening of anisotropic materials for directional dark matter detection.
}

\maketitle

% -----------------------------------------------------------------------------
\section{Introduction}
% -----------------------------------------------------------------------------

The existence of dark matter (DM) is one of the clearest indications that the Standard Model is incomplete. Its gravitational effects are observed over a wide range of astrophysical and cosmological scales, from the dynamics of galaxies and galaxy clusters to the anisotropies of the cosmic microwave background and the growth of large-scale structure~\cite{Zwicky:1933gu,Zwicky:1937zza,Rubin:1980zd,Faber:1979pp,Ostriker:1974ft,Clowe:2006eq,Komatsu:2010fb,Aghanim:2018eyx}. Despite the overwhelming evidence for its existence, the microscopic identity of DM remains ever elusive. 

Direct detection experiments aim to address this question by searching for the small energy deposits produced when dark matter scatters in a terrestrial target~\cite{Goodman:1984dc,Lewin:1995rx,Freese:2012xd}.
For electroweak-scale dark matter masses, the kinetic energy available in a scattering event is sufficient to produce nuclear recoils at the keV scale, making large underground liquid detectors highly effective~\cite{Akerib:2016vxi,Aprile:2018dbl,Agnes:2018ves,PICO:2019vsc,DarkSide-50:2022qzh,LZ:2024zvo,XENON:2025vwd,PandaX:2025rrz}.
Nuclear recoils are much harder to observe for sub-GeV dark matter masses, especially for dark matter near the MeV scale, 
as the available kinetic energy is much smaller. Dark matter might instead be revealed by the electronic, vibrational, phononic, or other low-energy excitations it induces in an experiment. In particular, MeV-scale dark matter naturally motivates target materials with electronic transitions at the eV scale~\cite{Essig:2011nj,Essig:2015cda,Derenzo:2016fse,Essig:2019kfe,Blanco:2019lrf,SENSEI:2023zdf,DAMIC-M:2025ltz}. These low-threshold experiments often encounter larger background event rates~\cite{Baxter:2025odk}, which have in turn prompted a wide range of proposals for anisotropic detector designs that could use daily modulation to more robustly extract a dark matter signal from the Standard Model backgrounds~\cite{Bozorgnia:2011tk,Hochberg:2016ntt,Hochberg:2017wce,Griffin:2018bjn,Coskuner:2019odd,Geilhufe:2019ndy,Vahsen:2020pzb,Blanco:2021hlm,Hochberg:2021ymx,Blanco:2022pkt,Boyd:2022tcn,Stratman:2024sng,Abbamonte:2025guf,Blanco:2026kda,Sherpa:2026tgy}. 
Rather than relying on time variation in the local lab frame dark matter distribution $\rho_\chi g_\chi(\vec v)$, as in the recent modulation analyses from DAMIC-M and SENSEI~\cite{Arnquist:2023llv,SENSEI:2025qvp}, 
the daily modulation signal from anisotropy relies on an intrinsic property of the detector material, and the phase and amplitude of the dark matter signal can be modified simply by rotating the experimental apparatus in the laboratory.

\begin{figure}[t]
    \centering
    \includegraphics[width=0.76\textwidth,trim={1.0cm 3.2cm 1.0cm 3.2cm},clip]{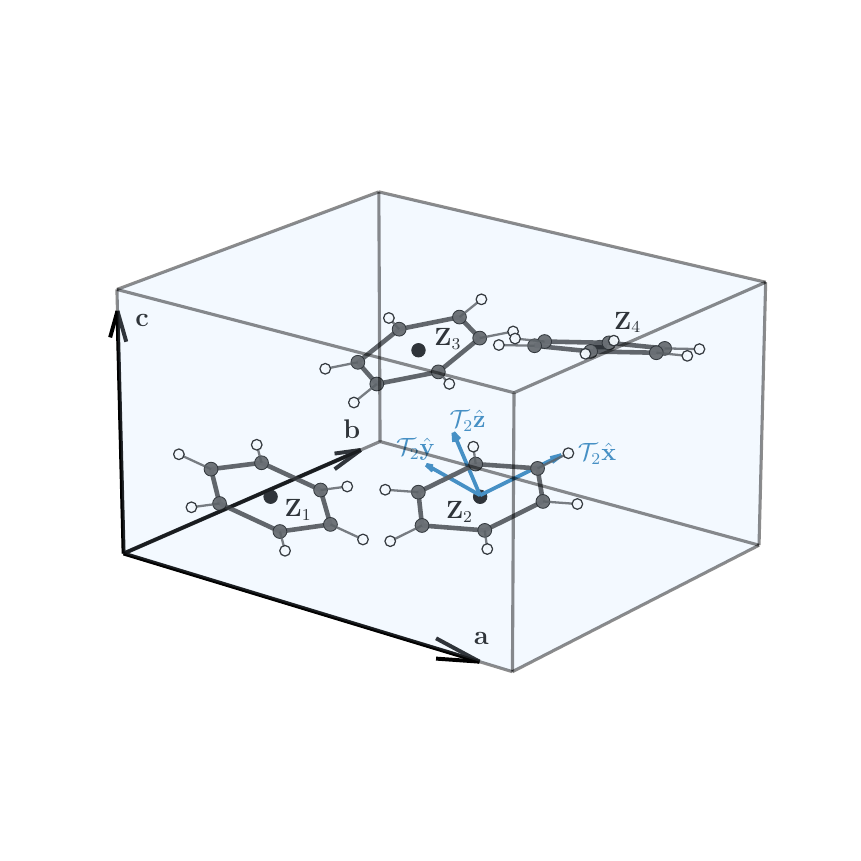}
    \caption{Schematic unit cell for a benzene crystal. The orthorhombic unit cell is spanned by lattice vectors $\mathbf{a}$, $\mathbf{b}$, and $\mathbf{c}$. Within the unit cell, the molecules are centred on the positions $\vec{Z}_i$, whilst their orientations relative to the crystal frame are described by Cartesian orientation matrices $\Qrot_i \equiv \mathcal{R}_i \Qrot$, with $\Qrot$ the internal-to-crystal rotation matrix, and $\mathcal{R}_i$ the point-group symmetry operation that maps one molecule onto its symmetry images. The blue axes show the rotated molecular-frame directions $\Qrot_2\xhat$, $\Qrot_2\yhat$, and $\Qrot_2\zhat$ for molecule $i=2$.
    The relative misalignment of the different molecules in the unit cell will modestly suppress the anisotropic response of the crystal. 
    }
    \label{fig:unit_cell_schematic}
\end{figure}

Organic molecular materials are especially attractive in this context~\cite{Blanco:2019lrf,Blanco:2021hlm}, as their electronic excitations often lie at energies of order a few eV, whilst their electronic wave functions can be highly anisotropic. As an added benefit, the candidate molecule space is also enormous: extended $\pi$-conjugated systems, aromatic molecules, donor--acceptor structures, and charge-transfer systems all feature distinct classes of electronic transitions, each of which can be housed by molecules with a wide range of geometries, ranging from highly symmetric polycyclic compounds to long chiral chains. Broadly speaking,
\begin{center}
    \emph{the space of all chemicals is vast enough that there is likely a molecule for almost any desired electronic response.}
\end{center}
Due to the vastness of the chemical space, however, finding the optimal detector material is highly challenging, 
particularly if we must perform expensive electronic structure calculations~\cite{Blanco:2025sgv,Dreyer:2026bmz} for each new material.

To that end, we develop in this paper a set of symmetry-based metrics for identifying anisotropic candidate materials \emph{without} initially calculating the electronic structure, using only information about the geometry and discrete symmetries of a crystal.
Our first main result consolidates the 230 three-dimensional space groups into five \emph{quadrupole survival classes}, each of which has a different average degree of anisotropy in its spin-independent dark matter scattering rates. 
Translational symmetry reduces the 230 space groups to 32 point groups, while the evenness of the scalar squared response groups these into 11 Laue classes. Using properties of the partial rate matrix~\cite{Lillard:2023qlx}, we can rank the Laue classes according to how they act on the $\ell = 2$ spherical harmonics, which ultimately collapses them into our five quadrupole survival classes, $\mathsf{Q}_k$, for $k = 0, 1, 2, 3, 5$.
This result is not specific to molecular crystals, but can be applied more broadly to electronic systems that share the parity symmetry identified in Section~\ref{sec:oddell}.

Specialising to molecular crystals, we also conduct a comprehensive analysis of how crystal and molecular symmetries determine the scale of DM daily modulation signals in organic molecules. Specifically, we address how each of the 230 crystal space groups, together with the internal molecular symmetries of the individual molecules, acts to preserve or suppress the anisotropy of the individual molecular response. 
Using tools from group theory, linear algebra, and complex analysis, Section~\ref{sec:internal} derives a proxy for the residual anisotropy of the crystal form factor, assuming that the molecular excited states are not \emph{generally} aligned with the crystal lattice.
Figure~\ref{fig:unit_cell_schematic} illustrates the central physical problem: even when each molecule has a strongly anisotropic response, the differently-oriented molecular images in the unit cell can partially cancel that anisotropy in the crystal response. 
The more symmetric the crystal, the more likely it is that the preferred directions of the individual molecules are averaged away, severely dampening any daily modulation signal.
Our group-theoretic estimator quantifies how much anisotropy is typically lost this way, for any combination of molecular and crystal point group symmetries.

Section~\ref{sec:modulationProxy} converts this measure of anisotropy into an estimate of the fractional modulation signal amplitude in a dark matter direct detection experiment. 
Both methods depend solely on the discrete symmetries of the underlying molecule and crystal structure. 
For cases where the geometry of the molecule and its alignment within the crystal lattice are both known, Section~\ref{sec:coordproxy} derives a ``coordinate-aware estimator'' to further refine our anisotropy estimate.
The input data is still quite simple: in addition to the point groups for the molecule and the crystal, this final proxy uses a geometry-dependent tensor that is approximately the molecule's moments of inertia.
Despite the basic nature of these proxies, we find that the coordinate-aware estimator ($\Lambda_\text{coord}$) is directly proportional to the true modulation amplitude. Section~\ref{sec:demonstrations} demonstrates this correspondence  for twenty-five molecular crystals, using the quantum chemistry code \texttt{SCarFFF}~\cite{Blanco:2025sgv} and the dark matter rate calculation \texttt{vsdm}~\cite{Lillard:2023qlx,Lillard:2023cyy,Lillard:2025aim} to calculate the orientation-dependent scattering rate directly. For convenience, we also make these symmetry-based screening tools available in the pip-installable Python package \texttt{symmscreen}\footnote{\url{https://github.com/jdshergold/symmscreen}}, which computes the symmetry projectors and modulation signal survival estimators introduced in this work.

In summary, we present
% This classification leads naturally to 
a hierarchy of screening tools for new materials, based on the amount of available information. Symmetry alone provides simple design rules when the precise geometrical structure of a crystal is unknown, indicating which point groups are more likely to yield particularly anisotropic electronic structures.
In molecular crystals with known geometry, our coordinate-aware estimator can be used to rank molecular crystals \emph{before} performing the full suite of electronic structure and dark matter rate calculations.
With these tools in hand, we can begin our explorations of the landscape of possible materials with the materials that are most likely to be highly anisotropic.

%-----------------------------------------------------------------------------
\subsection{Scattering rates and form factors}
\label{sec:form_factors}
% -----------------------------------------------------------------------------
Our discussion begins with the scattering rate, which is ultimately the object that determines the sensitivity of a direct-detection experiment. 
For a transition from the ground state to a discrete excited state $s$ with transition energy $E_s$, the rate can be written as
\begin{align}\label{eq:rateFull}
R_s(t) &= N_T \rho_\chi \bar\sigma_e \int\! d^3v \,d^3q\, g_\chi(\vec v) \, \delta\!\left( E_s + \frac{q^2}{2 m_\chi} - \vec q\cdot\vec v\right) \frac{F_\text{DM}^2(q)}{4 \pi \mu_\chi^2 m_\chi} \mathcal D    (t) \cdot |f_s(\vec q)|^2
\end{align}
where $N_T$ is the number of microscopic scattering targets, $\rho_\chi$ and $m_\chi$ are the DM energy density and mass, respectively, $\bar{\sigma}_e$ is the free electron scattering cross-section, $g_\chi(\vec v)$ is the DM velocity distribution, and $\mu_\chi$ is the reduced mass of the combined DM-electron system. Here, $N_T$ may count either the number of unit cells in the crystal, or the number of molecules in the crystal, depending on how $|f_s(\vec q)|^2$ is normalised. The relative orientation of the detector-wind system is encoded by the time-dependent active rotation operator, $\mathcal{D}(t)$, which can be e.g.~be expressed as a $3\times 3$ matrix (see Appendix~\ref{app:rotationS}). In this work, we will consider DM form factors of the form
\begin{equation}\label{eq:dmFormFactor}
    F_\mathrm{DM}(q) = \left(\frac{\alpha_\mathrm{EM} m_e}{q}\right)^n,
\end{equation}
with $\alpha_\mathrm{EM}$ the fine-structure constant, $m_e$ the electron mass, and $n$ an integer. Specifically, we will consider both the $n = 0$ and $n = 2$ cases in our analyses, corresponding to heavy and light mediators, respectively.
In the examples of Section~\ref{sec:demonstrations} we use a Standard Halo Model (SHM) velocity distribution for $g_\chi(\vec v)$, but the precise form of $g_\chi$ is not important for our main results.

Key to understanding the effect of symmetries of the system is the molecular form factor
\begin{equation}
    f_s(\vec{q}) = \bra{\Psi_s}\tilde n_e(-\vec{q})\ket{\Psi_g},
    \label{eq:molecular_form_factor}
\end{equation}
with $\tilde{n}_e(-\vec{q})$ the Fourier transform of the electron density operator~\cite{Blanco:2019lrf,Blanco:2025sgv}, whilst $\ket{\Psi_g}$ and $\ket{\Psi_s}$ denote the molecular ground state $g$ and excited state $s$, respectively. This can alternatively be written in terms of the transition density, $\Phi^{g\to s}$, as
\begin{equation} \label{eq:fsFromTDM}
    f_s(\vec{q}) = \int d^3r \, \Phi^{g\to s}(\vec{r}) e^{i\vec{q} \cdot \vec{r}},
\end{equation}
which encodes the overlap between the initial and final state electron densities. The squared form factor determines the probability of exciting an electron when a momentum transfer $\vec{q}$ is deposited by \textit{e.g.} a photon, or DM particle, and importantly, encodes the entire material dependence of the scattering rate. 
In terms of the dynamic structure factor of~\cite{Trickle:2019nya}, the crystal form factor can also be written via the sum over indistinguishable final states: 
\begin{align} \label{eq:dynamicS}
S(\vec q, \omega) &= \frac{2\pi}{V_\text{cell} } \sum_{s} |f_s(\vec q)|^2 \delta(E_{ s} - \omega),
\end{align}
where $ E_{s}$ is the excitation energy, 
and $V_\text{cell} = (\vec a \times \vec b) \cdot \vec c$ is the volume of the crystal unit cell. 

Calculating the finite-momentum material response function $f_s^2(\vec q)$ or $S(\vec q, \omega)$ generally requires expensive electronic structure calculations~\cite{Blanco:2025sgv,Dreyer:2026bmz}. 
Previous methods for high-throughput material searches use
quantities such as the oscillator strength~\cite{Cook:2024cgm} or the dielectric tensor~\cite{Griffin:2025wew},
to approximate the momentum form factor by extrapolation from $|\vec q| \rightarrow 0$ data.
This can provide a good proxy for the brightness or anisotropy of a material, if its behaviours at $|\vec q| \sim \alpha m_e$ and $q \rightarrow 0$ are set by the same dynamics. 
Unfortunately, there are many examples (e.g.~\cite{Blanco:2021hlm}) where the finite-momentum response relevant for dark matter scattering cannot be predicted from the $q \approx 0$ behaviour of a material.

Fortunately, recent progress has made this problem far more tractable. The recently released $\texttt{SCarFFF}$ package~\cite{Blanco:2025sgv} computes the full three-dimensional form factor that encodes molecular anisotropy, using time-dependent density functional theory (TD-DFT) from \texttt{PySCF}~\cite{Sun:2018lpq,Sun:2020jul} for the electronic structure calculations, followed by three efficient, GPU-accelerated methods for turning these into the 3d form factor. With \texttt{SCarFFF}, a single molecule can be screened in $\sim 1\ \mathrm{minute}$, or a few seconds excluding the TD-DFT. These form factors can then be interfaced with \texttt{vsdm}~\cite{Lillard:2023qlx,Lillard:2023cyy,Lillard:2025aim}, which combines the form factors with arbitrary DM velocity distributions to compute scattering rates in seconds, whilst simultaneously optimising modulation signals over detector orientations.

Machine learning has also begun to play a key role in searches for ideal molecular scintillators~\cite{Cook:2024cgm}, due to its ability to ingest huge molecular datasets and approximate expensive electronic calculations, speeding up screening efforts by orders of magnitude. Nevertheless, both direct electronic-structure calculations and the machine learning models trained to approximate them remain expensive at scale, the latter requiring large, high-quality datasets before they become predictive.
This motivates simpler proxies that can identify promising candidates before either route is pursued in detail.

\subsection{Angular mode decomposition and the partial rate matrices}
The scattering rate as written in \eqref{eq:rateFull} is cumbersome to integrate, and must be re-evaluated for every combination of dark matter particle model, velocity distribution, material form factor, and detector orientation $\mathcal D \in SO(3)$. 
The partial rate matrix formalism of~\cite{Lillard:2023qlx} greatly reduces the computational difficulty, 
while also illuminating the interplay between the angular structures of the form factor and the velocity distribution. 
As the first step of the rate calculation, then, we 
% It is therefore far more natural to 
decompose the form factor and velocity distribution onto a basis of spherical harmonics,
\begin{align}
    |f_s(\vec{q})|^2 &= \sum_{\ell,m} f_{s,\ell m}^2(q) \,Y_{\ell m}(\hat q),
    &
    g_\chi(\vec v) &= \sum_{\ell ,m} g_{\ell m}(v) \, Y_{\ell m}(\hat v) ,
\end{align}
where $Y_{\ell m}$ are the \emph{real} spherical harmonics. 
This is a natural choice for any function that must later be rotated, as the real spherical harmonics transform as representations of $SO(3)$. 

To avoid notational overload, we will suppress the $s$ label on the $f_{s,\ell m}^2$ going forward. In places, we will also write this as a vector of length $2\ell +1$,
\begin{equation} \label{eq:emVector}
    \vec{f}_{\ell}^2(q) = \left(f_{\ell,-\ell}^2(q), \dots, f_{\ell,\ell}^2(q)\right)^T,
\end{equation}
to simplify the notation in many of our symmetry metrics.
Now, if the form factor transforms under the action of some (improper) rotation, $\mathcal{R} \in O(3)$, these coefficients transform as
\begin{equation}\label{eq:radialRotation}
    \vec{f}_{\ell}^2(q) \xlongrightarrow{\mathcal{R}} p_\ell \, G^{(\ell)} (\widetilde{\mathcal{R}})  \vec{f}_{\ell}^2(q),
\end{equation}
where $p_{\ell} = (\det \mathcal{R})^\ell$ and $\widetilde{\mathcal{R}} = \det(\mathcal{R}) \mathcal{R} \in SO(3)$, and $G^{(\ell)}$ are the Wigner-G matrices~\cite{Lillard:2023cyy}, the real analogues of the Wigner-D matrices~\cite{Lillard:2023cyy}. This result follows directly from how spherical harmonics transform under rotations, and under parity:
\begin{equation}
    Y_{\ell m}(\widetilde{\mathcal{R}}^{-1} \hat q) = \sum_{m'} G^{(\ell)}_{mm'}(\widetilde{\mathcal{R}}) Y_{\ell m'}(\hat q), \qquad Y_{\ell m}(-\hat{q}) = (-1)^\ell Y_{\ell m}(\hat q).
\end{equation}
We can apply a similar decomposition to the velocity distribution, $g_{\chi}(\vec{v})$, the coefficients of which we will denote $g_{\ell m}(v)$.

These coefficients, especially those at small $\ell$, have a clear physical interpretation. The $\ell = 0$ mode describes the spherically symmetric power of the function. For example, in the absence of a DM wind, or equivalently in the frame where the mean DM velocity is zero, only the $g_{00}$ mode will be non-zero. The $\ell = 1$, or dipole modes, describe the leading anisotropy of a function along each of the three Cartesian directions, $\hat{y}\  (m=-1)$, $\hat{z}\  (m=0)$, or $\hat{x}\  (m = 1)$. As such, for a DM wind along the $\hat z$ direction, we expect that the dominant anisotropic mode will be $g_{10}$, whilst $g_{1\pm 1}$ will vanish identically due to the azimuthal symmetry of the Standard Halo Model. Moving onto the quadrupole modes at $\ell = 2$, these describe the anisotropy of the function involving pairs of the Cartesian axes. The physical interpretation of higher $\ell$ modes becomes increasingly obscure, but each corresponds to angular features on progressively smaller scales. 

After performing the angular decomposition, the spin-independent scattering rate can be written in the far more revealing form
\begin{equation}
    R(t) = N_T \rho_\chi \bar\sigma_e \sum_{\ell} \sum_{m,m'} G^{(\ell)}_{mm'}(\mathcal{D}(t)) K_{mm'}^{(\ell)},
    \label{eq:rateGK}
\end{equation}
where $\mathcal{D}(t)$ is the time-dependent detector-wind orientation matrix, which is discussed at length in Appendix~\ref{app:rotationS} and $K^{(\ell)}$ 
are the \emph{partial rate matrices} of~\cite{Lillard:2023qlx}, 
\begin{equation}\label{eq:partialRateSingle}
    K^{(\ell)}_{mm'} = \int_0^\infty q\,dq\int_{v_\mathrm{min}(q)}^{v_\mathrm{max}} v\,dv \,P_\ell\!\left(\frac{v_\mathrm{min}(q)}{v}\right) g_{\ell m}(v) \frac{F^2_\mathrm{DM}(q,v)}{2\mu_\chi^2 m_\chi} f_{\ell m'}^2(q), 
\end{equation}
where $P_\ell$ denotes the Legendre polynomial of degree $\ell$, and where 
\begin{align} \label{eq:vmin}
v_\text{min}(q) &\equiv \frac{E_s}{q} + \frac{q}{2 m_\chi}. 
\end{align}
Written in this form, the angular structure of the rate is far clearer. Only like-$\ell$ modes of the DM wind and molecular form factor couple, whilst their directions, the $m$ modes, are mixed by the relative orientation matrix $G^{(\ell)}(\mathcal{D})$. The $\ell > 0$ components of a DM wind will therefore go completely undetected if the detector has no anisotropy of its own. 
Likewise, the individual anisotropic components of the dark matter velocity distribution, e.g.~$g_{\ell,  0}$ for $\ell = 1, 2, \ldots$, can only induce an anisotropic scattering rate if the detector material response has some support at the matching $\ell$.

This decomposition of the scattering rate into angular modes $R^{(\ell)}$, 
\begin{equation}
    R(t) = \sum_{\ell = 0}^\infty R^{(\ell)}(t), 
    \label{eq:rateEll}
\end{equation}
is particularly convenient for quantifying the suppression of the modulation amplitude. For smooth functions $g_\chi(\vec v)$ and $f_{s}^2(\vec q)$, we show in Sec.~\ref{sec:modulationProxy} that $\ell$ modes above some finite $\ell_0$ become exponentially suppressed, so the expansion in $\ell$ can be safely truncated.

\subsection{Extension to molecular crystals}\label{sec:crystalExtension}

When a molecule is embedded in a crystal, the relative orientations of multiple molecules in the unit cell may dampen the anisotropy that was present in the single-molecule form factor,  suppressing the modulation signal and the statistical power of an experiment.
A complete solid-state treatment should also include the effects of intermolecular interactions, and the periodic structure of the extended crystal system. However, for localised molecular excitations in crystals with weak van~der~Waals interactions between neighbouring molecules, we expect these effects to be subleading, and so we neglect them throughout. Under these assumptions, the squared form factor for the unit cell of the crystal can be written as the incoherent sum of the rotated one-molecule squared form factors, explicitly: 
\begin{equation}\label{eq:incoherentSum}
    |f_{\mathrm{uc}}(\vec{q})|^2 \simeq \sum_{i=1}^{Z} |f(\mathcal{R}_i^{-1} \hat q)|^2,
\end{equation}
where $Z$ is the number of molecules in the unit cell and $\mathcal{R}_i \in O(3)$ are the (improper) rotation matrices, which may include reflections, that map one molecule onto its images in the unit cell, up to translations. The only effect of translations is to introduce relative phases between the unsquared form factors, but this has no impact on the squared form factors or the scattering rate.

% {\color{red}{In practice...intermolecular effects *can* be important, and may distort the above picture. However, for the purposes of...the above...these cannot change the symmetries of the crystal. At the very least, the symmetry projection framework that we will develop remains exact at the crystal level. The main effect is to modify energy levels and transition dipoles...by mixing energy eigenstates of nearby levels into new crystal eigenstates...this is Davydov splitting.}}

In practice, intermolecular interactions can modify the molecular picture described above. Close packing, orbital overlap, and dipole-dipole coupling between neighbouring molecules can shift transition energies, redistribute oscillator strengths, and mix local molecular excitations into collective crystal eigenstates. This includes Davydov splitting, where nominally equivalent molecular transitions in the unit cell split into several crystal excitations with different transition dipoles and optical strengths. Such effects can change the detailed form factor, and in some cases may invalidate the approximation that the unit-cell response is an incoherent sum of isolated-molecule responses.

These effects, however, do not by themselves alter the crystallographic symmetry of the material, and so the full crystal response must still transform according to the same crystal point group. The symmetry-projection framework that we will develop here therefore remains exact at the level of the full crystal response. What becomes approximate is the use of isolated-molecule form factors as input to that projection. Strong intermolecular coupling should therefore be viewed as a regime in which the same symmetry framework should be applied to crystal-level electronic-structure calculations, rather than as a failure of the symmetry classification itself.

Decomposing the unit cell form factor into spherical harmonic modes,
\begin{equation}
    |f_{\mathrm{uc}}(\vec{q})|^2 = \sum_{\ell,m} f^2_{\mathrm{uc},\ell m}(q) Y_{\ell m}(\hat q),
\end{equation}
it follows immediately from \eqref{eq:radialRotation} that each coefficient $f^2_{\text{uc}, \ell m}$ for the unit cell can be rewritten in terms of the single-molecule coefficients:
\begin{equation}
    \vec{f}_\mathrm{uc,\ell}^2(q) = \left(\sum_{i=1}^{Z} p_{i,\ell}G^{(\ell)}(\widetilde{\mathcal{R}}_i)\right) \vec{f}_{\ell}^2(q).
\end{equation}
This can equivalently be written in terms of a projector matrix, $\Pi_{\ell}$, which acts to project the individual molecule form factor onto the unit cell form factor via 
\begin{equation}
     \vec{f}_\mathrm{uc,\ell}^2(q) = Z \Pi_{\ell}\vec{f}_{\ell}^2(q),
\end{equation}
where, explicitly
\begin{equation}\label{eq:projectorPhys}
    \Pi_\ell \equiv \frac{1}{Z} \sum_{i=1}^{Z}p_{i,\ell}G^{(\ell)}(\widetilde{\mathcal{R}}_i), \qquad \Pi_{\ell}^2 = \Pi_{\ell}.
\end{equation}
This will be the central object of our analysis. We can use these projection operators to determine the amount of anisotropy that survives in going from the individual molecular form factor to that of the crystal. However, the form given in \eqref{eq:projectorPhys} is not particularly enlightening for these purposes, as it is a coordinate-dependent construction and is not yet written in terms of the inherent symmetries of the crystal. In Sec.~\ref{sec:crystals}, we will reconstruct this operator from the point group symmetries of the crystal, and derive several coordinate-independent quantities that describe the level of anisotropy preservation.

\section{The death of odd-$\ell$ modes}\label{sec:oddell}
Before moving onto the group-theoretic analysis of molecular crystals, we would like to make a key point about the allowed angular modes of \emph{any} crystal: for a scalar, spin-independent crystal response, the odd-$\ell$ modes of $f_{\ell m}^2$ must vanish due to invariance of the squared form factor under $\vec{q} \to -\vec{q}$. We will prove the latter statement first. The squared form factor, written in terms of the transition density is
\begin{equation}
|f_s(\vec{q})|^2 = \int d^3x \int d^3x' \,\Phi^{g \to s}(\vec{x}) \Phi^{g \to s}(\vec{x}') e^{-i\vec{q}\cdot(\vec{x} - \vec{x}')}.
\end{equation}
where we have used the fact that, for a scalar, spin-independent electronic response with no additional time-reversal-breaking structure, the transition density may be chosen to be real. This is true \textit{e.g.} in the absence of an external magnetic field. We leave the exploration of these systems to future work. Now taking $\vec{q} \to -\vec{q}$, we have
\begin{equation}
    \begin{split}
    |f_s(-\vec{q})|^2 &= \int d^3x \int d^3x' \,\Phi^{g\to s}(\vec{x}) \Phi^{g\to s}(\vec{x}') e^{i\vec{q}\cdot(\vec{x} - \vec{x}')} \\
    &= \int d^3x' \int d^3x \,\Phi^{g\to s}(\vec{x'}) \Phi^{g\to s}(\vec{x}) e^{i\vec{q}\cdot(\vec{x}' - \vec{x})} \\
    &=  \int d^3x \int d^3x' \,\Phi^{g\to s}(\vec{x}) \Phi^{g\to s}(\vec{x}') e^{-i\vec{q}\cdot(\vec{x} - \vec{x}')} \\
    &= |f_s(\vec{q})|^2,
    \end{split}
\label{eq:fs2even}
\end{equation}
where in going from the first to the second line we relabelled the integral measures $\vec{x} \leftrightarrow \vec{x}'$.
    
The most important consequence of this is then that all odd-$\ell$ angular momentum modes vanish from the spherical harmonic coefficients. To see this, consider the expansions
\begin{align}
    |f(\vec{q})|^2 &= \sum_{\ell,m} f_{\ell m}^2 (q) Y_{\ell m}(\hat q), \\
    |f(-\vec{q})|^2 &= \sum_{\ell,m} f_{\ell m}^2 (q) (-1)^\ell Y_{\ell m}(\hat q).
\end{align}
    Equating the two expansions, and projecting out the $f_{\ell m}^2$ coefficients with a spherical harmonic, we arrive at
    \begin{equation}
        f_{\ell m}^2(q) = (-1)^\ell f_{\ell m}^2(q).
    \end{equation}
    The odd-$\ell$ modes of the squared form factor therefore \textit{always} vanish, irrespective of the underlying symmetries of the molecule. The vanishing of odd-$\ell$ modes is a general property of the squared Fourier transform of a real-valued function, and will also apply to more generic dynamic structure functions $S(\vec q, \omega)$ derived from a real-valued electronic transition density $n_e(\vec r)$. 
    
    This means that we should be a bit more careful in our symmetry measures, and not ``overweight'' or ``overnormalise'' them by modes that cannot possibly exist. Consequently, all symmetry-related measures in this paper use the sums
    \begin{equation}
        \sum_{\ell \in 2\mathbb{N}} \equiv \sum_{\ell = 2,4,\dots}^\infty, \qquad \sum_{\ell \in 2\mathbb{N}}^L \equiv \sum_{\ell = 2,4,\dots}^L
    \end{equation}
    with $L$ an even number, when considering their contribution to modulation signals.

    From a physics point of view, the vanishing of odd-$\ell$ modes has a simple intuition. The detector response is determined by the scalar electron-density response probed by the incident wave. In the absence of polarising fields, magnetic fields, or other time-reversal-breaking effects, this response has no inherent arrow. It may be spatially anisotropic, but it is reciprocal. That is to say, two waves with momentum transfers $\vec{q}$ and $-\vec{q}$ induce the same transition probability. As such, the squared form factor can distinguish axes, but not directions along those axes. This is the case, for example, for singlet--singlet transitions with scalar density coupling in the absence of applied magnetic fields.

    To make the distinction between axes and directions clear, consider a molecule with a preferred axis $\hat n$. This may be, for example, a planar, or highly axial molecule. The leading anisotropic scalar response has the schematic form
    \begin{equation}
        |f_s(\vec{q})|^2 = A_s(q) + B_s(q) |\hat{q} \cdot \hat n|^2 + \dots,
    \end{equation}
    which, importantly, depends strongly on the alignment of the momentum transfer with the preferred axis, but not on the direction along that axis. 
 
    The practical implication is that inversion symmetry in a crystal has no dampening effect on the spin-independent material response.  
    A centrosymmetric crystal removes all odd-$\ell$ modes, but these modes are already absent from $|f_s(\mathbf q)|^2$. Consequently, centrosymmetric crystals can still preserve a sizable modulation signal, provided that they retain the leading even-$\ell$ anisotropies, especially the quadrupolar ($\ell=2$) structure. In order for odd-$\ell$ modes to survive, we would require something that distinguishes $\mathbf q$ from $-\mathbf q$, such as the parity-odd spin-dependent operators of Ref.~\cite{Giffin:2025hdx}, or external magnetic fields. More generally, this can be anything that breaks time-reversal symmetry.

\section{Crystal space groups and anisotropy}\label{sec:crystals}

In the previous section we saw that, for the scalar spin-independent response considered in this work, the squared form factor contains only even-$\ell$ spherical harmonic modes $Y_{\ell m}$. We now ask which of these modes survive once the molecule is embedded in a crystal. The aim of this section is to rewrite the coordinate-level projector of \eqref{eq:projectorPhys} directly in terms of crystallographic point-group operations, and to use it to define simple, symmetry-only measures of anisotropy survival. 
Our conclusions in this section will apply not just to molecular crystals, but to more generic electronic excited state structures that are contained within a crystal lattice. 

\subsection{Point groups and Laue classes}

A crystal is defined here as a repeating structure that is invariant under translations of $\vec a$, $\vec b$, or $\vec c$, the three lattice vectors of the crystal. The unit cell is the smallest repeating element, with volume $(\vec a \times \vec b) \cdot \vec c$, and it may in principle contain any integer number of atoms or molecules. Appendix~\ref{sec:coords} reviews the conventions for representing generic crystal lattices by their unit cell lengths $(a, b, c)$ and opening angles $(\alpha, \beta, \gamma)$. 

In addition to translations of the unit cell, 
a crystal may also be invariant under a nontrivial space group, $S$, the elements of which are generally combinations of translations and rotations. Each operator $\mathcal S_j \in S$ acts on spatial positions $\vec r$ as:
\begin{align}
\mathcal S_j \cdot \vec r &= \mathcal R_j \vec r + \vec d_j .
\end{align}
Here $\mathcal R_j \in O(3)$ may include improper rotations, i.e.~the combination of a reflection (e.g.~central inversion, $\vec r \rightarrow - \vec r$) with a proper $SO(3)$ rotation. 
For ionic, molecular, or atomic crystals, for every atomic site $\vec r_a$ in the crystal lattice, 
its images under $S$
\begin{align}
\mathcal S_j \vec r_a = \vec r_{j, a}
\end{align}
identify other atomic sites in the lattice of the same type.

As already discussed in Sec.~\ref{sec:crystalExtension}, the translational component $\vec d_i$ has no effect on the scattering rate: it only introduces a relative phase between the molecular form factors $f(\vec q)$, which has no effect on the absolute value. We can therefore restrict our attention to the corresponding point group, $P(S)$, which consists only of the rotation operators. This effectively reduces the 230 possible crystal space groups to just 32 distinct point groups. 
Thus, the subset of distinct rotation operators $\mathcal R_i \in O(3)$ that appear in the space group $S$ are also representations of the discrete point group $P \subset O(3)$, satisfying the group axioms (e.g.~closure, $\mathcal R_i \mathcal R_j = \mathcal R_k \in P$). 

We can simplify the crystal space groups even further
by taking into account the results of Sec.~\ref{sec:oddell}. Since the odd-$\ell$ modes of the squared response vanish, only the point groups that differ in their action on the even-$\ell$ modes are distinct. This equivalence further narrows the classification from 32 point groups to just 11 distinct classes, known as the \textit{Laue classes}, which are shown in Table~\ref{tab:xiInfinityLaueGroups}. The Laue group, $\mathcal L(S)$, which represents the corresponding Laue class of each space group, $S$, is determined by adding inversion to each point group and then closing the group. For example, starting from the point group $C_{3h}=\{E,C_3,C_3^2,\sigma_h,S_3, S_3^5\}$, adjoining inversion $i$ generates the additional proper rotation $i\sigma_h=C_2$. The closed group therefore contains both $C_3$ and $C_2$, and hence the full $C_6$ axis, so that the Laue class is $\mathcal{L}(S) = C_{6h}\equiv 6/m$. We provide a discussion of the point groups, their operations, and notation in Appendix~\ref{app:point_groups}.

For parity-even quantities like the squared molecular form factor $|f_s(\vec q)|^2$, then, rather than summing over all molecules $i = 1 \ldots Z$ in the unit cell,    
we can define the projector $\Pi_\ell$ of \eqref{eq:projectorPhys} 
by instead summing over elements of the Laue group $\mathcal L$:
\begin{equation}\label{eq:crystalProjector}
    \Pi_\ell^{(\mathcal{L})} = \frac{1}{N_\mathcal{L}}\sum_{i \in \mathcal{L}(S)} p^{(\mathcal{L})}_{i,\ell} G^{(\ell)}(\widetilde{\mathcal{R}}_i)
    ~~\xrightarrow{\text{$\ell$ even}}~~
    \frac{1}{N_\mathcal{L}}\sum_{i \in \mathcal{L}(S)}  G^{(\ell)}(\widetilde{\mathcal{R}}_i) ,
\end{equation}
where $\widetilde{\mathcal R} \in SO(3)$ are proper rotations. 
For even values of $\ell$, recall that $p^{(\mathcal{L})}_{i,\ell} = 1$.  

Finally, we note that the rotation operators $\mathcal R_i$ do not depend on the lengths of the unit cell vectors, $(a, b, c)$. Specific representations of $\mathcal R_i$ (e.g.~as $3 \times 3$ matrices) are basis-dependent only to the extent that they depend on the orientations of the symmetry axes. This fact can be obscured in some presentations of the crystal geometry and space group. 
It is common, for example, to label positions within a unit cell using a nonorthogonal coordinate system based on the lattice vectors $(\vec a, \vec b, \vec c)$:
\begin{align} \label{eq:coordRot}
    \vec{r}_a &= A \vec u_a,
    &
    A &\equiv 
    \begin{pmatrix}
        a_x & b_x & c_x \\
        a_y & b_y & c_y \\
        a_z & b_z & c_z \\
    \end{pmatrix} ,
\end{align}
where $\vec u_a \equiv (u_a, v_a, w_a)$ is a unitless vector, and  the invertible matrix $A$ provides the mapping from $\vec u$ back to position space $\vec r$.
In the skewed lattice coordinate space $\vec u$, the representations of the space group $\mathcal U_i \in S$ act instead as
\begin{equation}\label{eq:oneSymop}
    \mathcal U_i \vec{u}_a = R_{i} \vec{u}_a + \boldsymbol{\tau}_i = \vec{u}_{i,a},
\end{equation}
where $R_i \in SL(3, \mathbb{R}) \times \mathbb{Z}_2$ are volume-preserving ($\det(R_i) = \pm1 $) but not necessarily orthogonal.
They are related to $\mathcal R_i \in O(3)$ via:
\begin{align} \label{eq:mcalR_A}
R_i &= A^{-1} \mathcal R_i A,
&
\mathcal R_i &= A R_i A^{-1}.
\end{align}
Despite being possible nonorthogonal, the $R_i$ are still equivalently valid representations of the point group $P$:
the mapping $\mathcal R_i \leftrightarrow R_i$ is bijective, 
and so
\begin{align}
R_i R_j &= A^{-1} \mathcal R_i A A^{-1} \mathcal R_j A = A^{-1} \mathcal R_k A = R_k
\end{align}
satisfies the point group multiplication rule.

For crystal geometries where $R_i$ and $A$ are provided, \eqref{eq:mcalR_A} provides a simple way to reconstruct the orthogonal matrices $\mathcal R_i \in O(3)$ that we require for evaluating $G^{(\ell)}_{m m'}(\mathcal R_i)$. 
We show in Appendix~\ref{app:invariances} that a valid set of $\mathcal R_i$ can be constructed directly from the skew-rotation operators $R_i$, without reference to any specific crystal geometry $A$. 

\subsection{Measuring anisotropy survival}\label{sec:survivalAnisotropy}

We now ask how to quantify the amount of molecular anisotropy that survives the crystal projection. A natural choice is to count the number of independent $f_{\ell m}^2$ combinations that remain after projection, as a fraction of the original $2\ell+1$ spherical harmonic modes for each $\ell$. The particular linear combinations that survive depend on the choice of crystal coordinates, but their number does not. We therefore define the fractional anisotropy survival parameter
\begin{equation}
    \kappa_\ell^{(\mathcal{L})} \equiv  \frac{\operatorname{rank}\!\left(\Pi_\ell^{(\mathcal{L})}\right)}{2\ell+1} = \frac{\operatorname{tr}\!\left(\Pi_\ell^{(\mathcal{L})}\right)}{2\ell+1} \in \left\{0,\frac{1}{2\ell+1},\ldots,\frac{2\ell}{2\ell+1},1\right\},
    \label{eq:kappaEllL}
\end{equation}
where the equality between rank and trace follows because $\Pi_\ell^{(\mathcal{L})}$ is a projector. Equivalently, $\kappa_\ell^{(\mathcal{L})}$ measures the fraction of the $\ell$-sector that remains invariant under the action of Laue group, $\mathcal{L}(S)$. %\BL{Traces are rotationally invariant.}
Note that because the trace of a matrix $\Pi_\ell$ is rotationally invariant, this $\kappa_\ell^{(\mathcal L)}$ is entirely basis-independent. 

Of course, the values of $f_{\ell m}^2(q)$ are not randomly distributed: they are determined by the electronic dynamics, and so it is possible that some of these coefficients may be accidentally small, even when they are not forbidden by a symmetry. However, the total scattering rate in the detector includes an integral over $q = |\vec q|$, as well as a sum over multiple final states $s$. It is reasonable to assume that most $f_{\ell m}^2$ modes that are not forbidden by a symmetry will be populated by at least some of the final states, which is why $\kappa_\ell^{(\mathcal{L})}$ is appropriate for quantifying the amount of anisotropy that is expected to survive crystallisation.

Larger values of $\ell$ correspond to smaller angular scales. One of the most powerful features of the partial rate matrix formulation in \eqref{eq:rateGK} is that the $f_{\ell m}^2(q)$ components of the material response couple only to the $g_{\ell' m'}(v)$ components of the velocity distribution with matching $\ell = \ell'$. Organic molecules often have strong features at $\ell \leq 6$, and an accurate picture of a material form factor may require $\ell \approx 12$ \cite{Blanco:2025sgv}. However, the Standard Halo Model velocity distribution does not have these small-angle features: instead, $g_{\ell m}(v)$ decreases dramatically for larger $\ell > 2$.
As a result, we find in Section~\ref{sec:demonstrations} that the contributions to $R(\mathcal D)$ from higher $\ell$ modes in \eqref{eq:rateEll} scale approximately as:
\begin{align}
R^{(\ell)}(\mathcal D) \sim \mathcal O(e^{- \ell / \ell_0}) \times R^{(0)} 
\end{align}
for some exponent $\ell_0 \sim 1$. 

As the odd-$\ell$ modes vanish for the squared response, the dominant anisotropic contribution is expected to come from the $\ell = 2$, or quadrupolar, modes. It is therefore particularly useful to group the Laue classes by the number of quadrupolar modes that they preserve. We refer to these groupings as the \emph{quadrupole survival classes},
\begin{equation}
    \mathsf{Q}_k, \qquad k = \operatorname{rank}\left(\Pi^{(\mathcal{L})}_2\right)
\end{equation}
For the 11 Laue classes, the allowed ranks are $k \in \{0,1,2,3,5\}$, such that there are just 5 quadrupole survival classes. As we will see in Sec.~\ref{sec:modulationProxy}, in most cases the quadrupole survival class alone is sufficient to determine the modulation signal loss due to symmetry. 

Notably, there is no $k = 4$ quadrupole class. The reason for this is simple. Quadrupole modes are built from the six combinations of axis pairs, $x^2, y^2, z^2, xy,xz,$ and $yz$, modulo the spherically symmetric combination $r^2 = x^2 + y^2 + z^2$, for a total of five independent quadrupole modes. At the same time, point groups are composed of mirror operations and rotations. A mirror operation takes \textit{e.g.} $x \to -x$, which changes at least two of the axis combinations, in this case $xy \to -xy$, $xz \to -xz$, leaving just three independent quadrupole modes. Similarly, a rotation must affect at least two axes, \textit{e.g.} the first application of a four-fold rotation about the $\hat z$-axis takes $x \to y$, $y \to -x$, again imposing two constraints and leaving just three independent quadrupole modes.

An additionally useful quantity is the fraction of \emph{all even angular modes} that survive up to a given $L$, which in the limit of equal weight assigned to each mode, gives a rough measure of the overall anisotropy-preserving quality of a Laue class. We define
\begin{equation}
    \xi_L^{(\mathcal{L})} = \frac{\sum_{\ell\in 2\mathbb{N}}^{L}\operatorname{tr}\!\left(\Pi_\ell^{(\mathcal{L})}\right)}{L(L+3)/2},
    \label{eq:totalSurvivalP}
\end{equation}
where
\begin{equation}
    \frac{L(L+3)}{2} = \sum_{\ell\in 2\mathbb{N}}^{L}(2\ell+1)
\end{equation}
is the total number of non-isotropic even-$\ell$ modes up to $L$.
In the $L\rightarrow\infty$ limit, this quantity has an exact expression after factoring the Laue group operations into the identity, $E$, inversion, $i$, and when present, all remaining
elements $\mathcal{L}'(S)$:
\begin{equation}
\begin{split}
    \xi_\infty^{(\mathcal{L})} &= \frac{1}{N_\mathcal{L}}\lim_{L\rightarrow\infty}\frac{1}{L(L+3)/2}\sum_{\ell\in 2\mathbb{N}}^{L}\sum_{i\in \mathcal{L}(S)} p_{i,\ell}^{(\mathcal{L})}\operatorname{tr}G^{(\ell)}(\widetilde{\mathcal R}_i) \\
    &= \frac{1}{N_\mathcal{L}}\lim_{L\rightarrow\infty}\left[2+\frac{1}{L(L+3)/2}\sum_{i\in \mathcal{L}'(S)}\sum_{\ell\in 2\mathbb{N}}^{L}\frac{\sin\!\left[(\ell+\frac{1}{2})\theta_i\right]}{\sin(\theta_i/2)}\right] \\
    &= \frac{2}{N_\mathcal{L}} .
\end{split}
\label{eq:xiInfty}
\end{equation}
In the second line we used $p_{i,\ell}^{(\mathcal{L})}=1$ for even $\ell$, and, for a nontrivial proper rotation through angle $\theta_i$,
\begin{equation}
    \operatorname{tr} G^{(\ell)}(\mathcal R_i) = \frac{\sin\left[(\ell+\frac{1}{2})\theta_i\right]}{\sin(\theta_i/2)}.
\end{equation}
For fixed non-zero $\theta_i$, the sine ratio remains bounded on $[-1,1]$ as $\ell$ increases, so that its contribution to the sum grows at most linearly with $L$, whereas the total number of even modes grows as $L^2$. As such, the contributions to $\xi_\infty^{(\ell)}$ from every nontrivial rotation vanish as $L \to \infty$.

Table~\ref{tab:xiInfinityLaueGroups} provides the values of $\xi_\infty^{(\mathcal{L})}$ for the 11 Laue groups,
% are given in Table~\ref{tab:xiInfinityLaueGroups}, 
along with their grouping into quadrupole survival classes, and their corresponding crystallographic point and space groups. We also show the relative frequency of each Laue group according to~\cite{CCDCSpaceGroupStats2026} in the final column. The three most common Laue classes, $2/m$, $\bar 1$, and $mmm$, together account for $94.48\%$ of the reported structures and have asymptotic survival fractions $\xi_\infty^{(\mathcal{L})}=1/2$, $1$, and $1/4$, respectively. This demonstrates that for the overwhelming majority of molecular crystals, the crystal structure does not catastrophically wash-out the modulation signal. On the other hand, for the small minority whose Laue group is cubic or hexagonal, the many symmetries tend to overwhelm any molecular anisotropies. 

\begin{table}
    \centering
    \setlength{\tabcolsep}{3.5pt}
    \renewcommand{\arraystretch}{1.08}
    \begin{adjustbox}{max width=\textwidth}
    \small
    \begin{tabular}{c|c|c|c|c|c}
    Survival class & Laue group & Point groups & Space group numbers & $\xi^{(\mathcal{L})}_\infty$ & CSD frequency \\
    \hline\hline
    \multirow[c]{2}{*}{$\mathsf{Q}_5$} & \textit{Triclinic} & \multirow[c]{2}{*}{$1$ $(C_1)$, $\bar{1}$ $(C_i)$} & \multirow[c]{2}{*}{$1$--$2$} & \multirow[c]{2}{*}{$1$} & \multirow[c]{2}{*}{$26.26\%$} \\
    & $\bar{1}$ $(C_i)$ & & & & \\
    \hline
    \multirow[c]{2}{*}{$\mathsf{Q}_3$} & \textit{Monoclinic} & $2$ $(C_2)$, $m$ $(C_s)$, & \multirow[c]{2}{*}{$3$--$15$} & \multirow[c]{2}{*}{$1/2$} & \multirow[c]{2}{*}{$51.32\%$} \\
    & $2/m$ $(C_{2h})$ & $2/m$ $(C_{2h})$ & & & \\
    \hline
    \multirow[c]{2}{*}{$\mathsf{Q}_2$} & \textit{Orthorhombic} & $222$ $(D_2)$, $mm2$ $(C_{2v})$, & \multirow[c]{2}{*}{$16$--$74$} & \multirow[c]{2}{*}{$1/4$} & \multirow[c]{2}{*}{$16.90\%$} \\
    & $mmm$ $(D_{2h})$ & $mmm$ $(D_{2h})$ & & & \\
    \hline
    \multirow[c]{12}{*}{$\mathsf{Q}_1$} & \textit{Tetragonal} & $4$ $(C_4)$, $\bar{4}$ $(S_4)$, & \multirow[c]{2}{*}{$75$--$88$} & \multirow[c]{2}{*}{$1/4$} & \multirow[c]{2}{*}{$1.05\%$} \\
    & $4/m$ $(C_{4h})$ & $4/m$ $(C_{4h})$ & & & \\
    \cline{2-6}
    & \textit{Tetragonal} & $422$ $(D_4)$, $4mm$ $(C_{4v})$, & \multirow[c]{2}{*}{$89$--$142$} & \multirow[c]{2}{*}{$1/8$} & \multirow[c]{2}{*}{$1.17\%$} \\
    & $4/mmm$ $(D_{4h})$ & $\bar{4}2m$ $(D_{2d})$, $4/mmm$ $(D_{4h})$ & & & \\
    \cline{2-6}
    & \textit{Trigonal} & \multirow[c]{2}{*}{$3$ $(C_3)$, $\bar{3}$ $(C_{3i})$} & \multirow[c]{2}{*}{$143$--$148$} & \multirow[c]{2}{*}{$1/3$} & \multirow[c]{2}{*}{$1.24\%$} \\
    & $\bar{3}$ $(C_{3i})$ & & & & \\
    \cline{2-6}
    & \textit{Trigonal} & $32$ $(D_3)$, $3m$ $(C_{3v})$, & \multirow[c]{2}{*}{$149$--$167$} & \multirow[c]{2}{*}{$1/6$} & \multirow[c]{2}{*}{$0.79\%$} \\
    & $\bar{3}m$ $(D_{3d})$ & $\bar{3}m$ $(D_{3d})$ & & & \\
    \cline{2-6}
    & \textit{Hexagonal} & $6$ $(C_6)$, $\bar{6}$ $(C_{3h})$, & \multirow[c]{2}{*}{$168$--$176$} & \multirow[c]{2}{*}{$1/6$} & \multirow[c]{2}{*}{$0.35\%$} \\
    & $6/m$ $(C_{6h})$ & $6/m$ $(C_{6h})$ & & & \\
    \cline{2-6}
    & \textit{Hexagonal} & $622$ $(D_6)$, $6mm$ $(C_{6v})$, & \multirow[c]{2}{*}{$177$--$194$} & \multirow[c]{2}{*}{$1/12$} & \multirow[c]{2}{*}{$0.23\%$} \\
    & $6/mmm$ $(D_{6h})$ & $\bar{6}m2$ $(D_{3h})$, $6/mmm$ 
    $(D_{6h})$ & & & \\
    \hline
    \multirow[c]{4}{*}{$\mathsf{Q}_0$} & \textit{Cubic} & \multirow[c]{2}{*}{$23$ $(T)$, $m\bar{3}$ $(T_h)$} & \multirow[c]{2}{*}{$195$--$206$} & \multirow[c]{2}{*}{$1/12$} & \multirow[c]{2}{*}{$0.26\%$} \\
    & $m\bar{3}$ $(T_h)$ & & & & \\
    \cline{2-6}
    & \textit{Cubic} & $432$ $(O)$, $\bar{4}3m$ $(T_d)$, & \multirow[c]{2}{*}{$207$--$230$} & \multirow[c]{2}{*}{$1/24$} & \multirow[c]{2}{*}{$0.42\%$} \\
    & $m\bar{3}m$ $(O_h)$ & $m\bar{3}m$ $(O_h)$ & & & \\
    \hline
    \end{tabular}
    \end{adjustbox}
    \caption{Asymptotic angular-mode survival fractions grouped by quadrupole survival class, $\mathsf{Q}_k$. The second column gives the representative Laue group, $\mathcal{L}(S)$, in Hermann--Mauguin notation, with the corresponding centrosymmetric Schoenflies symbol in parentheses. These are further subdivided into the 32 crystallographic point groups in the third column, and into the 230 space groups in the fourth column. The CSD frequencies are obtained by summing the 2026 CCDC space group counts~\cite{CCDCSpaceGroupStats2026} over the corresponding point-group ranges and normalising by the total number of entries. 
    Additionally, liquid crystals with a single symmetry axis (e.g.~materials polarised by a constant external electric field) would be included in $\mathsf{Q}_1$.}
    \label{tab:xiInfinityLaueGroups}
\end{table}

A strong suppression of the daily modulation signal can be readily seen in other searches where cubic symmetries are present, \textit{e.g.}~\cite{Dinmohammadi:2023amy,Stratman:2024sng}. Here the octahedral symmetry groups of the silicon and germanium crystals, or the tetrahedral symmetry of gallium arsenide, place them in the $\mathsf{Q}_0$ class and forbid any quadrupole anisotropies.
Consequently the typical modulation amplitude $f_\text{RMS}$ is negligible for most dark matter masses~\cite{Stratman:2024sng}.
Close to the kinematic threshold the remaining $\ell = 4$ anisotropy can give rise to $f_\mathrm{RMS} \sim 1$-$2\%$~\cite{Dinmohammadi:2023amy}, 
but the modulation signal becomes strongly dependent on the sharp cutoff at $|\vec v - \vec v_E| \leq v_\text{esc}$ in the SHM velocity distribution, and the overall scattering rate is greatly suppressed.

We expect that this situation is typical for atomic crystals, for which cubic space group symmetries are far more prevalent. By contrast, we will demonstrate in Sec.~\ref{sec:demonstrations} that molecular crystals with surviving quadrupole modes typically have modulation signal fractions of $f_\mathrm{RMS} \simeq 5$-$15\%$, depending on the mediator mass, and can be $20\%$ or more for some exceptional systems. Importantly, we will also show that this behaviour persists far from kinematic thresholds, and is therefore not particularly sensitive to the details of the galactic escape velocity.

Using simple properties of discrete groups, we have identified three measures of anisotropy for electronic states in crystals.
In dark matter direct detection experiments with an SHM velocity distribution, the quadrupole survival class $\mathsf{Q}_k$ is the most important quantity, as it gives the immediate estimate the anisotropy survival fraction for each Laue class:
\begin{align}
\kappa_{\ell = 2}^{(\mathcal L)} = \frac{k}{5}. 
\end{align}
For the tetrahedral and octahedral point groups of $\mathsf{Q}_0$ with $R^{(\ell = 2)} = 0$, the leading anisotropy comes instead from the $\ell = 4$ part of the electronic structure.
In this case, our second proxy $\xi_{L = 4}^{(\mathcal L)}$ counts the number of missing angular modes; however, it does not account for the relative suppression of the $\ell = 4$ mode in the SHM. 
Section~\ref{sec:modulationProxy} addresses this point, by deriving an appropriate $\ell$-dependent weighting function.
Finally, for systems with much smaller-scale angular features in $g_\chi(\vec v)$, we use the asymptotic limit $\xi_{\infty}^{(\mathcal L)}$ to quantify the surviving anisotropy. 
While $\ell \gg 4$ modes are irrelevant for SHM velocity distributions, $\xi_\infty$ may be useful for studies of boosted subcomponents of the local dark matter distribution: for example, solar-reflected dark matter~\cite{An:2021qdl}, which would be approximately collimated in a terrestrial laboratory. 

Our categorisation of materials according to their quadrupole survival class works even for materials with rotational order, such as liquid crystals,  that do not follow the rigid definition of ``crystal'' provided at the beginning of this section. For example, an otherwise isotropic material can be polarised by applying an external electric field $\vec E$, aligning the dipole moments of the microscopic constituents with $\vec E$ while permitting continuous rotations about the $\vec E$ axis. The point group in this case is $SO(2)$, and its quadrupole survival class is $\mathsf{Q}_1$.  

\section{Molecular crystals}\label{sec:internal}

    Our derivations of $\kappa^{(\mathcal{L})}_\ell$ and $\xi^{(\mathcal{L})}_\infty$ as measures of the crystal anisotropy assume that the form factor $|f_s(\vec q)|^2$ can contain all angular modes, or at least all modes that are even under $\vec q \rightarrow - \vec q$. Of course, this is not always the case: if a molecule is itself symmetric under a non-trivial point group, which we denote $H$, then there exists an ``internal'' coordinate system where some additional subset of the $(\ell, m)$ spherical harmonic modes vanish. In other cases, a molecule that is not strictly symmetric under $H$ may still have an electronic structure that closely approximates it. Either case can modify the story presented in Sec.~\ref{sec:crystals}: if there is significant overlap between the internal symmetries of the form factor and those of the crystal, then crystallisation does not cause much reduction in the anisotropy.
    This should not be interpreted as saying that more symmetric molecules are intrinsically better targets for modulation searches. 
    Rather, the point is that if those modes are already absent from the molecule, a crystal symmetry that is degenerate with some internal symmetry does not further suppress the modulation signal. 

    To quantify the anisotropy of a molecular crystal, then, we must take the crystal symmetries, internal molecular symmetries, and their relative alignment into account.
    We begin by noting that a molecule that is invariant under the symmetries of a point group $H$ will have $f_{\ell m}^2(q)$ coefficients that satisfy
    \begin{equation}\label{eq:internalProjection}
        \Pi_{\ell}^{(H)} \vec{f}^{2,\mathrm{int}}_\ell(q) = \vec{f}^{2,\mathrm{int}}_\ell(q),
    \end{equation}
    where the projection operator is constructed in an analogous way to \eqref{eq:crystalProjector}:
    \begin{equation}
        \Pi_{\ell}^{(H)} = \frac{1}{N_H} \sum_{i \in H} p^{(H)}_{i,\ell} G^{(\ell)}(\widetilde{h}_i), \qquad p^{(H)}_{i,\ell} = (\det h_i)^\ell, \qquad 
        \widetilde{h}_i \equiv \det(h_i) h_i.
    \end{equation}
    Here $N_H$ is the number of symmetry operations in $H$, $h_i \in O(3)$ is the matrix representation of the $i^\text{th}$ symmetry operation of the group. We use the superscript $\mathrm{int}$ to denote that the coefficients in  \eqref{eq:internalProjection} are those in internal coordinates, $\vec{x}_a$, where the symmetry operations $H$ hold. 

\begin{figure}[t]
    \centering
    \includegraphics[width=\linewidth]{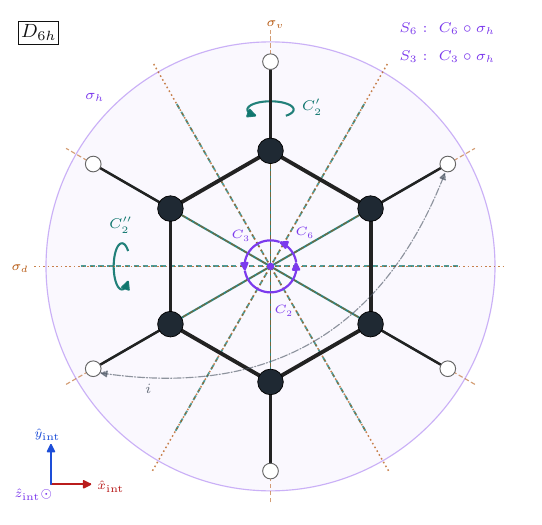}
    \caption{Internal symmetries of a benzene molecule, which is invariant under the actions of the $D_{6h}$ point group. This is composed of central inversion ($\vec x \rightarrow - \vec x$), denoted $i$, alongside several rotations, reflections, and rotoreflections. The rotations, denoted $C_n$, include $60^\circ$, $120^\circ$, and $180^\circ$ in-plane rotations about the centred $\hat{z}_\mathrm{int}$-axis, of $180^{\circ}$ rotation along each of the $H$-$C$ bond axes, and one about each of the axes bisecting each $C$-$C$ bond. The reflection operations, denoted $\sigma$, consist of the reflection across the molecular plane, $\sigma_h$, three vertical planes including the $H$-$C$ bonds, $\sigma_v$, and three which bisect the $C$-$C$ bonds, $\sigma_d$. Finally, there are two sets of rotoreflections, denoted $S_n$, which rotate the molecule about the principal axis, followed by a reflection in $\sigma_h$.}
    \label{fig:benzene_d6h}
\end{figure}

    Figure~\ref{fig:benzene_d6h} shows the example of a benzene molecule in the $x$--$y$ plane, which is invariant under the action of the point group $H = 6/mmm\,(D_{6h})$. For this specific embedding, with all atoms in the $x$--$y$ plane, $H$ contains several rotations about the principal $z$-axis by multiples of $60^\circ$; six $180^\circ$ degree rotations about perpendicular axes lying in the molecular plane, which itself is a mirror plane; central inversion; six mirror planes containing the $z$-axis; and finally two distinct sets of rotoreflections, i.e.~rotations about the principal $z$-axis followed by a reflection in the molecular $x$--$y$ plane. 

In this section we model the electronic structure of a crystal as an incoherent sum over the excited states on each molecule in the unit cell, following \eqref{eq:incoherentSum}. For the reasons discussed in Sec.~\ref{sec:crystalExtension}, the full crystal treatment may slightly distort this picture due to the effects of intermolecular interactions. Unlike the crystallographic point group, however, the molecular symmetry \textit{does not} remain exact under these perturbations. However, when full crystal effects are small, the molecular symmetry is only softly broken, and still approximately holds.

Figure~\ref{fig:density_slices} shows the full crystal transition density for the first excited states of benzene and 1,3,5-trichlorobenzene (1,3,5-TCB), as an isocontour at $38\%$ of the maximum value, defined by
\begin{equation}\label{eq:transition_density}
    \Phi^{g \rightarrow s}(\vec r) = \sqrt{2}\sum_{\mu,\nu} \phi _\mu^*(\vec{r}) T_{\mu\nu}^{(s)} \phi_\nu(\vec{r}),
\end{equation}
with $\phi_\mu$ the $\mu^\mathrm{th}$ molecular orbital, and $T_{\mu\nu}^{(s)}$ the transition density matrix coupling orbital $\mu$ to $\nu$ for excited state $s$. These are computed using \texttt{PySCF}~\cite{Sun:2017pyscf,Sun:2018lpq,Sun:2020jul,Sun:2020pyscf} with periodic boundary conditions, using the PBE0-D3BJ exchange-correlation functional, GTH-DZVP-MOLOPT-SR basis set, and the GTH-PBE effective core potential. We also define the intermolecular density, $\Phi^{g \to s}_\mathrm{inter}$, according to~\eqref{eq:transition_density}, with the transition density matrix containing only terms coupling orbitals from different molecules. For each molecule, we also show both the full and intermolecular transition density in the plane containing the maximum amount of intermolecular transition density, that is, the plane maximising the integral
\begin{equation}\label{eq:intermolecularDensity}
    \int |\Phi^{g \to s}_\mathrm{inter}(\vec{r})|\,dA.
\end{equation}
For both molecules, it is clear that the effects of intermolecular interactions are incredibly small, such that our incoherent sum approximation holds. For benzene, the maximum intermolecular transition density is just $1.58\%$ of the maximum total transition density, falling to $0.52\%$ for 1,3,5-TCB.
\clearpage
\begin{figure}[t]
    \centering
    \includegraphics[width=\linewidth,trim={0 0.8cm 0 0},clip]{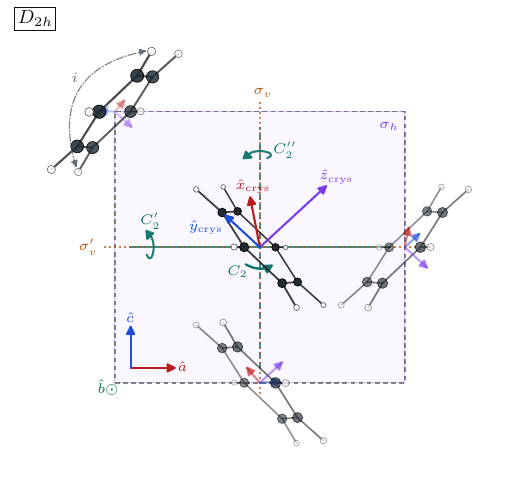}
    \caption{Point group symmetries of the benzene unit cell as viewed along the $\hat{b}$ axis, which is invariant under the actions of the $D_{2h}$ point group. This point group is composed of inversion, denoted $i$, alongside several rotations, and reflections. The rotations, each denoted $C_2$, are by $180^\circ$ about each of the three crystal axes, $\hat{a},\hat{b}$, and $\hat{c}$. The reflection planes, $\sigma_v$ and $\sigma_h$, bisect the unit cell along each axis, whilst the last remaining symmetry, inversion, is exactly degenerate with the internal inversion symmetry of benzene, mapping each atom onto its internal parity-partner. The rotation matrix, $\Qrot$, can be any of those that map the internal axes, $(\hat{x}_\mathrm{int},\hat{y}_\mathrm{int}, \hat{z}_\mathrm{int})$ from Figure~\ref{fig:benzene_d6h} onto one of the four crystal axis sets, $(\hat{x}_\mathrm{crys},\hat{y}_\mathrm{crys}, \hat{z}_\mathrm{crys})$ shown here. To simulate depth, atoms further along the $\hat{b}$ axis have a lower opacity, and are smaller than those at small $b$. The experimentally determined atom coordinates and symmetry group used here are taken from~\cite{Katrusiak:2010benzene,CCDC:757059}.}
    \label{fig:benzene_unit_cell_symmetry}
\end{figure}
\clearpage

\begin{figure}
    \centering
    \includegraphics[width=\linewidth]{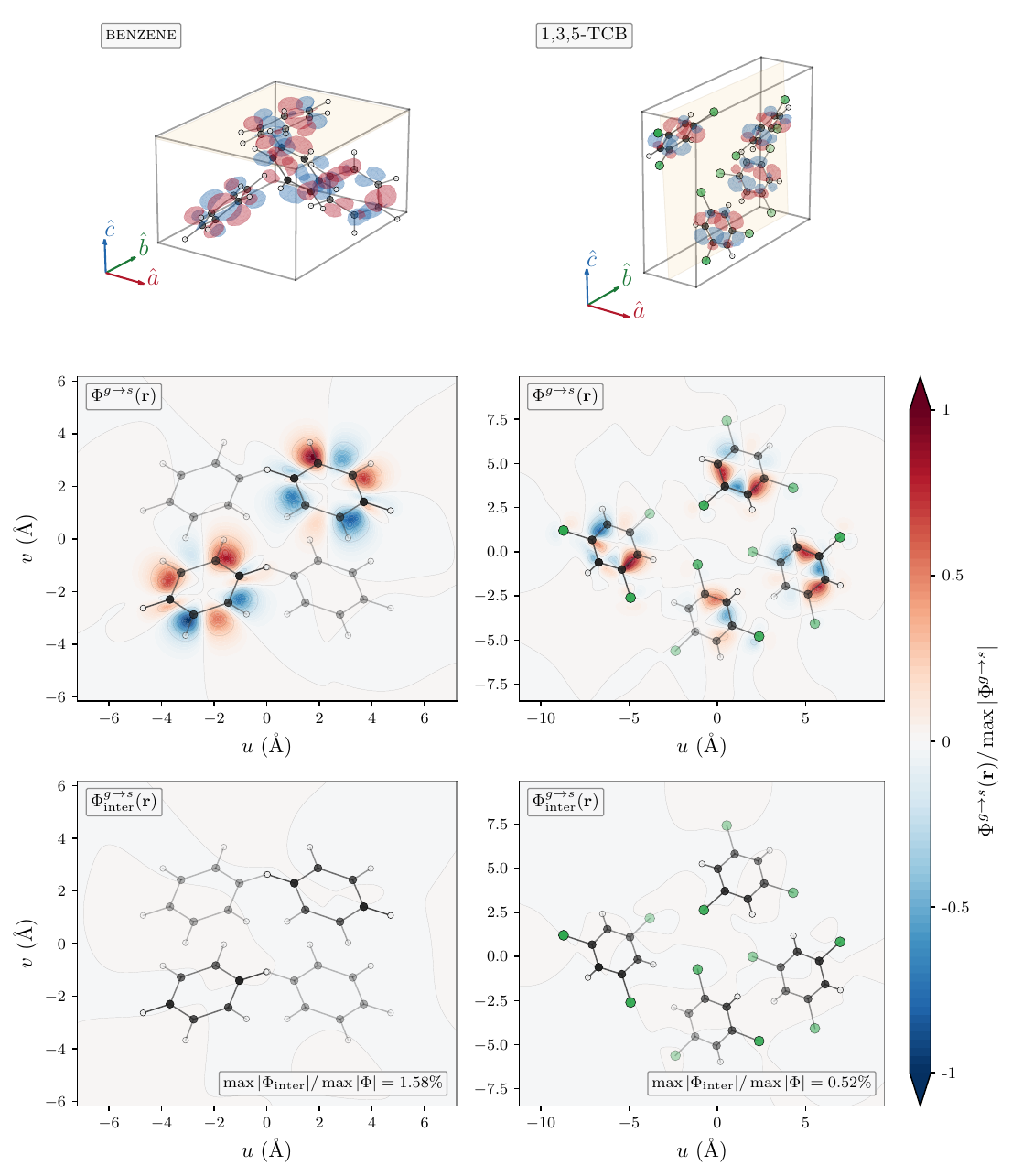}
    \caption{Full crystal transition density, $\Phi^{g \to s}(\vec{r})$, for benzene (left panels) and 1,3,5-trichlorobenzene (right panels), computing using periodic TD-DFT. The top row shows the full unit cell, along with the transition density isocontour at $38\%$ of maximum value, and the plane containing the most intermolecular transition density, as defined in~\eqref{eq:intermolecularDensity}. The second and third rows show, in turn, the full transition density in this plane, and the intermolecular transition density. In both cases, it is clear that the fractional intermolecular contribution is incredibly small. The crystal coordinates used for these computations are taken from~\cite{Woinska:2016CCDCcc1jspdd, Woinska:2016HydrogenAtoms,Hursthouse:2003CSDCommunication}.}
    \label{fig:density_slices}
\end{figure}
\clearpage
\noindent In both cases, it is also clear that the internal molecular symmetries are extremely well preserved. We can also estimate the fractional effect of the intermolecular interactions by computing their contribution to the total transition density
\begin{equation}
    \frac{||\Phi^{g\to s}_\mathrm{inter}||_2}{||\Phi^{g\to s}||_2} \simeq \sqrt{\frac{\int d^3r \, \left(\Phi_\mathrm{inter}^{g \to s}(\vec{r})\right)^2}{\int d^3r \, \left(\Phi^{g \to s}(\vec{r})\right)^2}}.
\end{equation}
For benzene, this ratio is $1.75\%$, whilst for 1,3,5-TCB it is $0.50\%$. More generally, we demonstrate in Appendix~\ref{app:degenerateLevels} that when the intermolecular transition density is small, coherent mixing between nearby crystal excitations only gives a correspondingly small correction to the summed form factor.
\subsection{Combined anisotropy survival}

The internal coordinates of a molecule need not align with the planes of symmetry of the crystal: generally, they will be related by some rotation $\mathcal{T} \in SO(3)$ satisfying
\begin{equation}
    \Qrot \, \vec{r}_{\mathrm{int,a}} = (\vec{r}_{\mathrm{crys,a}} -\bar{\vec{r}}_\mathrm{crys}),
\end{equation}
with $\bar{\vec{r}}_\mathrm{crys}$ the centroid of the molecular coordinates in the crystal frame. 
    
Figure~\ref{fig:benzene_unit_cell_symmetry} shows the corresponding embedding of benzene within its experimentally realised crystal unit cell, viewed along the crystallographic $\hat b$ axis. Although the isolated molecule has internal $D_{6h}$ symmetry, the crystal is invariant only under the lower-symmetry $D_{2h}$ point group, generated by three mutually perpendicular $C_2$ rotations, reflection planes, and central inversion. The four molecular orientations in the unit cell are related by these crystal operations, while the rotation $\Qrot$ maps the internal molecular axes $(\hat x_{\rm int},\hat y_{\rm int},\hat z_{\rm int})$ onto the corresponding axes of a chosen molecular image in the crystal frame. This distinction between the internal molecular symmetry and its realised orientation within the crystal must be taken into account in order to estimate the combined molecule-crystal anisotropy survival.
    
    The crystal frame form factor coefficients are related to those in internal coordinates by
    \begin{equation}
        \begin{split}
        \vec{f}^2_\ell(q) &= G^{(\ell)}(\Qrot) \vec{f}^{2,\mathrm{int}}_{\ell}(q) \\
        &= G^{(\ell)}(\Qrot) \Pi_\ell^{(H)} G^{(\ell)}(\Qrot^{-1}) \vec{f}^2_\ell(q),
        \end{split}
    \end{equation}
    from which it follows that the natural projector acting on $\vec{f}^2_{\ell}$ that correctly takes into account the internal symmetries of the molecule, as well as the relative orientation with respect to the crystal frame, is:
    \begin{equation}
        \Pi_{\ell,\Qrot}^{(H)} = G^{(\ell)}(\Qrot) \,\Pi_\ell^{(H)} \, G^{(\ell)}(\Qrot^{-1}) = \frac{1}{N_H} \sum_{i\in H} p^{(H)}_{i,\ell} G^{(\ell)}\left(\Qrot \widetilde{h}_i \Qrot^{-1}\right).
    \end{equation}
    Combining this with the point group projector, we find the combined anisotropy survival operator
    \begin{equation} \label{eq:mcalC}
        \mathcal{C}^{(\ell)}\!\left(\mathcal{L}(S),\Qrot,H\right) = \Pi_\ell^{(\mathcal{L})} \Pi_{\ell,\Qrot}^{(H)} = \frac{1}{N_\mathcal{L} N_H} \sum_{i \in \mathcal{L}(S)} \sum_{j \in H} \mathcal{P}_{i,\ell} G^{(\ell)}\left(\widetilde{\mathcal{R}}_i \Qrot \widetilde{h}_j \Qrot^{-1}\right),
    \end{equation}
    where $\mathcal{P}_{i,\ell} = \det(\mathcal{R}_i h_i)^\ell$. 
Because the combined survival operator $\mathcal C^{(\ell)}$ is \emph{not} generally a projector, it is the trace rather than the rank of $\mathcal C^{(\ell)}$ that captures the anisotropy survival fraction.
For molecular crystals, then, we modify $\kappa_\ell^{(\mathcal L)}$ of the crystal Laue group to include the orientation $\Qrot$ and the point group $H$ of the molecule:
\begin{equation}\label{eq:kappaFull}
    \kappa_\ell(\Qrot) = \frac{\mathrm{tr}\!\left[\mathcal{C}^{(\ell)}(\mathcal L, \Qrot, H)\right]}{2\ell+1} \in [0,1].
\end{equation}

Similarly, in analogy with \eqref{eq:totalSurvivalP}, we can define the combined survival fraction up to a given maximum harmonic mode $\ell \leq L$  as
    \begin{equation}
        \xi_L(\Qrot) = \frac{\sum_{\ell \in 2\mathbb{N}}^L \mathrm{tr}(\mathcal{C}^{(\ell)})}{L(L+3)/2}.
    \end{equation}
Rather than taking a single value for each type of point group, the asymptotic value of $\xi_L(\Qrot)$ now depends on the relative rotation between the molecule and the crystal lattice, making it a material-specific quantity. While we can no longer derive an exact value for $\xi_\infty(\Qrot)$, we can derive upper and lower bounds based on the symmetry groups $\mathcal L$ and $H$. 
The total number of angular modes surviving the combined projection cannot exceed the number surviving either of the projectors individually.
The minimum bound, on the other hand, is saturated if the angular modes removed by $\xi_\infty^{(\mathcal L)}$ are perfectly complementary to those removed by $\xi_\infty^{(H)}$.
Thus,
\begin{equation}
    \xi_{\infty}^{(\mathcal{L})} \xi_\infty^{(H)} \leq \xi_\infty(\Qrot) \leq \min\left(\xi_\infty^{(\mathcal{L})},\xi_\infty^{(H)}\right).
\label{eq:xiInftyRange}
\end{equation}
Saturating the upper bound requires $\Qrot$ to take special values, e.g.~$\Qrot \simeq \mathbb{I}$.

Returning to our previous example of benzene, the combination of the internal $D_{6h}$ and crystal $D_{2h}$ symmetries yields $1/48 \leq \xi_{\infty} \leq 1/12$, unsurprisingly suggesting that highly symmetric molecules such as benzene are not particularly well suited to modulation searches. On the other hand, a far less symmetric molecule such as 1,3,5-triphenylbenzene (1,3,5-TPB), with internal $D_3$ and crystal $C_{2v}$ symmetries instead has $1/24 \leq \xi_{\infty}(\Qrot) \leq 1/6$, if we were to use the gas-phase $D_3$ molecular symmetry.
    
For 1,3,5-TPB, however, there is a subtlety. In its crystalline form, intermolecular interactions modify the geometry of the individual molecules within the unit cell, by rotating one of the phenyl groups relative to the others. This breaks the $D_3$ symmetry group of a gas-phase 1,3,5-TPB molecule down to the much smaller $C_2$ group, and consequently sets the bound $1/8 \leq \xi_{\infty}(\Qrot) \leq 1/4$, which preserves far more anisotropy. In this way, crystallisation can sometimes augment rather than suppress the modulation signal. In other cases, crystallisation may create new internal symmetries---for example, by flattening an otherwise asymmetric molecule, and in so doing adding a $\sigma_h$ mirror plane. 

\begin{figure}[t]
    \centering
    \begin{subfigure}{0.495\linewidth}
        \centering
        \includegraphics[width=\linewidth]{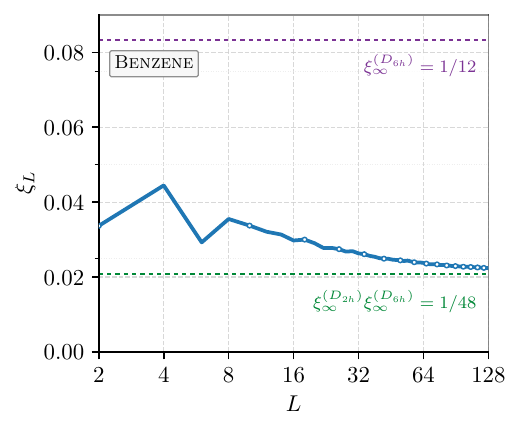}
    \end{subfigure}
    \hfill
    \begin{subfigure}{0.495\linewidth}
        \centering
        \includegraphics[width=\linewidth]{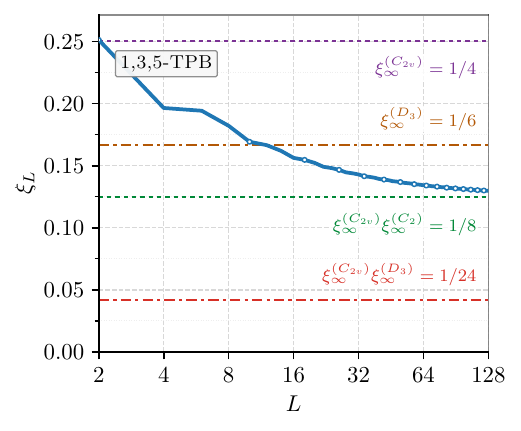}
    \end{subfigure}
    \caption{Convergence of $\xi_{L}$ with increasing $L$ for left: benzene, and right: 1,3,5-triphenylbenzene (1,3,5-TPB). The dotted lines in both panels alongside their asymptotic $L \to \infty$ bounds. For 1,3,5-TPB, we also show with the dot-dashed curves the asymptotic range predicted if the gas-phase $D_3$ symmetry is instead assumed, rather than the $C_2$ symmetry realised in the crystal. The crystal coordinates, lattice vectors, and symmetry groups used for this analysis are taken from~\cite{COD:7238223, Nayak:2010benzene} for benzene, and~\cite{CCDC:867818,Prasad:2013triarylbenzenes} for 1,3,5-TPB.}
    \label{fig:xi_L}
\end{figure}
     
Group theory arguments alone cannot predict how crystallisation will alter a molecule's geometry; that said, if the molecular orientation $\Qrot$ is known, then we can generally expect to know the solid-phase molecular geometry, which will reveal the correct $H$ for the phase of matter in question.

Figure~\ref{fig:xi_L} shows the convergence of $\xi_{L}$ with increasing $L$ for benzene and 1,3,5-TPB using their true crystal structures, \textit{i.e.} for their physically realised value of $\Qrot$, according to the solid-phase molecular geometries. 
In both cases, $\xi_L(\Qrot)$ asymptotes to the minimum end of its allowed range from \eqref{eq:xiInftyRange}. As we demonstrate in Appendix~\ref{app:Qaverage}, this lower bound also corresponds to the $\mathcal{T}$-averaged value of $\xi_\infty$. 

This indicates that the molecules do not take special, crystal-symmetry-aligned orientations within the unit cell. Instead, $\Qrot$ is such that the small scale $\ell \gtrsim 16$ angular features in the single-molecule form factor are often misaligned with the allowed modes of the crystal symmetry groups. In the 1,3,5-TPB example, we also show the range of $\xi_\infty(\Qrot)$ that one would expect from the $D_3$ gas-phase molecular symmetry. Whilst the asymptotic value of the true $\xi_L$ does lie within the allowed region for $D_3$, it is clearly the solid state $C_2$ symmetry that controls the asymptotic behaviour. 
For applications to dark matter direct detection, the anisotropy is usually dominated at $L= 2$ or $L = 4$, at least for SHM-like velocity distributions. In our two examples we find a relatively small $L = 2$ value of $\xi_2 \approx 1/25$ for benzene, and a relatively large $\xi_2 \approx 1/4$ for 1,3,5-TPB. 
    
In the following section, we will use these measures derived from $\mathcal{C}^{(\ell)}$ to estimate the size of the modulation signal, and the statistical significance preserved in a DM direct detection experiment, 
from just the geometry and symmetry group of a given molecular crystal.

\section{Modulation signals in dark matter detectors}\label{sec:modulationProxy}

Up until now, we have only considered the effect of symmetries on the number of angular modes that survive. This tells only part of the story, as not all angular modes contribute equally to the modulation signal, and consequently, to the statistical significance of a dark matter search. Ultimately, the anisotropic performance of a crystal is quantified by how much the scattering rate $R(t)$ from \eqref{eq:rateGK} and \eqref{eq:rateEll} changes as the detector is rotated relative to the dark matter halo.

Anisotropic daily modulation experiments (e.g.~\cite{Bozorgnia:2011tk,Hochberg:2016ntt,Hochberg:2017wce,Griffin:2018bjn,Coskuner:2019odd,Geilhufe:2019ndy,Vahsen:2020pzb,Blanco:2021hlm,Hochberg:2021ymx,Blanco:2022pkt,Boyd:2022tcn,Stratman:2024sng,Abbamonte:2025guf,Blanco:2026kda,Sherpa:2026tgy}) allow the detector apparatus to rotate in phase with the Earth, so that the expected dark matter scattering rate $R(t)$ is periodic over the sidereal day, $T_\mathrm{sid} = 86164.1\,\mathrm{s}$, or approximately $\frac{365.24}{366.24}$ of a solar day.
This modulating dark matter signal can be detected even in the presence of large irreducible backgrounds: the frequency $T_\text{sid}^{-1}$ is uncorrelated with any of the expected terrestrial sources of noise, and the phase and shape of $R(t)$ can be predicted from electronic structure calculations~\cite{Blanco:2025sgv,Dreyer:2026bmz}.

As demonstrated in~\cite{Blanco:2026kda}, the discovery significance $N_\sigma$ for a modulating signal in the presence of an unknown background scales as
\begin{equation}
    N_\sigma \sim f_\mathrm{RMS} \sqrt{T_\text{exp} \langle R \rangle_T},
\end{equation}
with $T_\text{exp}$ the total observation time, $\langle R \rangle_T$ the average scattering rate, and $f_\mathrm{RMS}$ the root-mean-square amplitude of the fluctuating signal components, normalised by the average event rate. Explicitly, this is given by
\begin{equation}
        f_\mathrm{RMS} = \frac{\sqrt{\left\langle \left(R - \langle R\rangle_T \right)^2\right\rangle_T}}{\langle R\rangle_T}
        = \sqrt{ \int_0^T\! \frac{dt}{T} \left(\frac{R(t) }{\langle R \rangle_T}  - 1 \right)^2    } ,
\end{equation}
where $\langle \dots \rangle_{T}$ denotes the time average of a quantity over a period $T$, which for our purposes is the sidereal day length. The crystal scattering rate is
\begin{equation}
    R(t) = N_\mathrm{cell} \rho_\chi \bar\sigma_e \sum_{\ell} \sum_{m,m'} G^{(\ell)}_{mm'}(\mathcal{D}(t)) K_{\mathrm{uc},mm'}^{(\ell)},
\end{equation}
with $N_\mathrm{cell}$ the number of unit cells in the detector volume, and where $K_\mathrm{uc}^{(\ell)}$ is the $\ell^\text{th}$ partial rate matrix for the unit cell, given by  \eqref{eq:partialRateSingle} under the replacement $f_{\ell m}^2 \to f^2_{\mathrm{uc},\ell m}$. 
Qualitatively, $K^{(\ell)}_{mm'}$ represents the contribution from the $f^2_{\ell m'}(q)$ part of the material form factor, weighted by the relative importance of the matching $\ell,m$ component of the dark matter velocity distribution.

Using the anisotropy survival operator $\mathcal C^{(\ell)}$ of \eqref{eq:mcalC}, the molecular crystal scattering rate can be written in terms of the single molecule partial rate matrices as
\begin{equation}
    \begin{split}
    R(t) &= ZN_\mathrm{cell} \rho_\chi \bar\sigma_e \sum_{\ell} \mathrm{tr}\left[(\mathcal{C}^{(\ell)})^{T} G^{(\ell)}\!\left(\mathcal{D}(t)\right)^T K^{(\ell)}\right], \\
    \end{split}
\end{equation}
with $Z$ the number of molecules per unit cell.
Defining the deviation from the mean signal at each $\ell$ as
\begin{equation}
    \delta R^{(\ell)}(t) = R^{(\ell)}(t) - \langle  R^{(\ell)}\rangle_T,
\end{equation}
with $R^{(\ell)}$ the contribution from each $\ell$-mode as defined in~\eqref{eq:rateEll}, the squared RMS fluctuation can be decomposed as:
\begin{equation}\label{eq:frmsAngularDecomposition}
    f^{2}_\mathrm{RMS} = \sum_{\ell} \mathcal{F}_{\ell \ell} + 2\sum_{\ell < \ell'} \mathcal{F}_{\ell \ell'}, \qquad \mathcal{F}_{\ell \ell'} = \frac{\left\langle\delta R^{(\ell)}
        \delta R^{(\ell')}\right\rangle_T}{\left(\left\langle\sum_L R^{(L)}\right\rangle_T\right)^2}.
\end{equation}
The diagonal $\mathcal F_{\ell \ell}$ term in  \eqref{eq:frmsAngularDecomposition} is strictly positive, and describes the contribution to the modulation signal from each individual angular momentum channel. The off-diagonal terms, however, need not be positive, and describe the correlations between modulations at different $\ell$. 

Of interest to us is the ratio of $f_\mathrm{RMS}$ for a given crystal point group, as compared to the value for the individual molecule. 
%, modulo the number of molecules per cell. 
Equivalently this is the ratio between $f_\mathrm{RMS}$ for the realised crystal, and that of one with the trivial crystal point group symmetry, $C_1$. We therefore define
\begin{equation}
    \Lambda^{(\mathcal{L})} = \frac{f_\mathrm{RMS}(\mathcal{L}(S),H,\Qrot)}{f_\mathrm{RMS}(C_1,H,\mathbb{I}_3)},
\end{equation}
with, as before, $\mathcal{L}(S)$ the point group corresponding to the crystal space group, $S$, $H$ the internal molecular point group, and $\Qrot$ the relative orientation between the internal molecular and crystal frames. This gives us a direct measure of the sensitivity to DM lost by crystallisation, as compared to an ideal ``free molecule'' based detector. If a physical crystal has a ratio $\Lambda^{(\mathcal{L})} < 1$, then to achieve the same discovery significance as the free molecule detector after a time $T_\mathrm{mol}$, one would need to expose for a time
\begin{equation}
    T_\mathrm{crys} = \frac{T_\mathrm{mol}}{\left(\Lambda^{(\mathcal{L})}\right)^2}.
\end{equation}
As a first approximation, one might expect that $\Lambda^{(\mathcal{L})} \sim \mathcal{O}\left(\xi_\infty^{(\mathcal{L})}\right)$, in which case a molecule such a benzene is expected to suffer a 16-fold reduction in sensitivity when crystallised. In practice, this is not quite the case, as $\xi^{(\mathcal{L})}_\infty$ alone fails to take into account how the importance of angular modes changes with $\ell$. Typically, the modes at lower $\ell$ have a larger contribution to the signal than others as there are very few features on very small angular scales, those at large $\ell$. We will shortly derive an estimator for $\Lambda^{(\mathcal{L})}$ that takes this into account, and show that it can accurately estimate the true $f_\mathrm{RMS}$ ratio obtained from electronic structure calculations.

A more hypothetical ratio that is also of interest to us is that ratio of $f_\mathrm{RMS}$ for our combined molecular crystal system, to that of the ``ideal'' molecule with no internal symmetries, and whose crystal structure has point $C_1$. This is defined as
\begin{equation}
    \Lambda = \frac{f_\mathrm{RMS}(\mathcal{L}(S),H,\Qrot)}{f_\mathrm{RMS}(C_1,C_1,\mathbb{I}_3)}.
\end{equation}
Unlike $\Lambda^{(\mathcal{L})}$, the comparison to the ideal molecule encoded by $\Lambda$ is not physical, and so cannot truly be calculated. That is to say that whilst we can embed \textit{e.g.} a benzene molecule in a crystal with any symmetry, arbitrarily changing the symmetries of a benzene molecule by moving its atoms changes its electronic structure, and so it ceases to be a benzene molecule. However, using the pure-group-theoretic estimator for $\Lambda^{(\mathcal{L})}$ that we will shortly derive, we will also be able to estimate $\Lambda$. This should then be interpreted as an absolute estimate of the quality of any molecular crystal: whilst $\Lambda^{(\mathcal{L})}$ may be small due to a less than ideal crystal packing, a small $\Lambda$ can arise due to either a bad crystal packing or a highly symmetric molecule.

\subsection{Group-theoretic estimator} \label{sec:Gestimate}

    We now seek a cheap way to estimate $\Lambda^{(\mathcal{L})}$, and by extension $\Lambda$, that will allow us to quickly rank molecular crystals by their modulation signals without the need for expensive electronic structure calculations. Ideally, this estimator should depend only on $\ell$, and our group theory metrics such as $\mathcal{C}^{(\ell)}$. Our starting point is the rate contribution at a given $\ell$:
    \begin{equation}
        R^{(\ell)}(t) \propto \mathrm{tr}\left[\mathcal{C}^{(\ell)} G^{(\ell)}\left(\mathcal{D}(t)^T\right) K^{(\ell)}\right].
    \end{equation}
    Importantly, we care about the $\ell$ dependence of each of these objects, and how it contributes to their magnitude. Starting with the partial rate matrix for a single molecule,
    \begin{equation}
        K^{(\ell)}_{mm'} = \int_0^\infty q\,dq\int_{v_\mathrm{min}(q)}^{v_\mathrm{max}} v\,dv \,P_\ell\left(\frac{v_\mathrm{min}(q)}{v}\right) g_{\ell m}(v) \frac{F_\mathrm{DM}(q,v)}{2\mu_\chi^2 m_\chi} f_{\ell m'}^2(q).
        \label{eq:partialrateK}
    \end{equation}
    the first $\ell$-dependent scaling that we can extract comes from the Legendre polynomial. The argument, $v_\mathrm{min}/v \in [0,1]$, and so we can use the asymptotic expansion
    \begin{equation}
        P_\ell(\cos\theta) = \sqrt{\frac{2}{\pi \ell \sin\theta}}\cos\left[\left(\ell+\frac12\right)\theta-\frac{\pi}{4}\right] + \mathcal{O}(\ell^{-3/2}), \qquad 0<\theta<\pi.
    \end{equation}
    From this we see that the magnitude is damped as $\ell^{-\frac{1}{2}}$, and so we use this as our first scaling factor. The remaining $\ell$ scaling of the partial rate matrices, that of the $g_{\ell m}$ and $f_{\ell m}^2$ coefficients, is a little more challenging to extract, and is derived in Appendix~\ref{app:angularDecay}.
    In summary here, we find that the asymptotic scaling of $K^{(\ell)}$ with $\ell$ is described by
    \begin{equation}
        K^{(\ell)} = K^{(0)} w_{\ell}^{(H)} S^{(\ell)}, \qquad w_\ell^{(H)} =\frac{1}{\sqrt{\ell}} \exp\left[-\frac{(\ell_0(H)+1)\sqrt{\ell(\ell +1)}}{\ell_0(H)}\right],
    \end{equation}
    with $w_0^{(H)} \equiv 1$, where $\ell_0$ is the angular mode at for which the squared factor becomes exponentially, inversely proportional to the characteristic angular size of important features, $\Delta\theta$. 
    As a first estimate, we take $\ell_0$ to be the first even $\ell > 0$ for which $\operatorname{tr}(\Pi^{(H)}_\ell) \neq 0$. 
    Here $K^{(0)}$ is simply the isotropic part of the partial rate matrix, which determines the isotropic average rate $R^{(\ell = 0)}$. 
    As we will soon see, the exact choice of $\ell_0$ in the weighting function makes little difference, as the suppression of higher $\ell$ modes turns out to be dominated by the scaling of the SHM coefficients. 
    The matrix $S^{(\ell)}$ is a dimensionless, $\mathcal O(1)$ \emph{shape matrix} in $m$, proportional to 
    \begin{align}
    K^{(\ell)}_{m m'} \propto S^{(\ell)}_{m m'},
    \end{align}
    which can only be calculated exactly if we know both $g_\chi(\vec v)$ and $f_s^2(\vec q)$. 
    Even if the electronic structure $f_s^2$ is unknown, however, we can still estimate the shape of $S^{(\ell)}$.  
    First, for the axisymmetric SHM lab frame velocity distribution, $K^{(\ell)}_{m m'} = 0$ for all $m \neq 0$, so we can specialise to
    \begin{equation}
        S^{(\ell)} \approx \hat e_0 \vec{v}_\ell^T, \qquad v_{\ell,m} \propto K^{(\ell)}_{0m},
    \end{equation}
    with $\hat e_0$ the unit vector along the $m = 0$ direction, corresponding to a wind that is axisymmetric about the $\hat z$ axis.
    In this section, we will begin with an ansatz for $\vec v_\ell$ that depends only on the molecular symmetry group $H$, and the molecular orientation $\Qrot$, and derive an estimate for the $f_\text{RMS}$ modulation amplitude of a dark matter signal. In Section~\ref{sec:coordproxy} we refine this estimate of $\vec v_\ell$ using the locations of the atoms within the molecule, if they are known. 
    
    To estimate the modulation amplitude in $R(t)$ for a typical detector orientation, 
    first recall from \eqref{eq:frmsAngularDecomposition} that the $f_\mathrm{RMS}$ can be written as a sum over $\mathcal{F}_{\ell \ell'}$, with 
    \begin{equation} \label{eq:FllCdGdG}
        \mathcal{F}_{\ell \ell'} = \frac{\left\langle\delta R^{(\ell)}
        \delta R^{(\ell')}\right\rangle_T}{\left(\left\langle\sum_L R^{(L)}\right\rangle_T\right)^2}.
    \end{equation}
    Let us assume that the denominator is dominated by the $\ell = 0$ mode, in which case the time-averaged rate $\langle R \rangle_T$ is approximately the isotropic average rate, $\langle R \rangle_T \approx R^{(0)}$. 
    Consequently, defining $\delta G^{(\ell)}(t)$ as the deviation from the average value of $G^{(\ell)}(\mathcal{D}(t))$, this can be rewritten as
    \begin{equation}
        \mathcal{F_{\ell \ell'}} \approx w_\ell^{(H)} w_{\ell'}^{(H)}\sum_{b,c} \mathcal{C}^{(\ell)}_{bc} v_{\ell,c} \sum_{b',c'}  \mathcal{C}^{(\ell')}_{b'c'}v_{\ell',c'} \langle \delta G^{(\ell)}_{0b} \delta G^{(\ell')}_{0b'}\rangle_T .
    \end{equation}
    
     To handle the time average, we note that the rotational average of two Wigner-G matrices over the full sphere satisfies
    \begin{equation}
        \langle G^{(\ell)}_{ab}(R) G^{(\ell')}_{a'b'}(R)\rangle_{R \in SO(3)} = \frac{1}{2\ell + 1} \delta_{\ell \ell'} \delta_{aa'} \delta_{bb'}.
    \end{equation}
    Although $f_\text{RMS}$ and the time average in \eqref{eq:FllCdGdG} depend on the initial orientation of the detector, $\mathcal D_0$, and the subsequent cycle $\mathcal D(t)$ through $SO(3)$, we expect that the maximum possible $f_\text{RMS}(\mathcal D_0)$ will be within an order-1 factor of its value for an arbitrarily chosen initial orientation. 
    Therefore we approximate the expected performance of a material by averaging $\delta G^{(\ell)} \delta G^{(\ell)}$ over the full sphere:
    \begin{equation}
        \langle\delta G^{(\ell)}_{0b} \delta G^{(\ell)}_{0b'}\rangle_T \simeq \frac{\gamma_\ell}{2\ell +1} \delta_{\ell \ell'} \delta_{bb'},
    \end{equation}
    with nuisance parameter $\gamma_\ell$ denoting the fraction of all directions captured by the time average over $\mathcal D(t)$. In general, we will assume that the sky coverage fraction is independent of $\ell$, such that $\gamma_\ell = \gamma$. This yields
    \begin{equation}
        \mathcal{F}_{\ell \ell'} \approx  \frac{\delta_{\ell\ell'} \gamma}{(2\ell+1)} w_\ell^{(H)} w_{\ell'}^{(H)} \mathrm{tr}\left[(\mathcal{C}^{(\ell)})^{T} \mathcal{C}^{(\ell')} \vec{v}_{\ell'}\vec{v}_{\ell}^T\right].
    \end{equation}
    Absorbing the additional $2\ell+1$ factor into the combined weight coefficient by defining $W_\ell^{(H)} = (w_{\ell}^{(H)})^2/(2\ell+1)$, we sum over the allowed $\ell$ modes to get $f_\mathrm{RMS}^2$ as
    \begin{equation}\label{eq:fRMSsq}
        f_\mathrm{RMS}^2(\mathcal{L}(S),H,\Qrot) \approx \gamma \sum_{\ell \in2\mathbb{N}} W_{\ell}^{(H)} \left|\mathcal{C}^{(\ell)}(\Qrot) \vec{v}_\ell\right|^2.
    \end{equation}

In the absence of any information about the geometry of the molecules in the crystal, we pick a flat prior, $\vec{v}_\ell = \vec{v}_{\ell,\mathrm{flat}}$, defined by:
\begin{equation} \label{eq:vFlat}
    \vec{v}_{\mathrm{flat},\ell}(\Qrot) = \frac{1}{\sqrt{2\ell+1}} \sum_{i=-\ell}^{\ell}\hat{e}_i(\Qrot),
\end{equation}
with $\hat e_i$ the orthonormal basis vectors that diagonalise $[\mathcal{C}^{(\ell)}]^{T}\mathcal{C}^{(\ell)}$. 
This choice for $\vec v_\ell$ may include nonzero values for $(\ell, m)$ coefficients that are forbidden by the internal molecular symmetry.
However, $\mathcal C^{(\ell)}(\Qrot)$ satisfies
$\mathcal{C}^{(\ell)} \Pi_{\ell,\Qrot}^{(H)} = \mathcal{C}^{(\ell)}$; that is, it
contains the information about which linear combinations of $(\ell, m)$ modes are forbidden by the internal and crystal symmetries.
Consequently, the unphysical modes in $\vec v_{\text{flat}, \ell}$ are projected out by $\mathcal C^{(\ell)} \vec v_{\text{flat}, \ell}$, 
and we can estimate $f_\text{RMS}^2$ from
\begin{equation}
    f_\mathrm{RMS}^2(\mathcal{L}(S),H,\Qrot) \approx \gamma \sum_{\ell \in2\mathbb{N}}W_{\ell}^{(H)} \left|\mathcal{C}^{(\ell)} \vec{v}_{\mathrm{flat},\ell}\right|^2.
\end{equation}
We can describe this choice for $\vec v_\text{flat}$ as an $H$-informed flat prior for the shape of the $f^2_{\ell m}(q)$ electronic structure: effectively, this choice assumes that all spherical harmonics modes that are not forbidden by symmetry are similarly populated. This becomes an increasingly valid assumption as more higher-energy excited states $s$ are included in the rate calculation.

Now that we have an ansatz for $\vec v_\ell \propto K^{(\ell)}_{0m }$, we can rewrite
\begin{equation}
    \begin{split}
        \left|\mathcal{C}^{(\ell)}(\Qrot) \vec{v}_{\mathrm{flat},\ell}(\Qrot)\right|^2 &= \frac{1}{(2\ell+1)} \mathrm{tr}\left[(\mathcal{C}^{(\ell)})^{T}\mathcal{C}^{(\ell)} \sum_{i,j} \hat e_i \hat e_j^T\right] \\
        &= \frac{1}{(2\ell+1)} \sum_{i} \lambda_i^2(\Qrot) \\
        &= \frac{1}{(2\ell+1)} \mathrm{tr}\left[\mathcal{C}^{(\ell)}(\Qrot)\right] \\
        &= \kappa_\ell(\Qrot) ,
    \end{split}
\end{equation}
where the $\lambda_i$ are the eigenvalues of the combined survival operator, or equivalently the fraction of each orthogonal $m$-direction lost to the symmetry, and $\kappa_\ell$ is the fractional survival at each $\ell$, defined in  \eqref{eq:kappaFull}. Our final estimate for the reduction in modulation signal due to crystallisation is therefore given by the ratio of weighted $\kappa$ sums
    \begin{equation}
        \Lambda^{(\mathcal{L})}(\Qrot) \approx \sqrt{\frac{\sum_{\ell \in 2\mathbb{N}} W_{\ell}^{(H)} \kappa_{\ell}(\Qrot)}{\sum_{\ell \in 2\mathbb{N}} W_{\ell}^{(H)} \kappa^{(H)}_\ell}},
    \end{equation}
    where $\kappa^{(H)} = \mathrm{tr}\left(\Pi_\ell^{(H)}\right)/(2\ell+1)$ is the fraction of angular modes that survive internal molecular symmetry projection at each $\ell$. It follows immediately that the ratio of modulation signal from a given molecular crystal to that of the ideal molecule with no internal symmetry, i.e.~$\kappa^{(H)} = 1$, can be approximated by
    \begin{equation}
        \Lambda(\Qrot) \approx \sqrt{\frac{\sum_{\ell \in 2\mathbb{N}} W_{\ell}^{(H)} \kappa_\ell(\Qrot)}{\sum_{\ell \in 2\mathbb{N}} W_{\ell}^{(C_1)}}}.
    \end{equation}

The quantity $\Lambda^{(\mathcal L)}$ is a useful way to estimate the projected loss of anisotropy for an arbitrary molecule placed with a given orientation $\Qrot$ within a crystal with Laue group $\mathcal L$. However, if $\Qrot$ is known, then the coordinates of the atomic positions for each molecule are (almost by definition) known as well. If we know the shape of the molecule, then we can improve upon $\Lambda^{(\mathcal L)}$ by letting the molecular shape inform our choice for $\vec v_\ell$. We do exactly this in Section~\ref{sec:coordproxy}, to derive a \emph{coordinate aware estimator} $\Lambda_\text{coord}$. 

On the other hand, for molecules that have not yet been synthesised, crystallised and studied in the lab, the molecular packing arrangement $\Qrot$ within the crystal may be unknown. 
In this case, we can still quantify the average anisotropic performance of a Laue class by averaging over our ignorance of $\Qrot$:
\begin{align}    
\langle \Lambda^{(\mathcal{L})} \rangle_\Qrot &\equiv \sqrt{\left\langle\left(\Lambda^{(\mathcal{L})}\right)^2\right\rangle_\Qrot} = \sqrt{\frac{\sum_{\ell \in 2\mathbb{N}} W_{\ell}^{(H)} \kappa^{(\mathcal{L})}_\ell \kappa_\ell^{(H)}}{\sum_{\ell \in 2\mathbb{N}} W_{\ell}^{(H)}\kappa_\ell^{(H)}}}, 
\\
\langle \Lambda \rangle_\Qrot &\equiv \sqrt{\left\langle\Lambda^2\right\rangle_\Qrot} = \sqrt{\frac{\sum_{\ell \in 2\mathbb{N}} W_{\ell}^{(H)} \kappa^{(\mathcal{L})}_\ell \kappa^{(H)}_\ell}{\sum_{\ell \in 2\mathbb{N}} W_{\ell}^{(C_1)}}}.
\end{align}
We give the derivation of the $\mathcal{T}$-averaged value of $\mathcal{C}^{(\ell)}$ in Appenfix~\ref{app:Qaverage}.

Before moving onto some representative molecules, we will briefly comment on the exponential weighting function $W_{\ell}^{(H)}$. 
For dark matter scattering rates with SHM velocity distributions, we find that $W_\ell$ at $\ell \geq 2$ is strongly suppressed by the lack of small-scale angular features in the SHM $g_\chi(\vec v)$. Consequently, the weighting function decays so rapidly that it effectively picks out the first non-zero $\ell$ mode. We show this explicitly in Section~\ref{sec:demonstrations}, by plotting the $K^{(\ell)}_{m m'}$ coefficients for trans-stilbene for two types of dark matter model. Empirically, $K^{(\ell)} \sim e^{- \ell / 1} K^{(0)}$, so  the $\ell = 4$ part of the scattering rate is suppressed by an order of magnitude relative to $\ell = 2$. 

As a result, the anisotropy estimator $\Lambda^{(\mathcal L)}$ is well approximated by one of three forms:
    \begin{equation}\label{eq:differentGroupAgnostic}
        \langle \Lambda^{(\mathcal{L})}\rangle_\Qrot \simeq \begin{cases}
            \sqrt{\kappa^{(\mathcal{L})}_2}, &\qquad \text{non-cubic }\mathcal{L}\text{ and } H, \\
            \sqrt{\kappa_4^{(\mathcal{L})}}, & \qquad \text{cubic } H, \\\sqrt{\frac{W_4^{(H)}\kappa_4^{(\mathcal{L})}\kappa_4^{(H)}}{W_2^{(H)}\kappa_2^{(H)}}} \simeq0, & \qquad \text{cubic }\mathcal{L}, \text{ non-cubic } H.
        \end{cases}
    \end{equation}
    with the first form, $\langle\Lambda^{(\mathcal{L})}\rangle_\Qrot \simeq \sqrt{\kappa_2^{(\mathcal{L})}}$, by far the most common given the rarity of cubic Laue group symmetries. The estimator therefore suggests that, with the exception of uncommon cubic crystal symmetries, the retained modulation amplitude due to crystallisation typically takes one of five values,
    \begin{equation}\label{eq:agnosticEstimator}
        \langle \Lambda^{(\mathcal{L})} \rangle_\Qrot\simeq \sqrt{\kappa_2^{(\mathcal{L})}} \in \left\{0, \sqrt{\frac{1}{5}}, \sqrt{\frac{2}{5}}, \sqrt{\frac{3}{5}},1 \right\},
    \end{equation}
    corresponding to the five possible values of $\kappa_2^{(\mathcal{L})}$ across all crystal point groups. These are precisely the five quadrupole survival classes, $\mathsf{Q}_k$, introduced in Sec.~\ref{sec:survivalAnisotropy}, suggesting that identifying the survival class of a crystal alone is sufficient to estimate the fraction of modulation signal retained by crystallisation.
    
    The ``ideal'' molecular crystal estimator, $\Lambda$, follows similarly, and is well approximated by the two cases
    \begin{equation}
        \langle \Lambda\rangle_\Qrot \simeq \begin{cases}
            \sqrt{\kappa^{(\mathcal{L})}_2 \kappa_2^{(H)}}, &\qquad \text{non-cubic }\mathcal{L}\text{ and } H, \\
            \sqrt{\frac{W_4^{(H)}}{W_{2}^{(H)}}\kappa_4^{(\mathcal{L})}\kappa_4^{(H)}} \simeq0, & \qquad \text{cubic }\mathcal{L} \text{ or } H,
        \end{cases}
        \label{eq:LambdaQ}
    \end{equation}
    with, as before, the non-cubic case the most common. It is particularly interesting to compare this to the asymptotic angular mode survival fraction, $\xi_\infty$. For example, a benzene crystal with $\mathcal{L}(S) = D_{2h}$ and $H = D_{6h}$ has a $\xi_\infty$ that approaches $1/48$. This is drastically different from the fractional RMS signal survival predicted by $\langle \Lambda\rangle_\Qrot$, which sits at $\sqrt{2}/5 \simeq 0.282$. For molecule-crystal pairs with any surviving quadrupole modes, the fractional modulation survival is therefore far less pessimistic than $\xi_\infty$ alone would suggest; in the worst case quadrupole scenario, with $\kappa^{(\mathcal{L})}_2 = \kappa^{(H)}_2 = 1/5$, the modulation signal is only expected to be $\sim 5$ times smaller than the best-case scenario, corresponding to a $\sim 25$ times longer exposure requirement to reach the same sensitivity.
    By contrast, if cubic symmetry appears anywhere in the combined system then the outlook is catastrophic, as the RMS modulation signal will be exponentially smaller than a less symmetric system.

\subsection{Coordinate-aware estimator} \label{sec:coordproxy}
    The estimator derived in~\eqref{eq:agnosticEstimator} is quite useful when we do not know any of the molecular crystal properties, as it collapses to one of five values that depend solely on the quadrupole survival class of the crystal. In many cases, however, we will have more information than this,
    particularly if the crystal structure has already been determined experimentally. In its simplest form, an experimentally derived crystal geometry is nothing more than a set of coordinates, from which the crystal Laue group, $\mathcal{L}_\mathrm{crys}$, the internal molecule symmetry point group, $H_\mathrm{crys}$, and the embedding in crystal frame $\Qrot$, can all be derived. 
    
    Additionally, the coordinates allow us to make a more informed choice of the $(\ell m)$ shape vector, $\vec v_{\ell}$, by taking into account the axiality and planarity of the molecule. This follows from the coordinate coordinate matrix
    \begin{equation}
        \Sigma = \frac{1}{N} (X^T X), \qquad X = \begin{pmatrix}
    x_1 & y_1 & z_1 \\
    & \vdots & \\
    x_N & y_N & z_N
        \end{pmatrix},
    \end{equation}
    with $x_1$ the centred $x$-coordinate of the $x$-coordinate of the of the first atom of one molecule in the crystal, and $x_n$ the centred $x$-coordinate of the last atom in the same molecule. The eigenvectors of $\Sigma$, $\hat n_1, \hat n_2,$ and $\hat n_3$, are then the principal directions of the molecule, and their ``length'' is determined by the corresponding eigenvalues, $\lambda_1, \lambda_2,$ and $\lambda_3$. Without loss of generality, we will take $\lambda_1 \geq \lambda_2 \geq \lambda_3$, such that $\hat n_1$ is the ``longest'' principal axis. A molecule for which $\lambda_1 \gg \lambda _2, \lambda_3$ can be considered as axial, or rod-like, as it is particularly longer along one direction that the other two, whilst a molecule with $\lambda_1 \simeq \lambda_2 \gg \lambda_3$ is planar. 

We have already seen that the value of $f_\mathrm{RMS}$ is dominated by the $\ell = 2$ contribution, and so we only need to convert these principal axes into the quadrupole shape vector, $\vec{v}_{2}^{\mathrm{coord}}$:
\begin{equation}
    \vec{v}_2^{\mathrm{coord}}(\vec n) = \frac{1}{\mathcal{N}} \begin{pmatrix}
    \sqrt{\frac{15}{4\pi}} n_x n_y \\
    \sqrt{\frac{15}{4\pi}} n_y n_z \\
    \sqrt{\frac{5}{16\pi}} \left(3n_z^2 - 1\right) \\
    \sqrt{\frac{15}{4\pi}} n_x n_z \\
    \sqrt{\frac{15}{16\pi}} \left(n_x^2 - n_y^2\right)
    \end{pmatrix} ,
\end{equation}
using the Condon-Shortley phase convention for spherical harmonics, where $n_x$, $n_y$ and $n_z$ are the $x$, $y$ and $z$ components of the ``important'' principal axis, and $\mathcal{N}$ is a normalisation factor. 

By using the shape of the molecule to inform our choice of $\vec v_\ell$, we are in effect suggesting that the electronic structure in momentum space will tend to align with the principal axes of the molecule. For a single excited state, this may not be a good approximation: however, when summing over a range of excited states of similar energies (e.g.~$E_s < 10$\,eV, or the first $\mathcal O(10)$ bound states), it becomes a better approximation. 
After explicitly calculating the electronic excited states for 25 molecules and performing the rate calculations, we show in Sec.~\ref{sec:demonstrations} that this choice for $\vec v_\ell$ works surprisingly well.

Some molecules have particularly extreme shapes, with most of the atomic sites lying along a particular axis or near to a particular plane. We quantify the axiality, and planarity of a molecule as 
\begin{equation}
    A = 1-\frac{\lambda_2}{\lambda_1}, \qquad P = \frac{\lambda_2 - \lambda_3}{\lambda_1},
\end{equation}
respectively, and set the value of $\vec{v}_2^{\mathrm{coord}}(\vec n)$ according to
\begin{equation}
    \hat n = \begin{cases}
        \hat n_1, &\qquad A \geq 0.6, \quad P\leq 0.2, \\
        \hat n_3, &\qquad P \geq 0.3, \\
        \hat n_{\mathrm{med}}, &\qquad \text{otherwise},
    \end{cases}
\end{equation}
with $\hat{n}_\mathrm{med}$ the element-wise median of the three principal directions. This is motivated by the qualitative observation that for axial molecules, much of the form factor lies along the principal axis, whilst for planar molecules, it is out of the plane. Otherwise, for molecules that are not obviously long or flat, the median direction captures something closer to our old shape-agnostic estimator $\vec v_\text{flat}$. 

Finally, we apply $\vec v_\ell(\hat n)$ to \eqref{eq:fRMSsq} 
to derive our coordinate-aware estimator for the RMS signal loss due to crystallisation:
\begin{equation}
    \Lambda^{(\mathcal{L}_\mathrm{crys})}_\mathrm{coord} \simeq \frac{\left|\mathcal{C}^{(2)}\vec{v}_{2}^{\mathrm{coord}}\right|}{\left|\Pi^{(H_\mathrm{crys})}_{2,\Qrot_\mathrm{crys}}\vec{v}_{2}^{\mathrm{coord}}\right|},
\end{equation}
for non-cubic $\mathcal{L}_\mathrm{crys}$, and analogous constructions to \eqref{eq:differentGroupAgnostic} otherwise, with $\mathcal{C}^{(\ell)}$ constructed using $\Qrot_\mathrm{crys}$, $\mathcal{L}_{\mathrm{crys}}$, and $H_\mathrm{crys}$. We can similarly define the coordinate-aware estimator for the comparison to the perfect molecular crystal system as
\begin{equation}\label{eq:lambdaCoord}
    \Lambda_\mathrm{coord} \simeq \left|\mathcal{C}^{(2)}\vec{v}_{2}^{\mathrm{coord}}\right |,
\end{equation}
again for non-cubic crystal Laue group. Naturally, this estimator is somewhat more textured than the agnostic one previously derived, and takes more than five values. Importantly, however, it still relies purely on symmetry and geometry, and does not require any electronic structure calculations to compute, making it far more efficient for screening purposes. 

\section{Demonstrations}\label{sec:demonstrations}

    We will now demonstrate that our estimators for the modulation loss accurately reproduce the modulation signal predicted from electronic structure calculations, and that as a result, they can be used to rapidly screen candidate molecular crystals for direct DM detection. To that end, we have chosen 25 molecules with a wide range of internal symmetry groups, from the highly symmetric $D_{nh}$ groups down to the symmetry-free $C_1$ group. We split the molecules into two groups. The first are representative molecules with a wide range of symmetry groups, and no particular shape or structure, designed to test the broad applicability of our estimators. We refer to the second set as symmetry-breaking molecules, which are compounds that are similar in structure to benzene, but with varying numbers of additional functional groups or substitutions. The goal of the second set is to test whether the estimator works well when the molecular geometry lowers the point group, but the electron density may retain remnants of the higher benzene-like point group symmetry. 
    
    For each molecule, we then make two comparisons. In the first, we compute the form factor for the isolated, or gas-phase, molecule using \texttt{SCarFFF}~\cite{Blanco:2025sgv} and assume that its crystal structure is not known. For each molecule, we compute the first 12 excited states with DFT and TD-DFT, using the $\omega\text{B97X-D4}$ exchange correlation functional and the def2-TZVP basis set. The form factor computation is done on 251 $q$ points on a uniform grid up to $25\,\mathrm{keV}$, up to a maximum angular momentum mode $\ell = 24$. We then embed the molecule in each of the 11 allowed crystallographic Laue groups and average over $\Qrot$, its orientation within the crystal. Each $f_\mathrm{RMS}$ value is then averaged over initial detector orientations, $\mathcal{D}_0$, and normalised by $f_\mathrm{RMS}$ for the trivial point group, also $\mathcal{D}_0$-averaged. This is the ratio
    \begin{equation}
        r_\mathrm{gas} = \frac{\langle f_\mathrm{RMS}(\mathcal{L}(S),H_\mathrm{gas},\Qrot)\rangle_{\Qrot,\mathcal{D}_0}}{\langle f_\mathrm{RMS}(C_1,H_\mathrm{gas},\mathbb{I}_3)\rangle_{\mathcal{D}_0}},
    \end{equation}
    for each Laue group $\mathcal{L}$, a set of DM masses $m_{\chi} \in \{5,10,25,50,100\}\,\mathrm{MeV}$, and for both the heavy and light-mediator cases, $n = 0$ and $n = 2$, respectively. The resulting ratios are then compared to our RMS estimator, $\langle \Lambda^{(\mathcal{L})}\rangle_\Qrot$, and we compute the mean absolute error 
    \begin{equation}
        \langle \Delta r_\mathrm{gas}\rangle_{\mathcal{L},m_\chi} = \left\langle \left|r_\mathrm{gas} - \langle\Lambda^{(\mathcal{L})}\rangle_\Qrot\right|\right\rangle_{\mathcal{L},m_\chi},
    \end{equation}
    on a per-molecule, and per-molecule-group basis.
    This is therefore a test of how good the estimator is as an early-stage screening tool: given a molecule whose gas-phase symmetries are known, but whose crystal structure is not, can we determine if there is any crystal structure that preserves enough modulation signal to be worth pursuing? 

    \begin{table}[t]
    \centering
    \setlength{\tabcolsep}{4pt}
    \renewcommand{\arraystretch}{1.05}
    \begin{adjustbox}{max width=\textwidth}
    \begin{tabular}{c|c|c|c|c|c|c|c|c}
    \multirow{2}{*}{Name} & \multirow{2}{*}{$H_\mathrm{gas}$} & \multirow{2}{*}{$H_{\mathrm{crys}}$} & \multirow{2}{*}{$P_{\mathrm{crys}}$} & \multicolumn{2}{c|}{$\langle \Delta r_\mathrm{gas}\rangle_{\mathcal{L},m_\chi}$} & \multicolumn{2}{c|}{$\langle \Delta r_{\mathrm{crys}}\rangle_{m_\chi}$} & \multirow{2}{*}{Source} \\
    \cline{5-8}
    & & & & $n=0$ & $n=2$ & $n=0$ & $n=2$ & \\
    \hline\hline
    Ibuprofen & $C_1$ & $C_1$ & $C_{2h}$ & 0.063 & 0.053 & 0.131 & 0.109 & \cite{CCDC:IBPRAC21, Kleemiss:2020SilaIbuprofen} \\ \hline
    Aspirin & $C_1$ & $C_1$ & $C_{2h}$ & 0.040 & 0.041 & 0.146 & 0.042 & \cite{Karuppannan:2024CSDCommunication} \\ \hline
    Paracetamol & $C_1$ & $C_1$ & $C_{2h}$ & 0.045 & 0.044 & 0.279 & 0.009 & \cite{Deere:2022CSDCommunication} \\ \hline
    Nicotinamide & $C_1$ & $C_1$ & $C_{2h}$ & 0.043 & 0.037 & 0.005 & 0.032 & \cite{Jarzembska:2014CCDC991918, Jarzembska:2014PyrazinamideNicotinamide} \\ \hline
    Acetanilide & $C_1$ & $C_1$ & $D_{2h}$ & 0.038 & 0.043 & 0.209 & 0.189 & \cite{Hathwar:2011CCDC852170,Hathwar:2011MultipoleTransferability} \\ \hline
    Lactic acid & $C_1$ & $C_1$ & $D_2$ & 0.053 & 0.058 & 0.103 & 0.104 & \cite{Yang:2021CCDCYILLAG01, Yang:2021LacticAcidPolymorphism} \\ \hline
    Indole-3-acetic acid & $C_1$ & $C_1$ & $C_{2h}$ & 0.045 & 0.040 & 0.140 & 0.071 & \cite{Nigovic:2000AuxinDerivatives, Nigovic:2000CCDCIndoleAceticAcid} \\ \hline
    Caffeine & $C_s$ & $C_s$ & $C_s$ & 0.041 & 0.040 & 0.037 & 0.031 & \cite{Lehmann:2007CCDC610381, Lehmann:2007BetaCaffeine} \\ \hline
    Hydrogen peroxide & $C_2$ & $C_2$ & $D_4$ & 0.034 & 0.020 & 0.147 & 0.130 & \cite{osti_1199816} \\ \hline
    trans-Stilbene & $C_{2h}$ & $C_i$ & $C_{2h}$ & 0.040 & 0.045 & 0.230 & 0.164 & \cite{Li:2021CCDCcc28bj38, Li:2021DeuteratedAlkenes}\\ \hline
    Biphenyl & $D_2$ & $D_{2h}$ & $C_{2h}$ & 0.053 & 0.048 & 0.125 & 0.090 & \cite{LanderosRivera:2021CCDCBiphenyl, LanderosRivera:2021BiphenylCrystal} \\ \hline
    Naphthalene & $D_{2h}$ & $D_{2h}$ & $C_{2h}$ & 0.041 & 0.048 & 0.049 & 0.061 & \cite{Capelli:2006CCDCNaphthalene, Capelli:2006NaphthaleneMotion} \\ \hline
    Anthracene & $D_{2h}$ & $D_{2h}$ & $C_{2h}$ & 0.034 & 0.043 & 0.114 & 0.206 & \cite{Nieger:2022CSDCommunication}\\ \hline
    Boric acid & $C_{3h}$ & $C_{3h}$ & $C_i$ & 0.041 & 0.040 & 0.000 & 0.000 & \cite{Wu:2022BoricAcidRTP, Wu:2022CCDCcc297p6g} \\ \hline
    Borazine & $D_{3h}$ & $D_{3h}$ & $D_4$ & 0.040 & 0.040 & 0.001 & 0.002 & \cite{MerinoGarcia:2022CCDCcc25w46f,MerinoGarcia:2022BenzeneBorazine} \\ \hline
    \end{tabular}
    \end{adjustbox}
    \caption{Representative molecules used to compare our estimators to the true modulation signal lost by crystallisation. The first two columns show the internal point group symmetry of the gas phase and crystallised molecule, respectively, whilst the third column shows the point group symmetry of the physically-realised crystal. The remaining columns show the RMSE between the estimator and the true modulation loss for the heavy ($n=0$) and light-mediator ($n=2$), averaged over DM masses, and, for the gas phase estimator, also over the 11 Laue groups. The final column shows the data sources for the crystal comparison. See the main text for exact symbol definitions.}
    \label{tab:representativeMols}
    \end{table}
    
    In the second comparison, we perform the same analysis using the atomic coordinates of the molecule in a known crystal polymorph, so that the crystal point group, molecular coordinates, and consequently $\Qrot = \Qrot_\mathrm{crys}$ are fixed. Additionally, as discussed in Sec.~\ref{sec:internal}, the internal symmetry group of the crystallised molecule may differ from that of the gas phase. This is the ratio
    \begin{equation}
        r_\mathrm{crys} = \frac{\langle f_\mathrm{RMS}(\mathcal{L}(S_\mathrm{crys}),H_\mathrm{crys},\Qrot_\mathrm{crys})\rangle_{\mathcal{D}_0}}{\langle f_\mathrm{RMS}(C_1,H_\mathrm{crys},\mathbb{I}_3)\rangle_{\mathcal{D}_0}},
    \end{equation}
    We then compare this to the value of the coordinate-aware estimator for a fixed orientation and point group using the mean absolute error
    \begin{equation}
        \langle \Delta r_\mathrm{crys}\rangle_{m_\chi} = \left\langle \left|r_\mathrm{crys} - \Lambda^{(\mathcal{L}_\mathrm{crys})}_\mathrm{coord}\right|\right\rangle_{m_\chi}.
    \end{equation}
    This is therefore a more physics-informed test, to check whether the estimator can accurately reproduce the fractional modulation signal survival once the full geometry of the system is known. 

We calculate $R(t)$ and $f_\text{RMS}$ using a Standard Halo Model (SHM) velocity distribution, $g_\chi(\vec v)$,
\begin{align}
g_\chi(\vec v) &= \frac{1}{N_0} \exp\left( - \frac{|\vec v + \vec v_E|^2 }{v_0^2} \right) \Theta(v_\text{esc} - |\vec v + \vec v_E|),
\label{eq:gXshm}
\end{align}
where $N_0$ is the normalisation constant that ensures $\int\! d^3 v\, g_\chi(\vec v) \equiv 1$. 
Following the recommendation of~\cite{Baxter:2021pqo}, we set the escape speed to $v_\text{esc} = 544\,\mathrm{km}\,\mathrm{s}^{-1}$, and we take both the velocity dispersion $v_0$ and the local standard of rest velocity $v_\text{LSR}$ to be $v_0 = v_\text{LSR} = 238\,\mathrm{km}\,\mathrm{s}^{-1}$. We set the lab-frame speed relative to the galactic centre of rest to $v_E = 250\,\mathrm{km}\,\mathrm{s}^{-1}$: this is close to the annual average, as well as the instantaneous value in February or September. 

\begin{figure}[t]
        \centering
        \begin{subfigure}{0.325\linewidth}
            \centering
            \includegraphics[width=\linewidth]{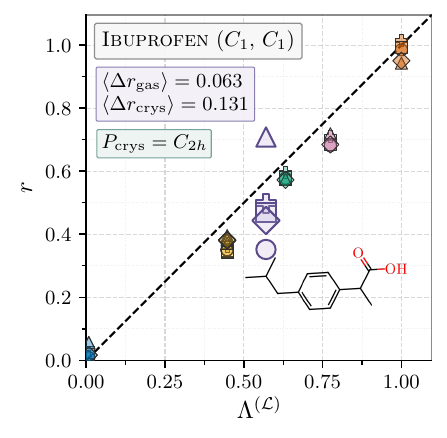}
        \end{subfigure}
        \hfill
        \centering
        \begin{subfigure}{0.325\linewidth}
            \centering
            \includegraphics[width=\linewidth]{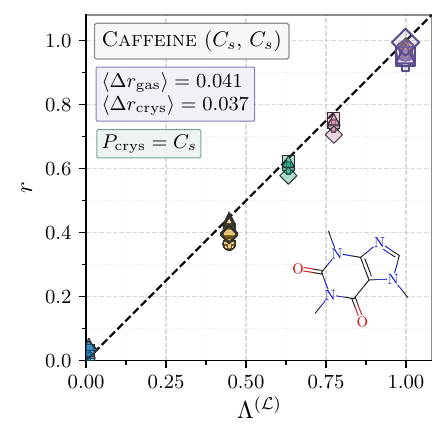}
        \end{subfigure}
        \hfill
        \begin{subfigure}{0.325\linewidth}
            \centering
            \includegraphics[width=\linewidth]{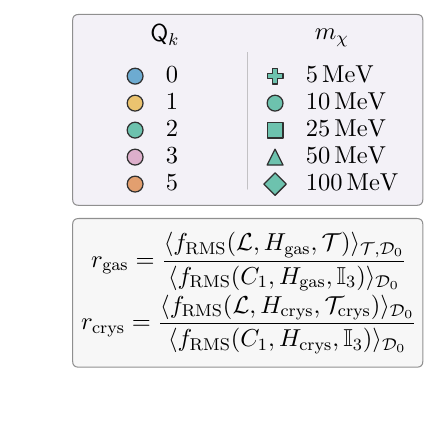}
        \end{subfigure}
        \hfill
        \begin{subfigure}{0.325\linewidth}
            \centering
            \includegraphics[width=\linewidth]{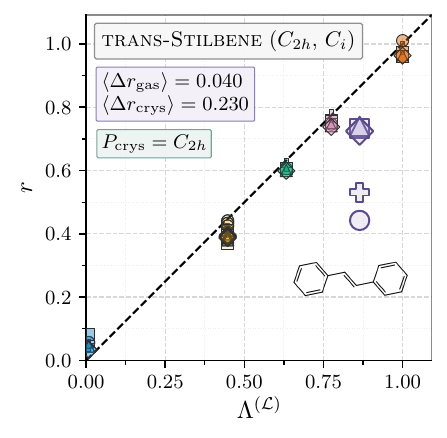}
        \end{subfigure}
        \hfill
        \begin{subfigure}{0.325\linewidth}
            \centering
            \includegraphics[width=\linewidth]{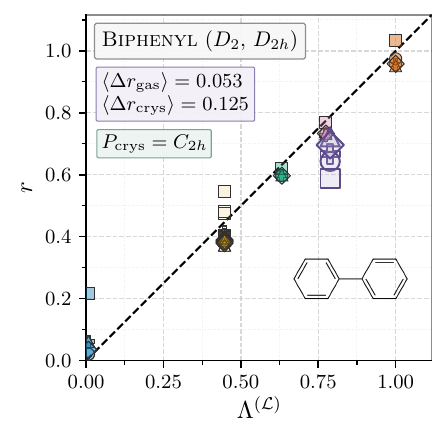}
        \end{subfigure}
        \hfill
        \begin{subfigure}{0.325\linewidth}
            \centering
            \includegraphics[width=\linewidth]{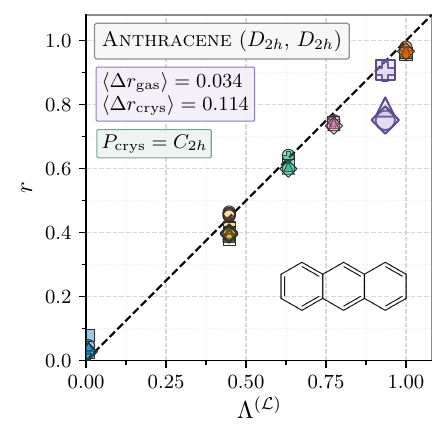}
        \end{subfigure}
        \hfill
        \begin{subfigure}{0.325\linewidth}
            \centering
            \includegraphics[width=\linewidth]{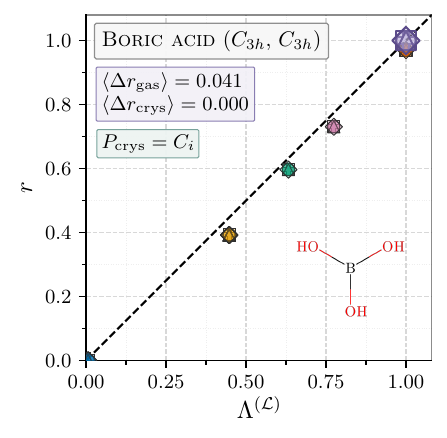}
        \end{subfigure}
        \hfill
        \begin{subfigure}{0.325\linewidth}
            \centering
            \includegraphics[width=\linewidth]{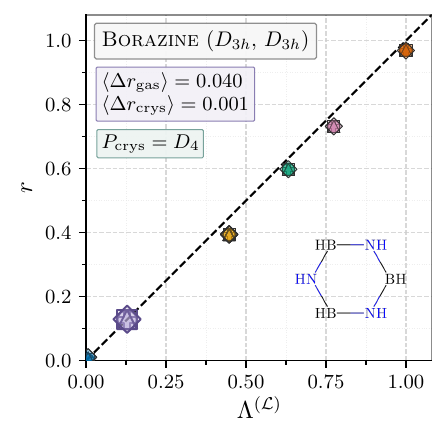}
        \end{subfigure}
        \hfill
        \begin{subfigure}{0.325\linewidth}
            \centering
            \includegraphics[width=\linewidth]{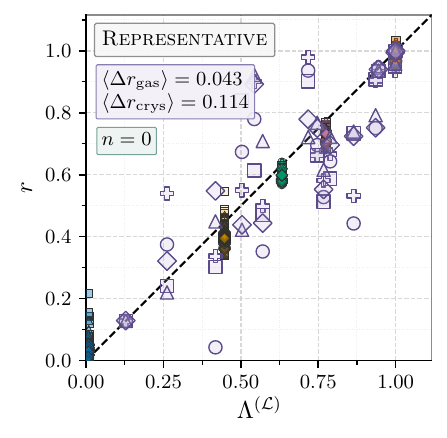}
        \end{subfigure}
         \caption{Comparisons of the estimators to the true $f_\mathrm{RMS}$ ratios for a subset of the molecules in Table~\ref{tab:representativeMols}, assuming a heavy mediator ($n=0$). The bracketed point group names after each molecule name correspond to the pair $(H_\mathrm{gas}, H_\mathrm{crys})$, the gas-phase and crystal-phase internal point groups, respectively. The smaller points show the comparison of the agnostic estimator, $\langle \Lambda^{(\mathcal{L})}\rangle_\Qrot$, to the orientation- and detector-averaged $f_\mathrm{RMS}$ ratio, for each of the 11 Laue classes $\mathcal{L}$, for a DM mass $m_\chi \in \{5,10,25,50,100\}\,\mathrm{MeV}$, and coloured according to their quadrupole survival class, $\mathsf{Q}_k$. The larger purple markers show the comparison of the coordinate aware estimator, $\Lambda_\mathrm{coord}^{(\mathcal{L}_\mathrm{crys})}$, to the physically-realised polymorph of each molecular crystal. The bottom right plot shows the results for all molecules in Table~\ref{tab:representativeMols} for the $n = 0$ case overlaid.}
        \label{fig:representativeSymmetriesn0}
    \end{figure}

\begin{figure}[t]
        \centering
        \begin{subfigure}{0.325\linewidth}
            \centering
            \includegraphics[width=\linewidth]{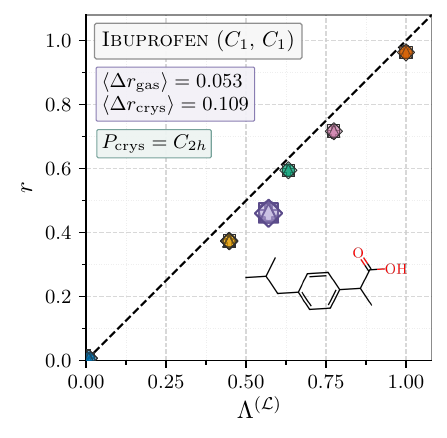}
        \end{subfigure}
        \hfill
        \centering
        \begin{subfigure}{0.325\linewidth}
            \centering
            \includegraphics[width=\linewidth]{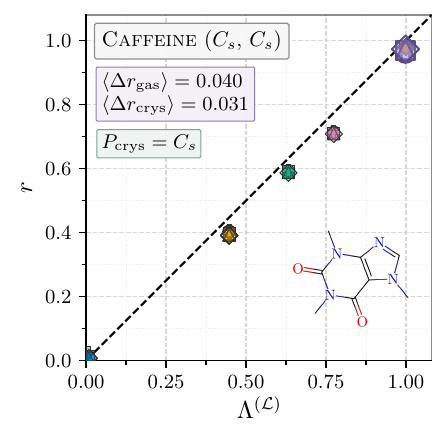}
        \end{subfigure}
        \hfill
        \begin{subfigure}{0.325\linewidth}
            \centering
            \includegraphics[width=\linewidth]{figures/proxy_comparisons/legend_combined.pdf}
        \end{subfigure}
        \hfill
        \begin{subfigure}{0.325\linewidth}
            \centering
            \includegraphics[width=\linewidth]{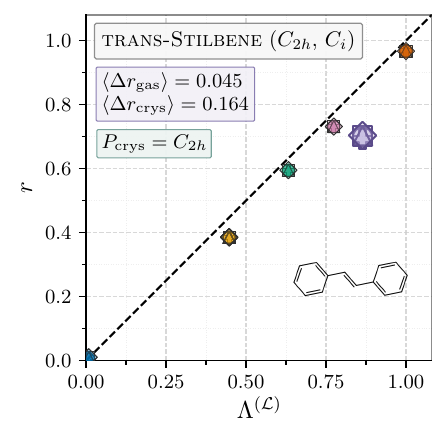}
        \end{subfigure}
        \hfill
        \begin{subfigure}{0.325\linewidth}
            \centering
            \includegraphics[width=\linewidth]{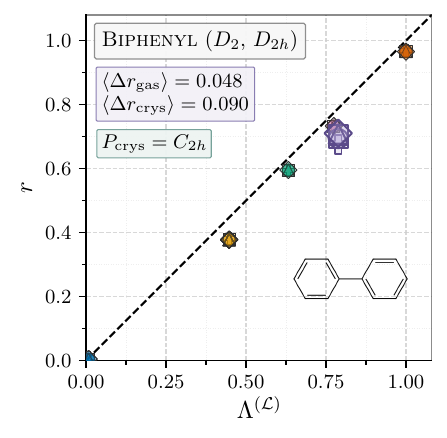}
        \end{subfigure}
        \hfill
        \begin{subfigure}{0.325\linewidth}
            \centering
            \includegraphics[width=\linewidth]{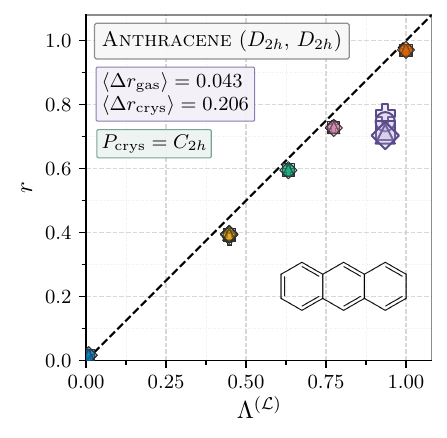}
        \end{subfigure}
        \hfill
        \begin{subfigure}{0.325\linewidth}
            \centering
            \includegraphics[width=\linewidth]{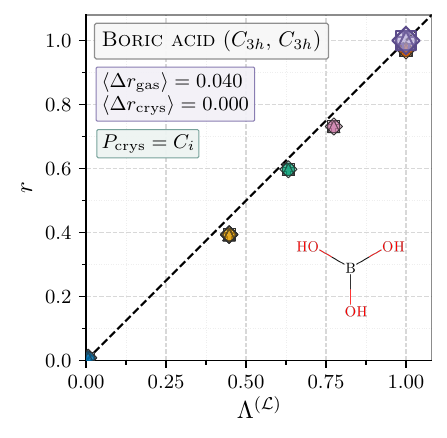}
        \end{subfigure}
        \hfill
        \begin{subfigure}{0.325\linewidth}
            \centering
            \includegraphics[width=\linewidth]{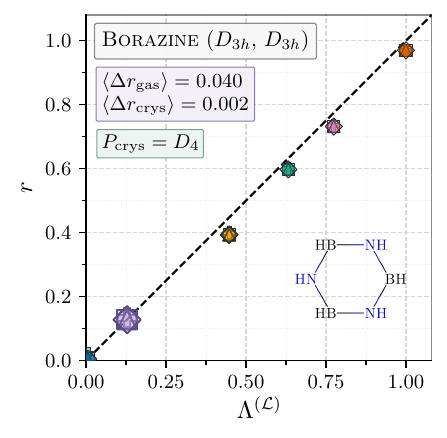}
        \end{subfigure}
        \hfill
        \begin{subfigure}{0.325\linewidth}
            \centering
            \includegraphics[width=\linewidth]{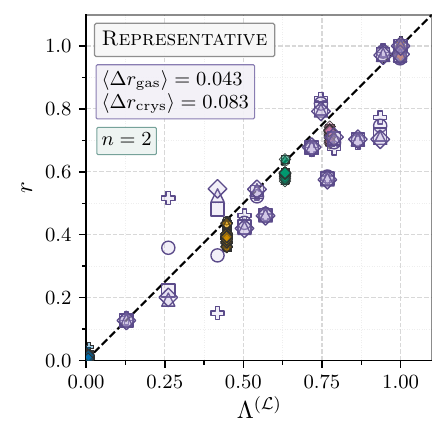}
        \end{subfigure}
          \caption{Comparisons of the estimators to the true $f_\mathrm{RMS}$ ratios for a subset of the molecules in Table~\ref{tab:representativeMols}, assuming a heavy mediator ($n=2$). The bracketed point group names after each molecule name correspond to the pair $(H_\mathrm{gas}, H_\mathrm{crys})$, the gas-phase and crystal-phase internal point groups, respectively. The smaller points show the comparison of the agnostic estimator, $\langle \Lambda^{(\mathcal{L})}\rangle_\Qrot$, to the orientation- and detector-averaged $f_\mathrm{RMS}$ ratio, for each of the 11 Laue classes $\mathcal{L}$, for a DM mass $m_\chi \in \{5,10,25,50,100\}\,\mathrm{MeV}$, and coloured according to their quadrupole survival class, $\mathsf{Q}_k$. The larger purple markers show the comparison of the coordinate aware estimator, $\Lambda_\mathrm{coord}^{(\mathcal{L}_\mathrm{crys})}$, to the physically-realised polymorph of each molecular crystal. The bottom right plot shows the results for all molecules in Table~\ref{tab:representativeMols} for the $n = 2$ case overlaid.}
        \label{fig:representativeSymmetriesn2}
    \end{figure}

    In Table~\ref{tab:representativeMols} we show the results for the representative molecules, and a subset of them are plotted in Figs.~\ref{fig:representativeSymmetriesn0} and~\ref{fig:representativeSymmetriesn2} for the heavy and light-mediator cases, respectively. For the gas phase comparison, where only the internal molecular symmetry is assumed to be known, the mean absolute error, $\langle \Delta r_\mathrm{gas}\rangle$, remains at or below $0.06$ for all molecules, with means of $0.043$ for both choices of mediator. This clearly demonstrates that the crystal-agnostic estimator, $\langle \Lambda^{(\mathcal{L})}\rangle_{\Qrot}$ does a good job of capturing the typical modulation signal loss due to crystallisation for a given crystal point group when the molecular orientation within said crystal is unknown, irrespective of the internal molecular point group symmetry, DM mass, and mediator mass.

    When the physically realised crystal is known, the coordinate-aware estimator, $\Lambda^{(\mathcal{L}_\mathrm{crys})}_\mathrm{coord}$, remains accurate, albeit slightly less so than the gas phase comparison. Concretely, across all fifteen representative molecules, we find a mean absolute error, $\langle \Delta r_\mathrm{crys}\rangle = 0.114$ for the heavy mediator case, and $\langle \Delta r_\mathrm{crys}\rangle = 0.083$ for the light mediator. 
    
    Of these, paracetamol shows the largest deviation of $\langle \Delta r_\mathrm{crys}\rangle = 0.279$ for the heavy mediator case. However, this is somewhat of an outlier, as for the light-mediator case the same ratio for paracetamol is predicted almost perfectly, with $\langle \Delta r_\mathrm{crys}\rangle = 0.009$. Notably, the performance does not seem to depend on the internal molecular point group, suggesting that the coordinate-aware estimator is broadly applicable to a wide range of molecules. Perhaps most importantly, the estimator does an excellent job of predicting the signal loss for molecules with no internal symmetry, $H_\mathrm{crys} = C_1$, which are by far the most common class of molecules, especially when larger molecules are considered. Across the seven representative $C_1$ molecules considered here, the mean absolute error is $0.145$ for the heavy mediator ($n = 0$), falling to $0.079$ for the light-mediator ($n=2$). This demonstrates that the estimator remains broadly applicable even in the absence of internal molecular symmetry, where there are no constraints on the number of quadrupole modes that can survive. As $C_1$ molecules are expected to be the most common class of large organics, our shape-aware estimator should therefore be applicable to a wide range of realistic targets.

    \begin{table}[t]
    \centering
    \setlength{\tabcolsep}{4pt}
    \renewcommand{\arraystretch}{1.05}
    \begin{adjustbox}{max width=\textwidth}
    \begin{tabular}{c|c|c|c|c|c|c|c|c}
    \multirow{2}{*}{Name} & \multirow{2}{*}{$H_\mathrm{gas}$} & \multirow{2}{*}{$H_{\mathrm{crys}}$} & \multirow{2}{*}{$P_{\mathrm{crys}}$} & \multicolumn{2}{c|}{$\langle \Delta r_\mathrm{gas}\rangle_{\mathcal{L},m_\chi}$} & \multicolumn{2}{c|}{$\langle \Delta r_{\mathrm{crys}}\rangle_{m_{\chi}}$} & \multirow{2}{*}{Source} \\
    \cline{5-8}
    & & & & $n=0$ & $n=2$ & $n=0$ & $n=2$ & \\
    \hline\hline
    Benzene & $D_{6h}$ & $D_{6h}$ & $D_{2h}$ & 0.042 & 0.042 & 0.003 & 0.004 & \cite{Woinska:2016CCDCcc1jspdd, Woinska:2016HydrogenAtoms} \\ \hline
    1,3,5-Trichlorobenzene & $D_{3h}$ & $D_{3h}$ & $D_2$ & 0.041 & 0.041 & 0.053 & 0.051 & \cite{Hursthouse:2003CSDCommunication} \\ \hline
    p-Dichlorobenzene & $D_{2h}$ & $D_{2h}$ & $C_{2h}$ & 0.032 & 0.041 & 0.178 & 0.010 & \cite{Estop:1997pDichlorobenzene, Estop:1997CCDCpDichlorobenzene} \\ \hline
    Fluorobenzene & $C_{2v}$ & $C_{2v}$ & $D_4$ & 0.038 & 0.041 & 0.112 & 0.010 & \cite{Thalladi:1998Fluorobenzenes} \\ \hline
    Chlorobenzene & $C_{2v}$ & $C_{2v}$ & $D_{2h}$ & 0.028 & 0.043 & 0.198 & 0.026 & \cite{Nath:2015Chlorobenzene, Nath:2015CCDCChlorobenzene} \\ \hline
    o-Dichlorobenzene & $C_{2v}$ & $C_{2v}$ & $C_{2h}$ & 0.038 & 0.041 & 0.080 & 0.023 & \cite{Boese:2001CCDCcc5jxtd,Boese:2001Dichlorobenzenes} \\ \hline
    m-Dichlorobenzene & $C_{2v}$ & $C_{2v}$ & $C_{2h}$ & 0.041 & 0.041 & 0.091 & 0.050 & \cite{Aniola:2016CCDCcc1lvkss,Aniola:2016HighPressureDichlorobenzeneDibromobenzene} \\ \hline
    Phenol & $C_{s}$ & $C_s$ & $C_2$ & 0.044 & 0.045 & 0.173 & 0.018 & \cite{Zavodnik:1988Phenol} \\ \hline
    Salicylaldehyde & $C_s$ & $C_s$ & $C_{2h}$ & 0.047 & 0.047 & 0.118 & 0.179 & \cite{Kirchner:2011CCDCSalicylaldehyde,Kirchner:2011AldehydeHydrogenBonds} \\ \hline
    Toluene & $C_s$ & $C_s$ & $C_2$ & 0.041 & 0.041 & 0.028 & 0.012 & \cite{Marciniak:2016Toluene,Marciniak:2016CCDCcc1k0j25} \\ \hline
    \end{tabular}
    \end{adjustbox}
    \caption{Symmetry-breaking molecules used to compare our estimators to the true modulation signal lost by crystallisation. The first two columns show the internal point group symmetry of the gas phase and crystallised molecule, respectively, whilst the third column shows the point group symmetry of the physically-realised crystal. The remaining columns show the RMSE between the estimator and the true modulation loss for the heavy ($n=0$) and light-mediator ($n=2$), averaged over DM masses, and, for the gas phase estimator, also over the 11 Laue groups. The final column shows the data sources for the crystal comparison. See the main text for exact symbol definitions.}
    \label{tab:symmetryBreakingMols}
    \end{table}
    
\begin{figure}[t]
        \centering
        \begin{subfigure}{0.325\linewidth}
            \centering
            \includegraphics[width=\linewidth]{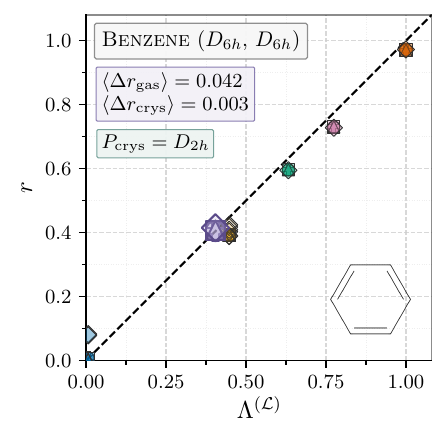}
        \end{subfigure}
        \hfill
        \centering
        \begin{subfigure}{0.325\linewidth}
            \centering
            \includegraphics[width=\linewidth]{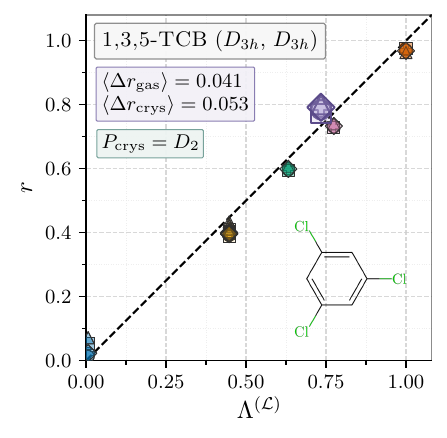}
        \end{subfigure}
        \hfill
        \begin{subfigure}{0.325\linewidth}
            \centering
            \includegraphics[width=\linewidth]{figures/proxy_comparisons/legend_combined.pdf}
        \end{subfigure}
        \hfill
        \begin{subfigure}{0.325\linewidth}
            \centering
            \includegraphics[width=\linewidth]{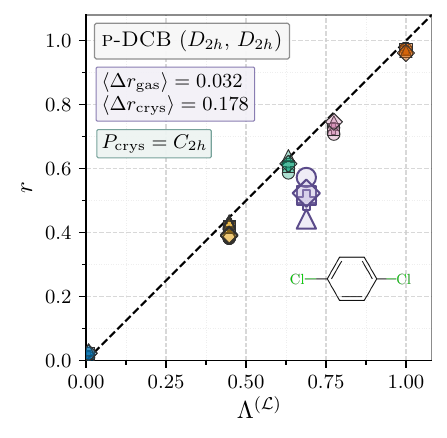}
        \end{subfigure}
        \hfill
        \begin{subfigure}{0.325\linewidth}
            \centering
            \includegraphics[width=\linewidth]{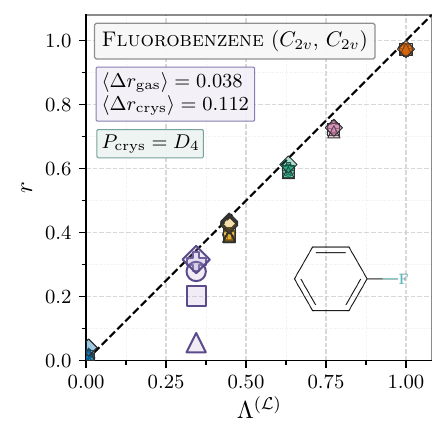}
        \end{subfigure}
        \hfill
        \begin{subfigure}{0.325\linewidth}
            \centering
            \includegraphics[width=\linewidth]{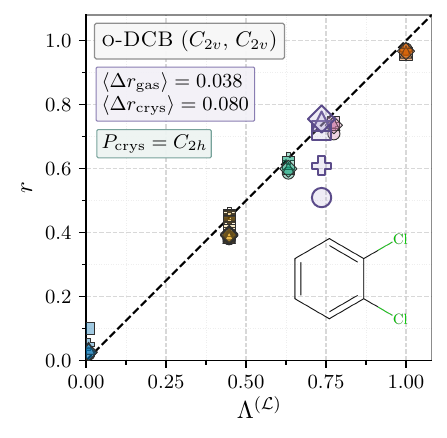}
        \end{subfigure}
        \hfill
        \begin{subfigure}{0.325\linewidth}
            \centering
            \includegraphics[width=\linewidth]{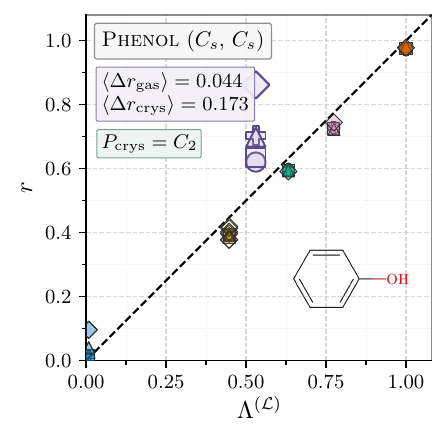}
        \end{subfigure}
        \hfill
        \begin{subfigure}{0.325\linewidth}
            \centering
            \includegraphics[width=\linewidth]{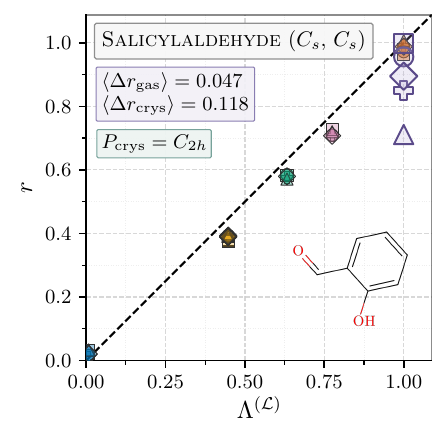}
        \end{subfigure}
        \hfill
        \begin{subfigure}{0.325\linewidth}
            \centering
            \includegraphics[width=\linewidth]{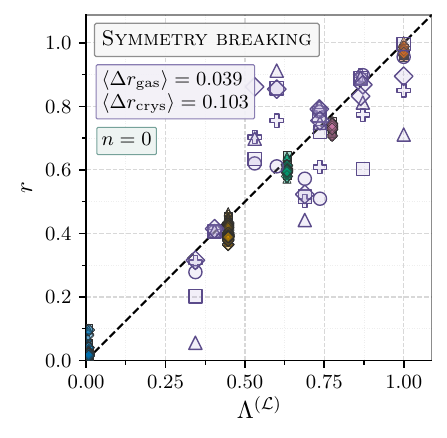}
        \end{subfigure}
         \caption{Comparisons of the estimators to the true $f_\mathrm{RMS}$ ratios for a subset of the molecules in Table~\ref{tab:symmetryBreakingMols}, assuming a heavy mediator ($n=0$). The bracketed point group names after each molecule name correspond to the pair $(H_\mathrm{gas}, H_\mathrm{crys})$, the gas-phase and crystal-phase internal point groups, respectively. The smaller points show the comparison of the agnostic estimator, $\langle \Lambda^{(\mathcal{L})}\rangle_\Qrot$, to the orientation- and detector-averaged $f_\mathrm{RMS}$ ratio, for each of the 11 Laue classes $\mathcal{L}$, for a DM mass $m_\chi \in \{5,10,25,50,100\}\,\mathrm{MeV}$, and coloured according to their quadrupole survival class, $\mathsf{Q}_k$. The larger purple markers show the comparison of the coordinate aware estimator, $\Lambda_\mathrm{coord}^{(\mathcal{L}_\mathrm{crys})}$, to the physically-realised polymorph of each molecular crystal. The bottom right plot shows the results for all molecules in Table~\ref{tab:symmetryBreakingMols} for the $n = 0$ case overlaid.}
        \label{fig:brokenSymmetriesn0}
    \end{figure}

    In Table~\ref{tab:symmetryBreakingMols} we show the analogous results for the symmetry-breaking molecules, and plots of a subset in Fig.~\ref{fig:brokenSymmetriesn0}. As with representative molecule set, the agreement between the crystal-agnostic estimator and the true modulation signal loss is incredibly strong, with a global MAE of $\langle \Delta r_\mathrm{crys}\rangle = 0.042$ for the heavy mediator cases, $0.039$ for the light-mediator case, and no single molecule with $\langle \Delta r_\mathrm{crys}\rangle > 0.05$. 
    The coordinate-aware estimator once again shows a larger deviation from the true modulation signal loss, with a global MAE of $\langle \Delta r_\mathrm{crys}\rangle = 0.103$ for the heavy mediator, falling to $\langle \Delta r_\mathrm{crys}\rangle = 0.038$ for the light-mediator case.
    
    Of the molecules considered, 
    estimating $r$ from $\Lambda_\text{coord}$ works the least well for chlorobenzene for the heavy mediator, with $\langle \Delta r_\mathrm{crys}\rangle = 0.198$, and salicylaldehyde for the light-mediator case, for which $\langle \Delta r_\mathrm{crys}\rangle = 0.179$. These results suggest that occasionally, the particular combination of crystal symmetry, molecular symmetry, and their relative alignment can be just so that the estimator does a bad job of predicting the modulation loss, even when some shape information is taken into account. Fortunately, this seems to be rare, as just 3 of the 10 symmetry-breaking molecules have $\langle \Delta r_\mathrm{crys}\rangle > 0.15$ for $n=0$, and only one for $n =2$. Across both molecule sets, these numbers change to 6/25 of and 4/25, respectively, suggesting that we can estimate the modulation signal loss accurately about 80\% of the time. Comparatively, 10 of the 25 molecules have $\langle \Delta  r_\mathrm{crys}\rangle < 0.1$ for $n = 0$, rising to 18/25 for the light-mediator, $n=2$ case. This once again highlights the improved performance in the light-mediator cases. Importantly, the estimates for the symmetry-breaking molecules do not appear to be significantly degraded, if at all, when compared to the representative molecules. This suggests that, in general, the true symmetry group of the molecule is in fact the one that drives the modulation signal, and not an approximate higher symmetry.

    Taken together, our results demonstrate that symmetry-based estimators are an incredibly strong predictive tool for the modulation signal loss due to crystallisation. The crystal-agnostic estimator captures the average modulation loss with consistently small errors, and correctly identifies that the 230 crystal space groups reduce to just 5 classes on average, determined by their quadrupole survival fractions. The physically-realised crystal, however, is not described by this mean, and instead corresponds to one particular point in that distribution, fixed by its crystal symmetry and molecular embedding. The role of the coordinate-aware estimator is therefore more demanding, since it must predict where the realised molecular crystal system lies within the symmetry-allowed distribution. The fact that it does so with small errors, with mean $\langle \Delta r_\mathrm{crys}\rangle = 0.088$ globally, and $\langle \Delta r_\mathrm{crys}\rangle < 0.1$ for the majority of systems demonstrates that the estimator retains predictive power even after the crystallographic degrees of freedom are fixed.

    \subsection{Dark matter daily modulation} \label{sec:demoTSB}

To illustrate in greater detail how the anisotropy of a molecular crystal depends on its geometry, we turn to trans-stilbene. 
Trans-stilbene is a relatively bright fluorescent scintillator, which has previously been proposed as an anisotropic material for sub-GeV dark matter direct detection~\cite{Blanco:2021hlm}. 
Its translationally invariant unit cell contains four molecules: $A$, $B$, $A'$, and $B'$, where $A'$ and $B'$ are the images of molecules $A$ and $B$ under a translation and $180^\circ$ rotation about the crystal $\hat b$ axis. The rotational geometry is similar to the diagram of crystalline benzene shown in Fig.~\ref{fig:unit_cell_schematic}, but with the long axes of the trans-stilbene molecules arranged in a herringbone pattern. Molecules $A$ and $B$ have slightly different geometries~\cite{Li:2021CCDCcc28bj38, Li:2021DeuteratedAlkenes}, causing slight differences in their electronic excited states: however, their isotropic average scattering rates are quite similar, with $\langle R_A \rangle \simeq \langle R_B \rangle$ to within $\leq 1\%$.

As the first step in the rate calculation, we calculate the partial rate matrix $K^{(\ell)}_{m' m}$ from \eqref{eq:partialrateK}, for the individual molecules $A$, $B$, etc., as well as the crystal unit cell. For simplicity, in this section we define the velocity coordinate system so that $\hat z$ points in the instantaneous $\vec v_E$ direction, so that the lab-frame DM wind origin is directly overhead. Because the SHM is azimuthally symmetric about the $\vec v_E$ axis, the partial rate matrix $K^{(\ell)}_{m' m}$ vanishes for all $m' \neq 0$.
The details of the Cartesian coordinate system used to define $|f_s(\vec q)|^2$ are provided in Appendix~\ref{sec:coords}.

\clearpage

    \clearpage

    \clearpage

    \clearpage

\begin{figure}
\centering
\includegraphics[width=\textwidth]{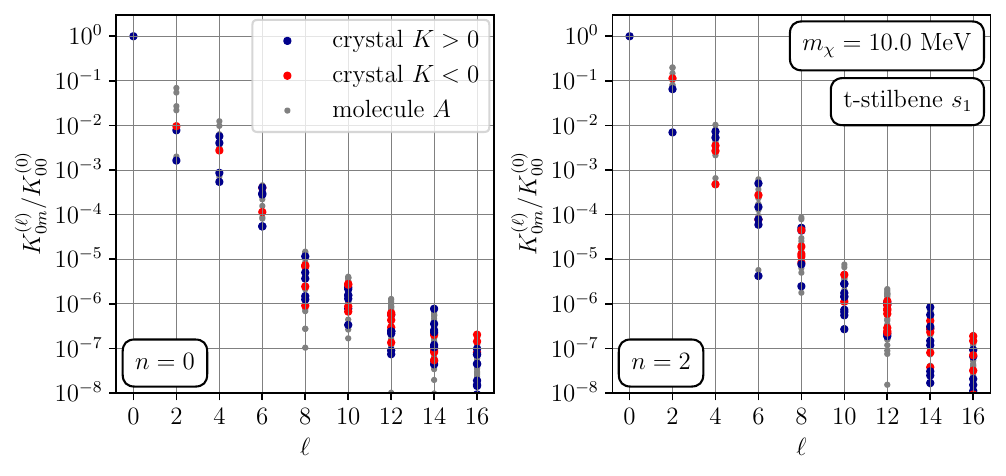}
\caption{Values of the partial rate matrix $K^{(\ell)}_{0m}$ for a 10 MeV dark matter candidate, normalised by the isotropic $\ell = 0$ component, for a transition from the ground state to the first excited state in crystalline trans-stilbene. All non-zero values of $|m|\leq \ell$ are shown in the same column. We show the result for spin-independent heavy mediator ($n=0$, left panel) and light mediator ($n = 2$, right panel) dark matter models, assuming a Standard Halo Model velocity distribution. The blue and red dots show positive and negative values of $K^{(\ell)}_{0m}$, respectively, for the crystal averaged form factor $f_{\ell m}^2$. In gray, we show the magnitude of $K^{(\ell)}_{0m}$ for one of the four constituent molecules ($A$) in the unit cell. In this coordinate system, the $C_{2h}$ point group causes the $m < 0$ components of the crystal $K^{(\ell)}_{0m}$ to vanish. Because the symmetry axes of molecule $A$ are not aligned with the crystal lattice, it has generally non-zero values for all $|m| \leq \ell$ in this coordinate frame, despite the fact that the isolated molecule is also $C_{2h}$ symmetric.}
\label{fig:mcalK}
\end{figure}

Figure~\ref{fig:mcalK} shows the values of $K^{(\ell)}_{0 m}$ for both heavy mediator ($n = 0$) and light mediator ($n= 2$) dark matter models. In this example we focus on transitions to the first excited state, $g \rightarrow s_1$, and we take the dark matter mass to be $m_\chi = 10$\,MeV. As a first demonstration of how crystallisation can cause destructive interference in the modulation signal, we show $K^{(\ell)}$ for both the crystal form factor (in blue and red) and for a single molecular component of the unit cell, $A$ (in gray), each normalised according to its isotropic component $K^{(0)}_{00}$. 
Recall from \eqref{eq:rateGK} that the scattering rate for a detector in orientation $\mathcal D \in SO(3)$ is proportional to the sum 
\begin{align}
R(\mathcal D) = \sum_{\ell} R^{(\ell)}(\mathcal D) \propto \sum_{\ell,m',m} G^{(\ell)}_{m' m}(\mathcal D) K^{(\ell)}_{m ' m},
\label{eq:rateDGK}
\end{align}
where $G^{(\ell)}(\mathcal D)$ is a $(2\ell + 1) \times (2\ell +1)$ real representation of $SO(3)$.
The degree of anisotropy is determined by the largest modes with $\ell \neq 0$, while $\ell =0$ sets the isotropic average rate. 
As we anticipated in Section~\ref{sec:Gestimate}, the higher $\ell$ modes drop off exponentially.
This is largely due to the SHM velocity distribution, which does not have any strong features at small angular scales. After combining the SHM $g_\chi(\vec v)$ with the material form factor $|f_s(\vec q)|^2$, we find that both types of mediator model $n$ are well fit by $K^{(\ell)} \sim e^{- \ell/\ell_0} K^{(0)}$ with $\ell_0 \simeq 1$.
Although the higher $\ell > 4$ modes have negligible effects on the scattering rates in our examples, this could in principle change if we were to replace the SHM $g_\chi(\vec v)$ with a velocity distribution with more features at small angular scales.

Before we conduct a detailed survey of the possible detector orientations $\mathcal D$, we can estimate the size of the anisotropy by inspecting the $\ell = 2$ components. 
In the $n= 0$, 10\,MeV model, the cancellation between the four members of the unit cell is unexpectedly large: 
although molecule $A$ has $\ell = 2$ components of size $10^{-1} K^{(0)}$, the orientations of the other three molecules conspire to reduce the largest $K^{(2)}$ component by an order of magnitude. 
In the light mediator model, $n = 2$, the cancellation is not nearly as extreme. Likewise, the strong $n = 0$ cancellation is specific to the first excited state: there are several transitions with slightly higher excitation energies where $K^{(2)} \sim 0.1 K^{(0)}$ for both molecule $A$ and the full crystal. 

\begin{figure}
\centering
\includegraphics[width=\textwidth]{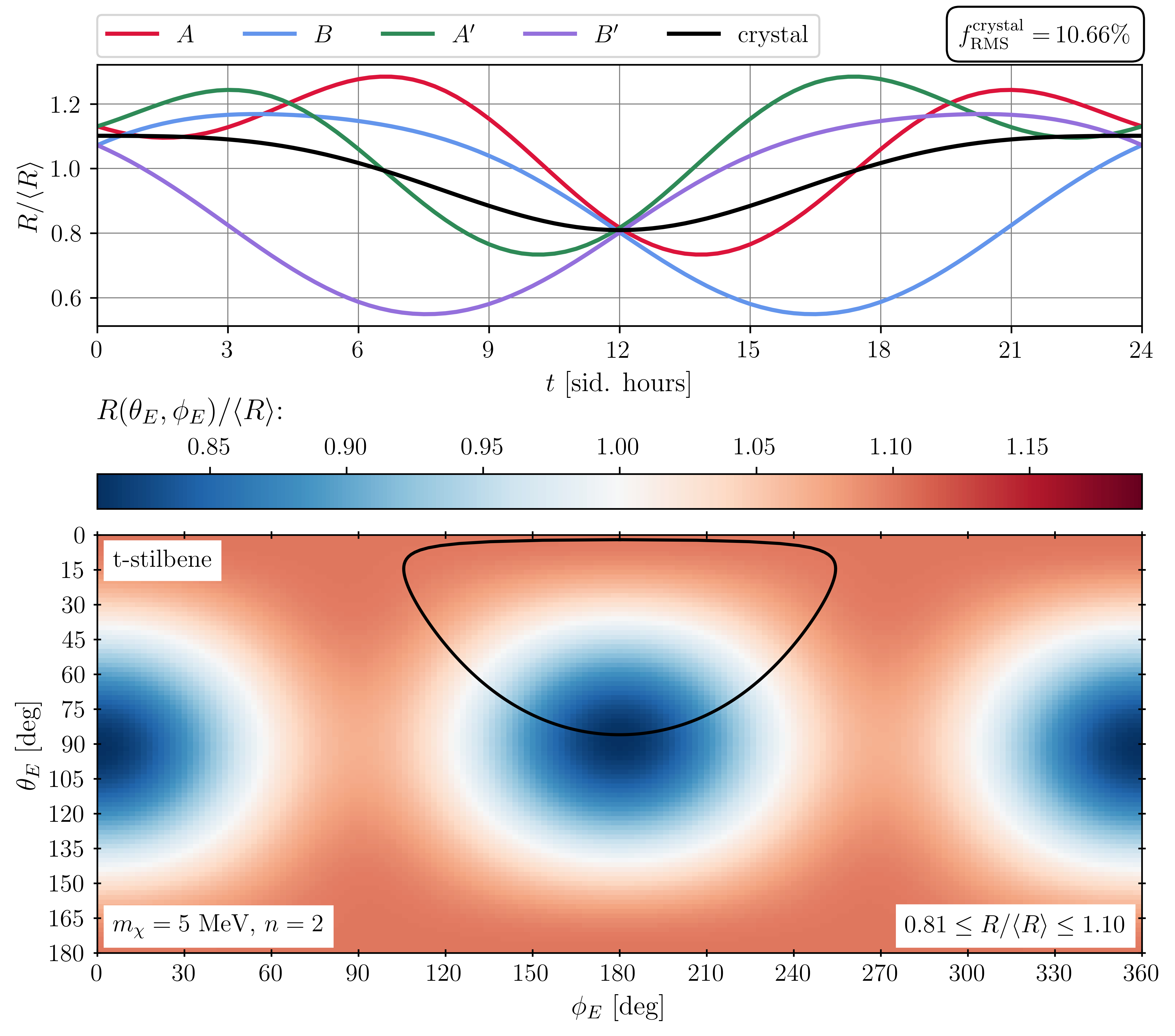}
\caption{
    An example of daily modulation in crystalline trans-stilbene, for a dark matter model with a light mediator ($n = 2$) and mass $m_\chi = 5$\,MeV. The lower panel shows the scattering rate $R(\theta_E, \phi_E)$ as a function of the Earth velocity vector direction $\hat v_E$, relative to the crystal lattice coordinate system. For this dark matter model, the rate is maximised when $\vec v_E \propto \pm \hat z$, and minimised when $\vec v_E \propto \pm \hat x$. The black cycle shows $\hat v_E(t)$ for one particular initial orientation, with the angle between the north pole and $\vec v_E$ ($\theta_N \simeq 42^\circ \pm 6^\circ$) set to its annual average. 
    The upper panel plots $R(t)$ for the chosen initial orientation, showing the total scattering rate in black, and the four  distinct single-molecule contributions in colour. Each rate is separately normalised by its isotropic average, $\langle R \rangle = R^{(\ell = 0)}$, and includes a sum over the first twelve excited states. For this detector orientation and dark matter model, the $f_\text{RMS}$ amplitude of the crystal, $R(t)$, is approximately $10.7\%$, while the $f_\text{RMS}$ scores of the individual $A,A'$ and $B,B'$ molecules are $16.2\%$ and $24.5\%$, respectively.
}
\label{fig:rate2d}
\end{figure}

Figure~\ref{fig:rate2d} shows an example of how trans-stilbene could be used in a daily modulation experiment. In this setup, we pick a particular initial detector orientation $\mathcal D_0$ at time $t=0$, and we allow the rotation of the Earth about its axis to let $\mathcal D(t)$ trace out a path through $SO(3)$. 
Equivalently, in a crystal-centric coordinate system, the Earth velocity vector is pointed in the initial direction 
\begin{align}
\vec v_E(t = 0) &= (\mathcal D_0^{-1} \hat z) |\vec v_E|;
\end{align}
then, as the Earth rotates, $\hat v_E$ traces out a circular path on the sphere, keeping the angle between $\vec v_E$ and the north pole fixed.
We use this frame of reference for the lower panel of Fig.~\ref{fig:rate2d}: rather than attempting to plot the rate $R$ as a function of $\mathcal D \in SO(3)$, a three dimensional space, we plot the conveniently two-dimensional $R(\mathcal D^{-1} \vec v_E)$, i.e.~the scattering rate as a function of Earth velocity direction relative to the crystal lattice. For a more complete explanation, see Appendix~\ref{sec:coords:Rsphere}. 
From Fig.~\ref{fig:rate2d}, we see that the rate is maximised when $\vec v_E \propto \pm\hat z$, and minimised when $\vec v_E \propto \pm \hat x$. To map these Cartesian directions onto the crystallographic $\vec a, \vec b, \vec c$ lattice vectors, see Appendix~\ref{sec:coords:crystal}. 

To maximise the daily modulation amplitude, we align the crystal such that the north pole is pointed along the $\theta_n = 44^\circ$, $\phi_n = 180^\circ$ crystal direction,  so that the DM wind closely approaches both the minimum and the maximum in $R(\theta_E, \phi_E)$ over the course of the day. This cycle in $\theta_E(t)$, $\phi_E(t)$, defined so that the Earth velocity unit vector $\hat n_E = (\theta_E, \phi_E)$ is at fixed angle $\theta_N = 42^\circ$ from the north pole, is indicated by the black curve on the 2d heat plot.
The corresponding $R(t)$ for the scattering rate in the crystal is shown in the upper panel, also in black;  its modulation amplitude is $f_\text{RMS} \simeq 10.7\%$. For more details about the geometric arrangement and coordinate system, see Appendix~\ref{sec:coords:Rsphere}. 

Figure~\ref{fig:rate2d} also shows the contribution to $R(t)$ from each of the four molecules, for the same detector orientation. Here the destructive interference is clearly visible: during the first half of the cycle, $t < 12$\,h, the contributions from $A$ and $B'$ are mostly opposed, while for $t > 12$\,h it is $A'$ and $B$ that mostly cancel each other. If the crystal contained only copies of molecule $A$ or $B$, the $f_\text{RMS}$ amplitude would instead have been $16.2\%$ or $24.5\%$, respectively. A different initial orientation $\mathcal D_0$ optimised for a single molecule could achieve an $f_\text{RMS}$ as large as $28.6\%$ for this 5\,MeV, $n=2$ dark matter model. The suboptimal arrangement of molecules in the crystal has reduced the maximum possible $f_\text{RMS}$ by nearly a factor of three.  

Both of our metrics identify trans-stilbene as a promising molecular crystal, more strongly so as we ascend the hierarchy of estimators. Without knowing the molecular coordinates, the orientation averaged estimator, $\langle \Lambda\rangle_\mathcal{T} \simeq 0.775$, already indicating strong survival of the leading anisotropic modes. The coordinate-aware estimator gives the refined estimate $\Lambda_\mathrm{coord} \simeq 0.864$. This is in line with $f_\mathrm{RMS} \simeq 10.7\%$, which as we show in Sec.~\ref{sec:fRMSprediction} is somewhat above average for the light mediator ($n = 2$) case. 

\subsection{Can we predict $f_\mathrm{RMS}$?}\label{sec:fRMSprediction}
We have so far demonstrated that our estimators are excellent at predicting the \textit{relative} loss of modulation signal due to crystallisation. A more demanding question, however, is whether our estimators are able to predict the \textit{absolute} quality of molecular crystals, at least relative to one another. In this section we address this question by comparing our absolute estimator, $\Lambda_\mathrm{coord}$, to the true values of $f_\mathrm{RMS}$. 

To make this comparison, we use the same molecular crystals considered in Tables~\ref{tab:representativeMols} and~\ref{tab:symmetryBreakingMols}, considering only the physically realised point group symmetries and embeddings. As before, we average over 100 initial detector orientations, $\mathcal{D}_0$, and consider five DM masses, $m_\chi \in \{5, 10, 25, 50, 100\}\,\mathrm{MeV}$. For each, we also consider both the heavy and light mediator cases, $n = 0$, and $n = 2$, respectively. 

    \begin{figure}[t]
        \centering
        \begin{subfigure}{0.495\linewidth}
            \centering
            \includegraphics[width=\linewidth]{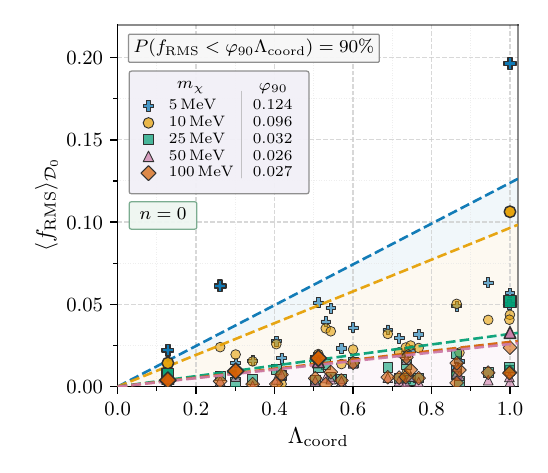}
        \end{subfigure}
        \hfill
        \begin{subfigure}{0.495\linewidth}
            \centering
            \includegraphics[width=\linewidth]{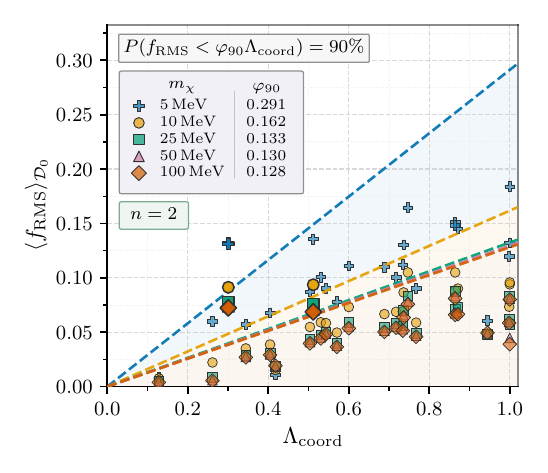}
        \end{subfigure}
        \caption{Scatter of the detector-averaged modulation signal amplitudes $\langle f_\mathrm{RMS}\rangle_{\mathcal{D}_0}$ for the physically realised crystals, as a function of their shape-aware modulation loss estimator, $\Lambda_\mathrm{coord}$ for left: the heavy mediator ($n= 0$) case, and right: the light mediator ($n = 2$) case. The dashed lines show the line $\langle f_\mathrm{RMS}\rangle_{\mathcal{D}_0} = \varphi_{90} \Lambda_\mathrm{coord}$, based on the log-normal fits to $\varphi$ below which approximately $90\%$ of points should lie.}
        \label{fig:proxyScatters}
    \end{figure}

\begin{table}[h]
    \centering
    \setlength{\tabcolsep}{7pt}
    \renewcommand{\arraystretch}{1.10}
    \begin{adjustbox}{max width=\textwidth}
    \begin{tabular}{c|ccc|ccc}
    \multirow{2}{*}{$m_\chi\,[\mathrm{MeV}]$} & \multicolumn{3}{c|}{Heavy mediator ($n=0$)} & \multicolumn{3}{c}{Light mediator ($n=2$)} \\
    \cline{2-7}
    & $\varphi_{90}$ & $\mu_\varphi$ & $\sigma_\varphi$ & $\varphi_{90}$ & $\mu_\varphi$ & $\sigma_\varphi$ \\
    \hline\hline
    $5$   & 0.124 & $-$2.94 & 0.664 & 0.291 & $-$1.90 & 0.520 \\ 
    \hline
    $10$  & 0.096 & $-$3.30 & 0.750 & 0.162 & $-$2.32 & 0.389 \\ 
    \hline
    $25$  & 0.032 & $-$4.31 & 0.678 & 0.133 & $-$2.54 & 0.405 \\ 
    \hline
    $50$  & 0.026 & $-$4.69 & 0.809 & 0.130 & $-$2.61 & 0.443 \\ 
    \hline
    $100$ & 0.027 & $-$4.52 & 0.713 & 0.128 & $-$2.64 & 0.457 \\
    \hline
    \end{tabular}
    \end{adjustbox}
    \caption{Fitted log-normal parameters for the residual prefactor $\varphi = \langle f_\mathrm{RMS}\rangle_{\mathcal{D}_0}/\Lambda_\mathrm{coord}$. For each DM mass and mediator choice, we show the fitted mean, $\mu_\varphi$, and width, $\sigma_\varphi$, of $\ln\varphi$, together with the corresponding $90\%$ value, $\varphi_{90}$.}
    \label{tab:alphaFits}
    \end{table}

We show our results in Figure~\ref{fig:proxyScatters}. Naturally, $\Lambda_\mathrm{coord}$ alone is unable to capture the scale of $f_\mathrm{RMS}$, which depends on the mediator model, DM mass, and other details of electronic structure. Importantly, however, we see that $\Lambda_\mathrm{coord}$ is able to strongly bound $f_\mathrm{RMS}$. That is, we find that in general that
\begin{equation}
    \langle f_\mathrm{RMS}\rangle_{\mathcal{D}_0} \lesssim \varphi(m_\chi) \Lambda_\mathrm{coord},
\label{eq:Lambda_fRMS}
\end{equation}
with $\varphi$ some unknown parameter depending on the non-symmetry components of the scattering rate. This is to be expected; comparing the estimator in  \eqref{eq:lambdaCoord} to $f_\mathrm{RMS}$ in  \eqref{eq:fRMSsq}, we see that
\begin{equation}
    \varphi \simeq  \sqrt{\gamma W_{2}^{(H)}}.
\end{equation}
The immediate consequence of this is that $\Lambda_\mathrm{coord}$ can still be used to rank the quality of candidate molecular crystals, with some uncertainty dependent on the distribution of $\varphi$ values. In particular, we find that $\varphi$ is log-normal distributed:
\begin{equation}
    \ln \varphi \sim \mathcal{N}(\mu_\varphi,\sigma_\varphi^2),
\end{equation}
with mean $\mu_\varphi$ and standard deviation $\sigma_\varphi$. Given this, the probability that molecular crystal $A$ has a bigger $f_\mathrm{RMS}$ that molecular crystal $B$, is
\begin{equation}\label{eq:pairwiseRanking}
    P\left(f_\mathrm{RMS}^A > f_\mathrm{RMS}^B\right) = \Phi\left[\frac{\ln\left(\Lambda^A_\mathrm{coord}/{\Lambda_\mathrm{coord}^B}\right)}{\sqrt{2}\,\sigma_\varphi}\right],
\end{equation}
where $\Phi$ denotes the CDF of the normal distribution. We give the values of $\mu_\varphi$ and $\sigma_\varphi$ for each $m_\chi$ and $n$ value in Table~\ref{tab:alphaFits}, along with $\varphi_{90}$, the envelope value that $90\%$ of $f_\mathrm{RMS}$ points lie below. Notably, $\varphi$ tends to take much larger values for the light mediator ($n = 2$) case, and with far less variance, suggesting that modulation searches are more sensitive to light mediators than heavy mediators. There is also a clear trend with the mass: lighter DM tends to produce a larger modulation signal, as they are closer to the kinematic threshold and so depend more strongly on directionality.

\begin{figure}[t]
        \centering
        \begin{subfigure}{0.495\linewidth}
            \centering
            \includegraphics[width=\linewidth]{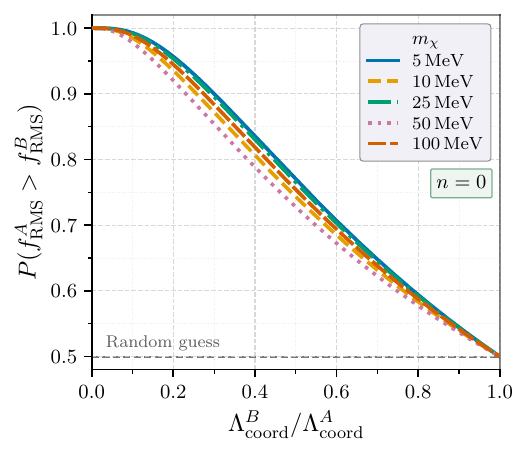}
        \end{subfigure}
        \hfill
        \begin{subfigure}{0.495\linewidth}
            \centering
            \includegraphics[width=\linewidth]{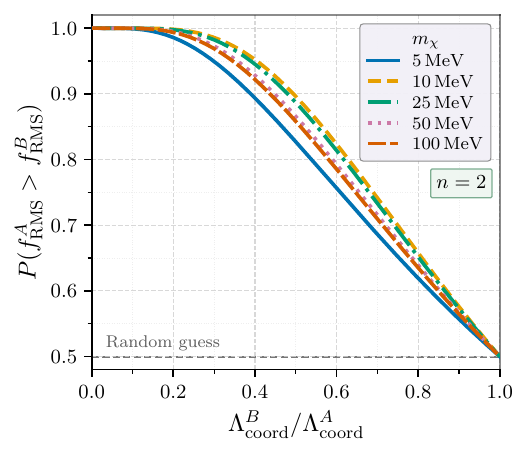}
        \end{subfigure}
        \caption{Pairwise ranking probability given in  \eqref{eq:pairwiseRanking} for two molecular crystals, $A$ and $B$, computed from the fitted log-normal scatter distribution of $\varphi$. Each curve shows the probability that molecule $A$ generates a larger modulation signal, $P(f_\mathrm{RMS}^A>f_\mathrm{RMS}^B)$, for given a shape-aware estimator ratio, $\Lambda^B_\mathrm{coord}/\Lambda^A_\mathrm{coord}$. The left panel shows the heavy mediator case ($n=0$), whilst the right panel shows the light mediator case ($n=2$).}       
        \label{fig:rankingProbability}
    \end{figure}

Figure~\ref{fig:rankingProbability} shows the pairwise ranking probability given in \eqref{eq:pairwiseRanking} as a function of $\Lambda^B_\mathrm{coord}/\Lambda_\mathrm{coord}^A$. As a result of the smaller spread in $\varphi$, we see that the ratio of $\Lambda_\mathrm{coord}$ more accurately predicts the ordering of the true modulation fractions for the light mediator case, $n=2$, than for the heavy mediator case, $n=0$. 

For $n=2$, even relatively modest separations in the estimator lead to useful ranking probabilities. For example, $\Lambda^B_\mathrm{coord}/\Lambda^A_\mathrm{coord}=0.5$, corresponding to a factor of two separation, gives a probability of $85$--$90\%$ that the larger-$\Lambda_\mathrm{coord}$ crystal also has the larger $f_\mathrm{RMS}$. For more extreme ratios, starting at around $0.2$ this probability tends to $100\%$. The heavy mediator case, $n=0$, shows the same qualitative behaviour, but is slightly less accurate on average due to the broader distribution of $\varphi$. In this case the corresponding probabilities are approximately $75$--$80\%$ for $\Lambda^B_\mathrm{coord}/\Lambda^A_\mathrm{coord}=0.5$, increasing to approximately $95\%$ at $\Lambda^B_\mathrm{coord}/\Lambda^A_\mathrm{coord}=0.2$.

These results clearly demonstrate how the estimator should be used. Small differences in $\Lambda_\mathrm{coord}$ should not be interpreted as a definitive ordering of molecular crystals, as the prefactor from electronic structure can still reverse the ranking. However, once two candidates differ significantly in their modulation signal loss estimates, the probability of such a reordering becomes small. The estimator is therefore best at identifying the tails of the distribution, allowing us to efficiently identify best and worst molecules, which are precisely the regimes that we care about. As such, our estimator should serve as an excellent screening tool, allowing us to efficiently restrict the candidate space to the most promising molecular crystals before performing expensive electronic-structure calculations and full rate computations.

%-----------------------------------------------------------------------------
\section{Conclusions}
\label{sec:conclusions}
% -----------------------------------------------------------------------------

Predicting from first principles how a material will respond to dark matter-electron scattering is a highly resource-intensive endeavour, and this makes it challenging to evaluate large lists of candidate detector materials. If we could estimate ahead of time which crystalline structures and molecular geometries are likely to provide the highest degrees of anisotropy, then we can direct our computational resources towards the most promising candidates for directional dark matter experiments.
In this work, we accomplish this by using simple properties of discrete symmetry groups to quantify the typical anisotropy of every possible crystal class. 

Our first result is that the 230 space groups can be grouped into just five \emph{quadrupole survival classes} $\mathsf{Q}_k$ listed in Table~\ref{tab:xiInfinityLaueGroups}, for $k = 0, 1, 2, 3, 5$. By decomposing the scattering rate $R$ into the contributions from each spherical harmonic mode $(\ell, m)$ in the material form factor and the Standard Halo Model dark matter velocity distribution, we have demonstrated that the relative anisotropy is well approximated by the $\ell = 2$ quadrupole moments. Crystals with smaller values of $k$ are likely to be less anisotropic, because the electronic structure is more symmetric. Notably, this $\mathsf{Q}_k$  ranking of the crystal space groups applies generally to molecular, ionic, or atomic crystals. For that matter, even liquid crystals can be categorised in this language: a liquid crystal created by applying a strong electric field $\vec E$, and thus aligning the microscopic components along an axis, would fall into the $\mathsf{Q}_1$ class, for example. 

As demonstrated in Sec.~\ref{sec:oddell}, the odd-$\ell$ spherical harmonic modes do not contribute to the spin-independent scattering rate. Parity-odd spin-dependent interactions could, on the other hand, induce an $\ell = 1$ contribution to the rate proportional to the degree of parity violation in the crystal. In addition to $\mathsf{Q}_k$, future studies of spin dependence in chiral materials should also consider the dipole survival classes, $\mathsf{D}_{j}$ ($j = 0, 1, 2, 3$).

Our other proxies for anisotropy use additional information about the crystal composition to make increasingly specific predictions for the potential daily modulation amplitude of a material.
Specialising to molecular crystals, Sec.~\ref{sec:internal} goes beyond $\mathsf{Q}_k$ to quantify how much anisotropy is lost when a molecule, symmetric under its own point group $H^\text{mol}$, is embedded within a crystal lattice with symmetry $H^\text{crys}$.  
It is usually easier to determine the molecular geometry and $H^\text{mol}$ than it is to predict \textit{ab initio} the crystal geometry. For these molecules with unknown crystal geometries, $\langle \Lambda \rangle_\Qrot$ of \eqref{eq:LambdaQ} provides a useful proxy for the average crystal $f_\text{RMS}$ daily modulation amplitude. More precisely, $\langle \Lambda \rangle_\Qrot$ is correlated with the \emph{ratio} between the crystal and single-molecule $f_\text{RMS}$ values. It can be used to predict how much anisotropy is lost due to crystallisation.

If the molecular and crystal geometries are both known, then we recommend the coordinate-aware proxy for anisotropy, $\Lambda_\text{coord}$ from \eqref{eq:lambdaCoord}, which can be used to estimate the absolute size of $f_\text{RMS}$. In Fig.~\ref{fig:proxyScatters}, we find that the calculated values of $f_\text{RMS}$ tend to fall within a range $0 < f_\text{RMS} \leq \varphi(m_\chi) \Lambda_\text{coord}$, for a mass-dependent proportionality constant $\varphi(m_\chi)$ that can be determined empirically from a relatively small set of electronic structure calculations.
More importantly, $\Lambda_\text{coord}$ by itself does a good job at ranking molecules according to their degree of anisotropy, as we show in Fig.~\ref{fig:rankingProbability}. The relation between $\Lambda_\text{coord}$ and the absolute value of $f_{\rm RMS}$ is necessarily probabilistic, because molecules with the same symmetry can populate their allowed angular modes differently. Nevertheless, when two estimator values differ by a factor
of approximately $1.5$--$2$, the correct ranking is obtained in roughly $75\%$ of cases, depending on the dark matter mass and mediator model. For ratios of order $3$ or larger, the ranking probability rises to about $90\%$ or better. The estimators are therefore especially effective at separating clearly promising from clearly unfavourable candidates, which is precisely the task required in large-scale screening.

Directional detection provides one of the most powerful ways to distinguish a dark matter signal from irreducible Standard Model backgrounds.
A high-throughput search for new direct detection materials will still need to perform electronic structure calculations in order to precisely predict the shape and phase of the daily modulating rate, $R(t)$. However, using $\Lambda_\text{coord}$, we can start our search with the crystal candidates that are most likely to have particularly large anisotropies. A strongly anisotropic molecule can still be a poor target if its crystal embedding projects out the relevant angular modes, while a more modest molecular anisotropy can remain useful if the crystal preserves its dominant directions. By reducing the crystallographic search space from 230 space groups to 11 even-response Laue classes and, at leading order, to 5 quadrupole survival classes, and by providing a hierarchy of increasingly material-specific estimators culminating in the coordinate-aware screening of resolved structures, the framework developed here makes symmetry a practical design variable for large-scale searches for directional molecular dark matter detector materials.

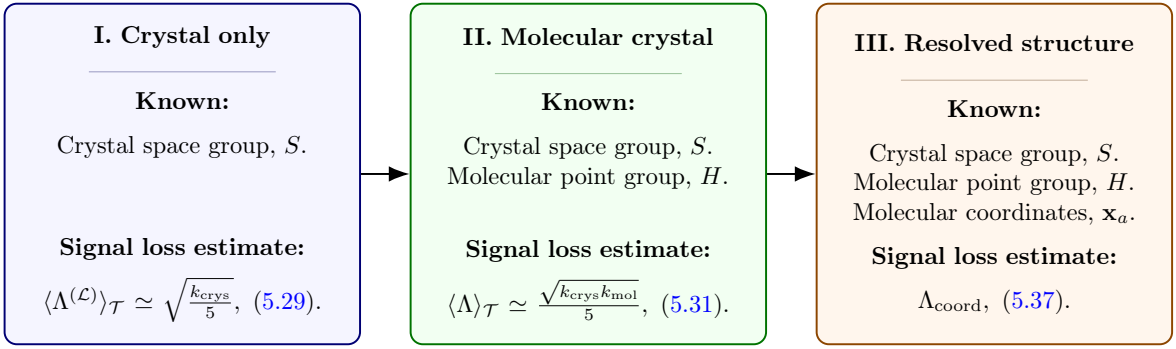
\begin{figure}[t]
    \centering
    \begin{adjustbox}{max width=\textwidth}
    \begin{tikzpicture}[
        font=\small,
        >=Latex,
        stage/.style={draw, rounded corners=5pt, thick, align=center, text width=0.29\textwidth, minimum height=4.9cm, inner sep=9pt},
        arrow/.style={-{Latex[length=3mm]}, thick},
    ]

    \node[stage, fill=blue!4, draw=blue!45!black] (L1) {
        {\bfseries I. Crystal only}
        \par
        {\color{blue!40!black!40}{\rule{0.6\linewidth}{0.2pt}}}
        \par\smallskip
        \textbf{Known:}\par\medskip
        Crystal space group, $S$.\par
        \phantom{Molecular point group, $H$.}\par
        \phantom{Molecular coordinates, $\vec{x}_a$.}
        \par\medskip
        \textbf{Signal loss estimate:}\par\medskip
        $\langle\Lambda^{(\mathcal L)}\rangle_\Qrot \simeq \sqrt{\frac{k_{\rm crys}}{5}}, \ \text{(\ref{eq:differentGroupAgnostic})}$.
    };

    \node[stage, fill=green!5, draw=green!45!black, right=7mm of L1] (L2) {
        {\bfseries II. Molecular crystal}
        \par
        {\color{green!40!black!40}{\rule{0.6\linewidth}{0.2pt}}}
        \par\smallskip
        \textbf{Known:}\par\medskip
        Crystal space group, $S$.\par
        Molecular point group, $H$.\par
        \phantom{Molecular coordinates, $\vec{x}_a$.}
        \par\medskip
        \textbf{Signal loss estimate:}\par\medskip
        $\langle\Lambda\rangle_{\Qrot} \simeq \frac{\sqrt{k_{\mathrm{crys}}k_{\mathrm{mol}}}}{5}, \ \text{(\ref{eq:LambdaQ})}$.
    };

    \node[stage, fill=orange!7, draw=orange!55!black, right=7mm of L2] (L3) {
        {\bfseries III. Resolved structure}
        \par
        {\color{orange!40!black!40}{\rule{0.6\linewidth}{0.2pt}}}
        \par\smallskip
        \textbf{Known:}\par\medskip
        Crystal space group, $S$.\par
        Molecular point group, $H$.\par
        Molecular coordinates, $\vec{x}_a$.
        \par\medskip
        \textbf{Signal loss estimate:}\par\medskip
        $\Lambda_{\mathrm{coord}},\ \text{(\ref{eq:lambdaCoord})}$.
    };

    \draw[arrow] (L1.east) -- (L2.west);
    \draw[arrow] (L2.east) -- (L3.west);
    \end{tikzpicture}
    \end{adjustbox}

    \caption{ \label{fig:strategy} Screening measures based on the amount of information known. When just the crystal space group, or equivalently its point group, Laue class, or quadrupole survival class, $\mathsf{Q}_{k_\mathrm{crys}}$, are known, the signal loss can be estimated using~\eqref{eq:differentGroupAgnostic}. This also serves as a loss estimator for non-molecular crystals. For molecular crystals, when the molecular point group or quadrupole survival class, $\mathsf{Q}_{k_\mathrm{mol}}$, is also known, the signal loss is captured by~\eqref{eq:LambdaQ}. Finally, if the molecular coordinates are also known, as is typically the case if one has a crystal information file (CIF), the coordinate-aware estimator of~\eqref{eq:lambdaCoord} should be used.}
    \label{fig:screening_hierarchy}
\end{figure}

Figure~\ref{fig:strategy} summarises how our framework can be used to effectively design directional dark matter detectors and screen candidate materials, depending on the information available and the complexity of the system. For convenience, we have implemented these symmetry-screening tools in the publicly available Python package \texttt{symmscreen},
\begin{center}
    {\large\texttt{pip install symmscreen}}
\end{center}
with source code available at \href{https://github.com/jdshergold/symmscreen}{\texttt{github.com/jdshergold/symmscreen}}. This package computes modulation signal loss estimators, and allows for the visualisation of the projection operators described in this work.

The most general use case requires only the discrete symmetry of the detector response, regardless of whether the target is a molecular crystal. For simple crystal  systems, where there is no additional substructure to resolve, the space group, or equivalently the point group, Laue class, or quadrupole survival class is entirely sufficient to determine the modulation signal loss due to symmetry. For molecular crystals specifically, it instead gives the strongest statement that can be made without using molecule-specific information. The estimates for molecular crystals can then be further refined when additional information is available. When the molecular point group is also known, but not the exact crystal embedding, the typical modulation loss can be estimated using the quadrupole survival classes of the crystal and molecule. This can be made material-specific once the molecular coordinates are known, which fix the alignment between the molecular and crystal symmetries, and give a simple estimate of the shape of the quadrupolar response. Going further would require electronic structure calculations.

In conclusion, we have demonstrated that crystal symmetries are key to understanding directional detection, and consequently, that taking them into account is not just an optional refinement, but a crucial part of detector design. In this work, we have shown that this can be done without a seemingly open-ended scan over molecular crystals, each requiring expensive electronic-structure and rate calculations. Rather, the problem can be reduced to a small set of symmetry classes, and when the exact embedding of a molecule is known, to a near-instantaneous calculation using symmetry and geometry alone. This opens the door to efficient mass screening of crystal structures for directional dark matter detection.

\acknowledgments
J.D.S would like to thank Henry McKenna for some useful discussions during the preparation of this manuscript. 
We thank Carlos Blanco for helpful conversations and suggestions.
We wish to acknowledge the use of the EPSRC funded Physical Sciences Data-science Service hosted by the University of Southampton and STFC under grant number EP/S020357/1. J.S. and J.D.S acknowledge support from the UK Research and Innovation Future Leader Fellowship~MR/Y018656/1. 

% \clearpage
\appendix

\section{Coordinate systems for crystals and galaxies}
\label{sec:coords}
Our most important proxies and measurements of anisotropy, e.g.~$\mathsf{Q}_k$, $\xi_L$, and $f_\text{RMS}$, are independent of the coordinate systems used to define the material form factor $f_s(\vec q)$ or the velocity distribution $g_\chi(\vec v)$. However, this is not the case for all of our intermediate results, such as the partial rate coefficients $K^{(\ell)}_{m'm}$ of Fig.~\ref{fig:mcalK}, or the scattering rate $R(\theta_E, \phi_E)$ in Fig.~\ref{fig:rate2d}. 
In this appendix, we describe our choices for the coordinate systems.

\subsection{Unit cell and crystal lattice}
\label{sec:coords:crystal}
A crystal lattice is defined by vectors $(\vec a, \vec b, \vec c)$, which define the orientations and lengths of the edges of the crystal unit cell. The unit cell is not generally orthogonal, and so the crystal lattice must also specify the opening angles $(\alpha, \beta, \gamma)$ between the different pairs of crystal axes, i.e.:
\begin{align}
\hat b \cdot \hat c &= \cos\alpha, 
&
\hat c \cdot \hat a &= \cos\beta, 
&
\hat a \cdot \hat b &= \cos\gamma. 
\end{align}
Tables~\ref{tab:representativeMols} and~\ref{tab:symmetryBreakingMols} provide references to each of the crystal geometries we used for the molecules in our analysis. 
For the most generic triclinic crystal structures, we define our Cartesian crystal coordinate system as:
\begin{align}
\vec a &= (a, 0, 0),
&
\vec b &= (b \cos\gamma,\, b \sin \gamma, 0), 
\end{align}
\begin{align}
\vec c &= \left( c \cos\beta,\, c \frac{\cos\alpha - \cos\beta \cos\gamma}{\sin\gamma} ,\, c \sqrt{ 1 - \cos^2\beta  - \left( \frac{\cos\alpha - \cos\beta \cos\gamma }{\sin\gamma} \right)^2 } \right) .
\end{align}
One should note that different entries for the same molecule in the Cambridge Structural Database may use different conventions for labelling the three crystal axes. 
For the trans-stilbene crystal in Sec.~\ref{sec:demoTSB}, for example, we follow the convention in Refs.~\cite{LanderosRivera:2021CCDCBiphenyl, LanderosRivera:2021BiphenylCrystal}, and define: 
\begin{align}
\vec a &= (a, 0, 0),
&
\vec b &= (0, b, 0), 
&
\vec c &= (\cos\beta, 0, \sin\beta),
\\
a &\simeq 15.49\,\text{\AA}, 
&
b &\simeq 5.67\,\text{\AA}, 
&
c &\simeq 12.28\,\text{\AA}, 
&
\beta &\simeq 111.95^\circ .
\end{align}
The trans-stilbene crystal lattice is invariant under a screw-rotation: specifically, a $180^\circ$ rotation about $\vec b$, combined with a translation of $\frac{1}{2} \vec b$. It is also symmetric under central inversion through a point, usually taken to be the origin. 

It should be noted that our previous study of trans-stilbene, Ref.~\cite{Blanco:2021hlm}, uses a different Cartesian coordinate system than~\cite{LanderosRivera:2021BiphenylCrystal}.
In Ref.~\cite{Blanco:2021hlm}, the $\vec b$ symmetry axis is aligned with $\hat b = \hat z$. Both~\cite{LanderosRivera:2021BiphenylCrystal} and~\cite{Blanco:2021hlm} align the long axis of the unit cell (in this work, $a = 15.49\,\text{\AA}$) with the $\hat x$ direction; however, Ref.~\cite{Blanco:2021hlm} identifies the 15.5\,\AA\ axis as $\vec c$.
So, when comparing Fig.~\ref{fig:rate2d} to the results of Ref.~\cite{Blanco:2021hlm}, one must keep in mind that the two works differ by a $90^\circ$ rotation about the $\hat x$ axis.

\subsection{Dark matter wind in the crystal coordinate system}
\label{sec:coords:Rsphere}
When working with a generic lab frame velocity distribution $g_\chi(\vec v)$, the scattering rate $R(\mathcal D)$ of \eqref{eq:rateFull} depends on the full three-dimensional detector orientation, $\mathcal D \in SO(3)$.
Because the Standard Halo Model (SHM) in the lab frame is symmetric with respect to rotations about the Earth velocity vector, $\vec v_E$, it is possible to specify the scattering rate by just two continuous parameters. That is, $SO(3) / SO(2) \cong S^2$ has the global structure of a 2d sphere. 
We use this feature in Fig.~\ref{fig:rate2d} to plot $R(\mathcal D) \rightarrow R(\theta_E,\phi_E)$ based on the relative orientation of the Earth velocity vector with the crystal lattice. 
In terms of the integrand of \eqref{eq:rateFull}, rather than acting with $\mathcal D$ on the detector, we apply the inverse rotation $\mathcal D^{-1}$ to the velocity distribution:
\begin{align}
g_\chi(\vec v, \vec v_E) \cdot \mathcal D \cdot |f_s(\vec q)|^2 
&= \left( \mathcal D^{-1} \cdot g_\chi(\vec v, \vec v_E) \right) |f_s(\vec q)|^2 
= g_\chi(\mathcal D \vec v, \vec v_E) |f_s(\vec q)|^2
\nonumber \\
% g_\chi(\vec v, \vec v_E) |f_s(\mathcal D^{-1} \vec q)|^2
&= g_\chi( \vec v, \mathcal D^{-1}\vec v_E) |f_s(\vec q)|^2 .
\end{align}
That $g_\chi(\mathcal D \vec v, \vec v_E) = g_\chi(\vec v, \mathcal D^{-1} \vec v_E)$ can be seen from the explicit form of the SHM $g_\chi(\vec v, \vec v_E)$ in \eqref{eq:gXshm}. 

In Fig.~\ref{fig:rate2d}, we work in the crystal-centric lab frame defined by the Cartesian coordinate system of Appendix~\ref{sec:coords:crystal}. 
In this frame, we define the angles $\theta_E$ and $\phi_E$ as the direction of the vector $\mathcal D^{-1} \vec v_E$: i.e., 
\begin{align}
\mathcal D^{-1} \vec v_E &= (|\vec v_E|, \theta_E, \phi_E), 
\end{align}
in spherical coordinates with polar angle $\theta_E$, $\cos\theta_E = \hat v_E \cdot \hat z$, and azimuthal angle $\phi_E$ (where $\phi_E = 0$ maps onto the $\hat x$ direction). 
In this way, the scattering rate for any azimuthally symmetric velocity distribution $g_\chi(\vec v)$ can be written as
\begin{align}
R(\mathcal D) &= R(\theta_E, \phi_E), 
\end{align}
where $\vec v_E$ is the axis of symmetry for the lab frame velocity distribution.

\subsection{Initial detector orientation for daily modulation}
\label{sec:coords:D0}
In a sidereal daily modulation experiment, the experiment is placed in an initial orientation $\mathcal D_0$ at time $t = 0$, and the rotation of the Earth causes $\mathcal D(t)$ to follow a path through $SO(3)$. If we ignore the small day-to-day changes caused by annual variation in the galactic-frame Earth velocity, $\mathcal D(t)$ is periodic over the sidereal day, 
\begin{align}
T_\text{sid} \simeq 86164.09\,\text{s}.
\end{align}
For any initial orientation $\mathcal D_0$, the angle between the Earth velocity and the north pole, $\theta_N$, is the same, and essentially constant over the course of a day. 
Thus, $\theta_E(t)$ and $\phi_E(t)$ trace out a cone on the $S^2$ sphere, with the north pole axis at the centre, and with an opening angle set by $\theta_N$. One such example is shown in the lower panel of Fig.~\ref{fig:rate2d}.
Note that $\theta_N$ varies by over $12^\circ$ over the course of a year, depending on the value one uses for the local standard of rest (LSR) velocity. With $v_\text{LSR} = 238\,\mathrm{km}\,\mathrm{s}^{-1}$, Ref.~\cite{Blanco:2026kda} finds that 
\begin{align}
36.0^\circ \leq \theta_N \leq 48.2^\circ,
\end{align}
with the minimum and maximum occurring around November~1 and April~15, respectively. 
In our calculations of $f_\text{RMS}$ in this work, we set $\theta_N \rightarrow 42^\circ$ to its approximate annual average.
Larger or smaller values of $\theta_N$ would increase or decrease the size of the $\theta_E(t)$, $\phi_E(t)$ circle shown in Fig.~\ref{fig:rate2d}, with a modest impact on the value of $f_\text{RMS}$.

An initial detector orientation $\mathcal D_0$ can therefore be parametrised by three degrees of freedom: the position of the north pole axis relative to the crystal, which we call $(\theta_n, \phi_n)$ in our crystal-centric spherical coordinates; and the initial rotation about the Earth's rotation axis, which is equivalent to shifting our definition of $t=0$ by a constant, $t_0$. 
Rotations about the north pole simply change the phase of the $R(t)$ signals as $t \rightarrow t - t_0$ without affecting the modulation amplitude $f_\text{RMS}$, so for our purposes these types of detector orientations are equivalent.

\begin{figure}
\centering
\includegraphics[width=\textwidth]{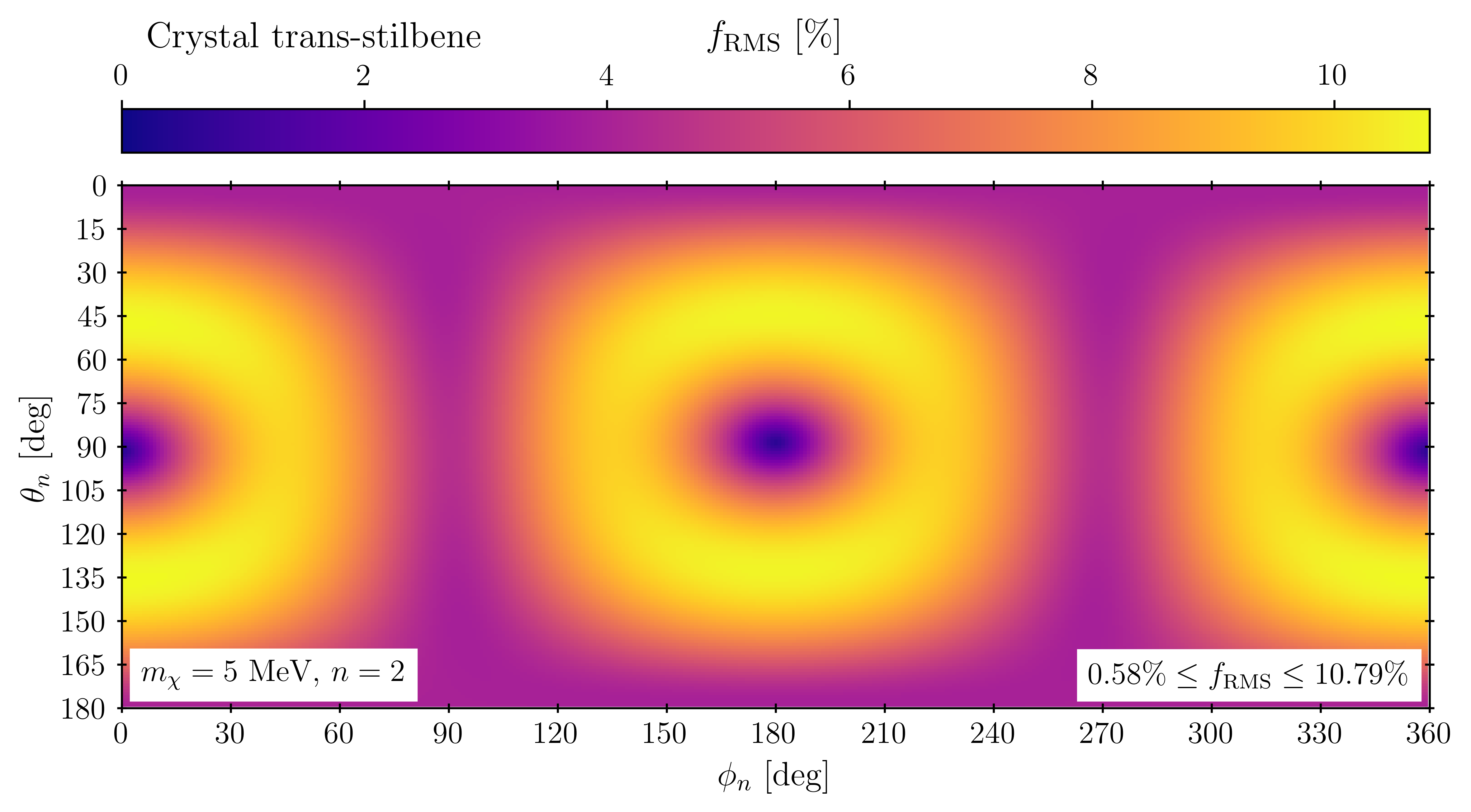}
\caption{The modulation amplitude $f_\text{RMS}$ as a function of initial detector orientation $\mathcal D_0 \rightarrow (\theta_n, \phi_n)$, parametrised by the orientation of the north pole relative to the crystal-centric coordinate system, for the same 5\,MeV, light mediator dark matter model of Fig.~\ref{fig:rate2d}. The modulation amplitude is largest at $\theta_n \simeq 136^\circ$, $\phi_n \simeq 0^\circ$. It is minimised for $(\theta_n, \phi_n) = \pm \hat x$, when the north pole is aligned with the crystal $\hat x$ direction. }
\label{fig:fRMSnorth}
\end{figure}

Figure~\ref{fig:fRMSnorth} shows the possible values of $f_\text{RMS}$ as a function of the initial detector orientation, $\mathcal D_0 \rightarrow (\theta_n, \phi_n)$.
As one would expect from Fig.~\ref{fig:rate2d}, different values of $(\theta_n, \phi_n)$ can lead to very different $R(t)$ profiles, and correspondingly large differences in $f_\text{RMS}$. For this 5\,MeV, $n=2$ dark matter model, the $f_\text{RMS}$ amplitude can be as large as $10.8\%$, or smaller than $0.6 \%$. In the latter case, the $(\theta_E(t), \phi_E(t))$ path orbits the global minimum in $R(\theta_E, \phi_E)$, lining up almost perfectly with a contour of fixed $R(\theta_E, \phi_E) \approx \langle R \rangle$. This special alignment suppresses the RMS amplitude for this dark matter model by a factor of about 18.5.

Incidentally, the $\langle f_\text{RMS} \rangle$ average over detector orientations that appears in \eqref{eq:Lambda_fRMS} is equivalent to an angular average of $f_\text{RMS}$ over $\theta_n$ and $\phi_n$, 
\begin{align}
\langle f_\text{RMS} \rangle_{\mathcal D_0} &= \frac{1}{4\pi}\int d\cos\theta_n\, d\phi_n \,f_\text{RMS}(\theta_n, \phi_n).
\end{align}

\subsection{Latitude-specific coordinate system}\label{app:rotationS}

The crystal-centric lab frame coordinate system is convenient for optimising the detector orientation, but in other contexts it can be more natural to work in a local coordinate system where ``up'' is one of the Cartesian axes.
In this section, we derive an expression for the detector orientation operator $\mathcal D$ this latitude dependent lab frame.

We begin by defining a celestial frame such that $\hat z_\mathrm{cel}$ is parallel to the rotation axis of the Earth, and where $\hat{x}_\mathrm{cel}$ and $\hat{y}_\mathrm{cel}$ lie in the equatorial plane of the Earth. In this frame, the azimuthal angle of the laboratory evolves as
    \begin{equation}
        \Phi(t) = \Phi_0 + \omega_\oplus t,
    \end{equation}
    with $\omega_\oplus = 2\pi/T_\mathrm{sid}$
the angular velocity of the Earth. As all of our results will averaged over the sidereal day, we can safely set $\Phi_0 = 0$. 

The local laboratory frame is then most naturally described in terms of its latitude $\lambda$, and orthonormal directions north, $\hat e_N$, east, $\hat e_E$ and ``up'', $\hat e_U$. Here, up is the local radial direction pointing away from the centre of the Earth, north is the projection of $\hat z_\mathrm{cel}$ onto the local tangent plane, or equivalently the direction along the surface of the Earth pointing toward the North Pole, whilst east satisfies $\hat e_E \times \hat e_N = \hat e_U$. Explicitly, with $\lambda =0$ corresponding to the equatorial plane, these are given by
    \begin{equation}
        \hat e_N(t) = \begin{pmatrix}
            -\sin\lambda \cos\Phi(t)\\
            -\sin\lambda \sin \Phi(t) \\
            \cos\lambda 
        \end{pmatrix}, \qquad 
        \hat e_E(t) = \begin{pmatrix}
            -\sin \Phi(t) \\
            \cos\Phi(t) \\
            0
        \end{pmatrix}, \qquad
        \hat e_U(t) = \begin{pmatrix}
            \cos\lambda \cos\Phi(t) \\
            \cos\lambda \sin\Phi(t) \\
            \sin\lambda
        \end{pmatrix}.
    \end{equation}
The laboratory frame components of a vector $\vec v$ defined in the celestial frame are then mapped onto their lab frame components by, \textit{e.g.} $v_U = \vec v \cdot \hat e_U$. Alternatively, in terms of the matrix $\mathcal{L}(t)$
    \begin{equation}
    \mathcal L(t) = \begin{pmatrix}
        -\sin\Phi(t) & \cos\Phi(t) & 0\\
        -\sin\lambda\cos\Phi(t) & -\sin\lambda\sin\Phi(t) & \cos\lambda \\
        \cos\lambda\cos\Phi(t) & \cos\lambda\sin\Phi(t) & \sin\lambda
    \end{pmatrix},
    \end{equation}
    whose rows are the unit laboratory frame unit vectors, we have
    \begin{equation}
        \vec v_\mathrm{lab}(t) = \mathcal{L}(t) \vec{v}_\mathrm{cel}.
    \end{equation}
    Over the course of a sidereal day, the DM wind direction is approximately constant in time, with fixed right ascension $\alpha_w$ and declination $\delta_w$ such that its celestial frame unit vector is given by
    \begin{equation}
        \hat w_\mathrm{cel} = \begin{pmatrix}
            \cos \delta_w \cos \alpha_w \\
            \cos\delta_w \sin\alpha_w \\
            \sin \delta_w
        \end{pmatrix}.
    \end{equation}
Note that declination is measured in degrees \emph{above} the celestial equatorial plane, i.e.~$\theta_N = 90^\circ - \delta_w$.
This is the direction of the source of the dark matter wind on the celestial sphere, which is also the instantaneous Earth velocity direction, $+ \hat v_E$.
We set $\hat w$ to its approximate annual average, $\alpha_w = 313^\circ$ and $\delta_w = 48^\circ$, which points towards the constellation Cygnus. 
Figure~2 of Ref.~\cite{Blanco:2026kda} shows more precisely how $\alpha_w$ and $\delta_w$ vary over the course of a year.

When expanding $g_\chi(\vec v)$ into spherical harmonic modes, $g_{\ell m}(v)$, it is most convenient to work in a coordinate system where the wind direction $\vec v_w = - \vec v_E$ is aligned with the $z$ axis. This useful choice ensures that for an axisymmetric $g(\vec v)$, only the $m = 0$ components at each $\ell$ survive. In a Cartesian coordinate system where $\hat z_w$ is aligned with $-\hat w_\text{cel}$, and where the $\hat x_w$ and $\hat y_w$ directions are defined as
\begin{align}
    \hat x_w &= \frac{\hat z_\mathrm{cel} \times \hat z_w}{|\hat z_\mathrm{cel} \times \hat z_w|}, & 
    \hat y_w &= \hat z_w \times \hat x_w,
\end{align}
the rotation matrix that converts between the wind frame and the celestial frame is:
\begin{align}
    \mathcal{W} &\simeq \begin{pmatrix}
    -0.731 & -0.682 & 0\\
    -0.507 & 0.544 & 0.669\\
    -0.456 & 0.489 & -0.743
    \end{pmatrix},
    &
\vec{v}_\mathrm{cel} &= \mathcal{W} \vec{v}_w.
\end{align}
    The last rotation that we need, $\mathcal{D_0}$, sets the initial detector orientation by mapping the crystal coordinates onto the laboratory frame coordinates. Explicitly, with our conventions above,
    \begin{equation}
        \hat{x}_\mathrm{crys} \to \hat e_E, \qquad \hat y_\mathrm{crys} \to \hat e_N, \qquad \hat z_\mathrm{crys} \to \hat e_U,
    \end{equation}
    or equivalently
    \begin{equation}
        \vec{v}_\mathrm{lab} = \mathcal{D}\vec{v}_\mathrm{crys}.
    \end{equation}
    For a real experimental setup, this is the only rotation matrix that we have the freedom to choose, as $h_i$, $\mathcal{R}_i$ and $\Qrot$ are all fixed by the crystal, whilst $\mathcal{W}$ and $\mathcal{L}(t)$ are fixed by the DM wind direction and the location of the laboratory on the Earth. Combining everything, we therefore find
    \begin{equation}
        \mathcal{D}(t) = \mathcal{W} \mathcal{L}(t)^T \mathcal{D}_0. 
    \end{equation}

\section{Point group symmetry operations}\label{app:point_groups}

    In this appendix we describe the various symmetry operations that can appear in point groups, along with their common notation. Each operation is an orthogonal transformation, and so can be represented by matrix $\mathcal{R} \in O(3)$. Below we give a description and a representative matrix for each operation, assuming that the principal axis is along the $\hat{z}$ direction.

    \begin{enumerate}
        \item[$E$:] The identity element. This leaves all points unchanged, and is represented by
        \begin{equation}
            E = \mathbb{I}_3, \qquad \det E = +1.
        \end{equation}
        \item[$C_n$:] A proper rotation by $2\pi/n$ about an axis. Choosing the principal axis, $\hat{z}$, this is represented by
        \begin{equation}
            C_n = \begin{pmatrix}
                \cos\left(\frac{2\pi}{n}\right) & -\sin\left(\frac{2\pi}{n}\right) & 0 \\
                \sin\left(\frac{2\pi}{n}\right) & \cos\left(\frac{2\pi}{n}\right) & 0 \\
                0 & 0 & 1
            \end{pmatrix}, \qquad \det C_n = +1.
        \end{equation}
        \item[$\sigma_v$:] A mirror reflection where the mirror plane includes the principal axis and one other axis. With the $z$-axis as the principal axis, one choice is the $x$-$z$ plane, in which case this can be represented by
        \begin{equation}
            \sigma_v =\begin{pmatrix}
                1 & 0 & 0 \\
                0 & -1 & 0 \\
                0 & 0 & 1 
            \end{pmatrix}, \qquad \det \sigma_v = -1
        \end{equation}
        \item[$\sigma_h$:] A mirror reflection where the mirror plane is normal to the principal axis. For the $z$-axis as the principal axis, the mirror plane is the $x$-$y$ plane, and is represented by
        \begin{equation}
            \sigma_h = \begin{pmatrix}
                1 & 0 & 0 \\
                0 & 1 & 0 \\
                0 & 0 & -1 \\
            \end{pmatrix}, \qquad \det \sigma_h = -1.
        \end{equation}
        \item[$\sigma_d$:] A diagonal mirror reflection, where the mirror plane includes the principal axis, but is rotated with respect to the other two axes. One choice is the plane for which $y = x$, with principal $z$-axis, in which case it the operation can be represented by
        \begin{equation}
            \sigma_d = \begin{pmatrix}
                0 & 1 & 0 \\
                1 & 0 & 0 \\
                0 & 0 & 1
            \end{pmatrix}, \qquad \det \sigma_d = -1.
        \end{equation}
        \item[$i$:] Inversion, which maps each point onto its parity partner, and is represented by
        \begin{equation}
            i = -\mathbb{I}_3, \qquad \det i = -1.
        \end{equation}
        \item[$S_n$:] A rotoreflection, which consists of a proper rotation $C_n$ about the principal axis, followed by a reflection in a perpendicular plane. With $\hat z$ as the principal axis, this is represented by
        \begin{equation}
            S_n = \sigma_h \circ C_n = \begin{pmatrix}
                \cos\left(\frac{2\pi}{n}\right) & -\sin\left(\frac{2\pi}{n}\right) & 0 \\
                \sin\left(\frac{2\pi}{n}\right) & \cos\left(\frac{2\pi}{n}\right) & 0 \\
                0 & 0 & -1
            \end{pmatrix}, \qquad \det S_n = -1.
        \end{equation}
        As a result, $S_1$ is just the reflection $\sigma_h$, whilst $S_2$ is equivalent to an inversion, $i$.
        \item[$\bar n$:] A rotoinversion, which consists of a proper rotation $C_n$ about the principal axis, followed by an inversion. Choosing $\hat z$ as the principal axis, this is represented by
        \begin{equation}
            \bar n = i \circ C_n = \begin{pmatrix}
                -\cos\left(\frac{2\pi}{n}\right) & -\sin\left(\frac{2\pi}{n}\right) & 0 \\
                \sin\left(\frac{2\pi}{n}\right) & -\cos\left(\frac{2\pi}{n}\right) & 0 \\
                0 & 0 & -1
            \end{pmatrix}, \qquad \det \bar n = -1.
        \end{equation}
        Consequently, $\bar 1$ is just an inversion, whilst $\bar 2$ is the mirror operation $\sigma_h$.
    \end{enumerate}

\section{Constructing projectors and anisotropy measures } \label{app:invariances}

\subsection{Orthogonal point group representations}\label{app:coordinateIndependence}

    In this appendix we demonstrate how to construct an orthogonal representation of a point group, $P$, starting from an arbitrary three-dimensional representation $\rho_i  \in GL(3,\mathbb{R})$. The need for an orthogonal representation follows from the fact that our projectors operate on coefficients defined on the surface of a sphere. To make this explicit, recall that we decompose the squared form factor as
    \begin{equation}
        |f_{s}(\vec{q})|^2 = \sum f_{\ell m}^2(q) Y_{\ell m}(\hat q),
    \end{equation}
    with $\hat q$ some direction, for which $|\hat q|^2 = \hat q^T \hat q = 1$. A point group representation appropriate for our construction should therefore map this $\hat q$ onto some other direction on the sphere, without changing its length and deforming the sphere into an ellipse. The appropriate point group representation should therefore satisfy
    \begin{equation}
        |\mathcal{R}_i \hat q|^2 = (\mathcal{R}_i \hat q)^T (\mathcal{R}_i\hat q) = 1,
    \end{equation}
    from which the requirement $\mathcal{R}_i^T \mathcal{R}_i = \mathbb{I}_3$ follows, \textit{i.e.} that $\mathcal{R}_i$ is some orthogonal representation. 

    In the frame where the arbitrary representation, $\rho_i$, acts, the set of directions, $\hat q_\mathrm{skew}$, need not necessarily satisfy $\hat q_\mathrm{skew}^T \hat q_\mathrm{skew} = 1$. More generally, they will satisfy
    \begin{equation}\label{eq:skewLength}
        \hat q_\mathrm{skew}^T M \hat q_\mathrm{skew} = 1,
    \end{equation}
    with $M$ some matrix, explicitly a \textit{metric tensor}. Under the action of the point group, this becomes
    \begin{equation}\label{eq:skewLengthRotated}
        (\rho_i\hat q_\mathrm{skew})^T M (\rho_i\hat q_\mathrm{skew}) = 1,
    \end{equation}
    such that the analogous constraint on $\rho_i$ and $M$ is that
    \begin{equation}\label{eq:metricConstraint}
        \rho_i^T M \rho_i = M.
    \end{equation}
    Now noting that for a symmetric, positive-definite matrix $M$, that we can always write
    \begin{equation}\label{eq:matrixSqrt}
        M = A^T A,
    \end{equation}
    it follows from Eqs.~(\ref{eq:skewLength}) and (\ref{eq:skewLengthRotated}) that
    \begin{equation}
        (A\hat q_\mathrm{skew})^T  (A\hat q_\mathrm{skew}) = 1, \qquad (A\rho_i\hat q_\mathrm{skew})^T  (A\rho_i\hat q_\mathrm{skew}) = 1,
    \end{equation}
    The first of these expressions allows us to identify $A \hat q_\mathrm{skew} = \hat q$, such that the second becomes
    \begin{equation}
        (A\rho_i A^{-1} \hat q)^T  (A\rho_i A^{-1}\hat q) = 1,
    \end{equation}
    and consequently
    \begin{equation}
        A\rho_i A^{-1} = \mathcal{R}_i \in O(3).
    \end{equation}
    That is to say, that provided we can find a valid metric for the skewed coordinate system, we can use it to construct an orthogonal representation of the point group.
    
    One such choice that depends only on the symmetry operations of the crystal is
    \begin{equation}
        M = \frac{1}{N_P} \sum_{i \in P(S)} \rho_i^T \rho_i,
    \end{equation}
    where $N_P$ is the order of the point group. Alternatively, one could construct this from the Laue group. To see that this does indeed satisfy \eqref{eq:metricConstraint}, we substitute in to find
    \begin{equation}
        \rho_i^T M \rho_i=\frac{1}{N_P} \sum_{j \in P(S)} \rho_i^T \rho_j^T \rho_j \rho_i = \frac{1}{N_P} \sum_{k \in P(S)} \rho_k^T \rho_k = M, 
    \end{equation}
    where in going from the second to third expressions we have used the closure of the group under multiplication, $\rho_j \rho_i = \rho_k$. The corresponding $A$ can then be found from \textit{e.g.} a Cholesky decomposition of the metric, or some other method of solving~\eqref{eq:matrixSqrt} for $A$. That is, if $M = YY^T$, and we define $Y = \mathrm{chol}(M)$, then
    \begin{equation}
        A = \mathrm{chol}(M)^T = \mathrm{chol}\left(\frac{1}{N_\mathcal{L}}\sum_{i\in \mathcal{L}(S)} R_i^T R_i\right)^T,
    \end{equation}
    for example. Thus we have found an expression for $A$ that depends only on the group symmetries, and subsequently a method of constructing an orthogonal representation of a point group, given any $3\times3$ matrix representation. 

\subsection{Invariance of anisotropy measures}\label{app:measureInvariance}
    In this paper we have defined several proxies for anisotropy; now let us demonstrate that $\kappa_\ell^{(\mathcal{\ell})}$ and $\xi^{(\mathcal{L})}_L$ are invariant under the choice of compatible crystal metric. Suppose that we have an alternative, equally valid metric for the same space group, $M' = A'^T A'$. The corresponding projector is
    \begin{equation}\label{eq:Gtrace}
        \Pi'^{(\mathcal{L})}_\ell = \frac{1}{N_\mathcal{L}}\sum_{i \in \mathcal{L}(S)} p'^{(\mathcal{L})}_{i,\ell} G^{(\ell)}(\widetilde{\mathcal{R}}'_i),
    \end{equation}
    with $\mathcal{R}'_i = A'R_i A'^{-1}$. Now, since $\mathcal{R}'_i$ and $\mathcal{R}_i$ are related by a similarity transformation, they share both a determinant and a trace. The first property sets $p_{i,\ell}' = p_{i,\ell}$, whilst the latter indicates that both rotate vectors by the same polar angle, satisfying $\mathrm{tr}(\widetilde{\mathcal{R}}'_i) = \mathrm{tr}(\widetilde{\mathcal{R}}_i) = 1 + 2\cos \theta_i$. Then since the trace of the Wigner-G matrices depends only on the polar angle, see  \eqref{eq:Gtrace}, the trace of the two projectors is the same, $\mathrm{tr}(\Pi_\ell') = \mathrm{tr}(\Pi_\ell)$. Our anisotropy parameters $\kappa^{(\mathcal{L})}_\ell$ and $\xi_L^{(\mathcal{L})}$ are therefore true, molecule- and specific-crystal-independent measures of anisotropy, depending only on the space group symmetries.    

\section{Angular mode decay}\label{app:angularDecay}
\begin{figure}[t]
        \centering
        \includegraphics[width=0.9\linewidth]{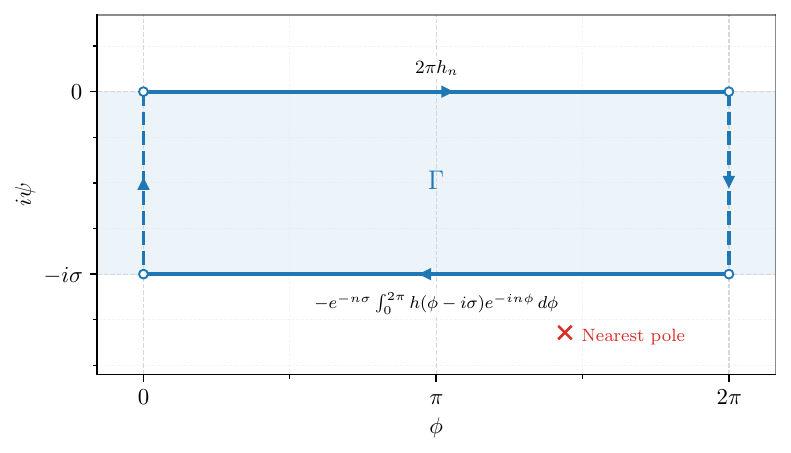}
        \caption{Contour integral used to demonstrate the exponential scaling of Fourier modes, $h_n$, with increasing $n$. The red cross shows the nearest pole of the function $h(z)$ below the real line, located just beyond $z = \phi + i\psi = \phi-i\sigma$.}
        \label{fig:contourIntegral}
    \end{figure}

    In this appendix we derive how the angular modes on a sphere decay with increasing $\ell$, which can be found by analogy with how Fourier modes decay. Let us consider the $n^\text{th}$ Fourier mode of some periodic function $h(\phi)$
    \begin{equation}\label{eq:fourierModes}
        h_n = \frac{1}{2\pi}\int_{0}^{2\pi} d\phi\, h(\phi)\, e^{-in\phi}.
    \end{equation}
    We extend this integral beyond the real line to the rectangular complex contour $\Gamma$ shown in Figure~\ref{fig:contourIntegral}, with vertices $z=\theta + i\psi =  0, 2\pi, 2\pi -i\sigma, -i\sigma$, where $\sigma$ denotes the furthest distance from the real line for which the contour $\Gamma$ does not enclose a pole of $h$. 
    As this contour contains no poles of $h$, we necessarily have that
    \begin{equation}
        \oint_\Gamma dz \, h(z)\, e^{-inz} = 0,
    \end{equation}
    from Cauchy's theorem. Splitting this integral up into its individual components, this becomes
    \begin{equation}
        2\pi h_n + i\int_{0}^{-\sigma} d\psi\, h(i\psi)\, e^{n\psi} -e^{-n\sigma}\int_{0}^{2\pi} d\phi\, h(\phi - i\sigma) e^{-in\phi} + i\int_{-\sigma}^{0}d\psi \, h(i\psi) \, e^{n\psi} = 0.
    \end{equation}
    The second and fourth terms cancel, and we are left with
    \begin{equation}
        h_n = \frac{e^{-n\sigma}}{2\pi} \int_{0}^{2\pi} d\theta \, h(\theta - i\sigma) e^{-in\theta}. 
    \end{equation}
    We are interested in the size of $h_n$, and so we now take the absolute value, and use the triangle inequality to find
    \begin{equation}\label{eq:exponentialDecayInt}
        |h_n| \leq \frac{e^{-n\sigma}}{2\pi}\int_0^{2\pi} d\theta \, |h(\theta-i\sigma)|.
    \end{equation}
    This demonstrates that the magnitude of the $h_n$ decay as $e^{- n\sigma}$, that is, that they exponentially decay with increasing $n$. Before continuing, we note that we could equally have extended the contour above the real line, in which case we would have arrived at the bound $h_n \lesssim  e^{n\sigma}$. However, this bound is not insightful, as it is completely superseded by the much stronger bound given in \eqref{eq:exponentialDecayInt}.
    
    Going back to our spherical harmonic modes, we have
    \begin{equation}
        f_{\ell m}^{2}(q) = \int d\Omega_q \, |f_s(\vec{q})|^2 Y_{\ell m}(\hat q),
    \end{equation}
    already looks similar to the expression for the Fourier modes in \eqref{eq:fourierModes}. Further noticing that the spherical harmonics satisfy an analogous eigenvalue equation to the Fourier modes, namely
    \begin{equation}
        \nabla_{\Omega_q}^2 Y_{\ell m}(\hat q) = - \ell (\ell + 1) Y_{\ell m}(\hat q) \longleftrightarrow \frac{\partial^2}{\partial\theta^2} e^{in\theta} = - n^2 e^{in\theta},
    \end{equation}
    we can make the identification $n^2 = \ell (\ell+1)$. That is, the $f_{\ell m}^2$ can be thought of as Fourier modes with angular frequency $\sqrt{\ell (\ell+1)}$. Then, from our previous arguments we find that the $f_{\ell m}^2$, and by the analogy, $g_{\ell m}$, decay as
    \begin{equation}
        f_{\ell m}^2(q) \sim e^{-\sigma_f(q) \sqrt{\ell (\ell +1)}}, \qquad g_{\ell m}(v) \sim e^{-\sigma_g(v) \sqrt{\ell(\ell +1)}}.
    \end{equation}
    Collecting everything together, we therefore expect that our partial matrix elements decay with $\ell$ as
    \begin{equation}
        K^{(\ell)} \sim \frac{1}{\sqrt{\ell}} \int dq\int dv\,\,e^{-C(q,v)\sqrt{\ell (\ell +1)}},
    \end{equation}
    where $C(q,v) = \sigma_f(q) + \sigma_g(v)$ is, in general, an unknown function. However, as this modulates the angular scale at which the function changes appreciably, we can comfortably estimate \textit{e.g.} $\sigma_f(q) \sim \Delta \theta(q) \sim 1/\ell_0(q)$, with $\Delta \theta(q)$ the characteristic angular size of features in the form factor at momentum transfer $q$. At large $q$, the form factor decays rapidly and so is dominated by noise, resulting in $\sigma(q) \to 0$ and effective polynomial scaling of the $f_{\ell m}^2$ modes with $\ell$. However, as the form factor is small in these regions, its contribution to the partial matrices is small, and so we can safely neglect it. More generally, we expect that the most important angular features at finite $q$ to appear at the first $\ell$ that is not forbidden by symmetry, as those at higher $\ell$ will decay exponentially. We therefore take $\ell_0$ to be the first non-zero $\ell$ that contributes to the squared form factor, and treat $\sigma_f$ as approximately constant in $q$. For the spherical halo model, which similarly decays exponentially with $v$, the dominant features instead begin at $\ell = 1$, and so in general we approximate $\sigma_g(v) \simeq 1$, resulting in
    \begin{equation}
        C(q,v) \simeq 1 + \frac{1}{\ell_0},
    \end{equation}
    such that the exponential scaling can be approximately pulled out of the integral.

\section{Angular average} \label{app:Qaverage}
In this appendix we briefly derive the orientation-averaged value of the combined survival operator, which we use in several places throughout this work. Let $X_\ell$ be any $(2\ell+1) \times (2\ell+1)$ matrix that acts on the angular momentum coefficients at each $\ell$. The rotational average of the rotated $X_\ell$ matrix, $X_\mathcal{\ell,Q}$, must itself be rotationally invariant, and so proportional to the identity. That is
    \begin{equation}
        \langle X_\mathcal{\ell,Q}\rangle_\Qrot = \int_{SO(3)} d\Qrot \, G^{(\ell)}(\Qrot) X_\ell G^{(\ell)}(\Qrot^T) = c\,\mathbb{I}_{2\ell+1},
    \end{equation}
    with $d\Qrot$ the normalised Haar measure, and $c$ some proportionality constant. We now take the trace of both sides, and note that the trace is invariant under rotations, yielding
    \begin{equation}
        \mathrm{tr}(X) = c(2\ell+1) \implies c = \frac{\mathrm{tr}(X)}{2\ell+1}.
    \end{equation}
    Consequently, the angular average of the combined survival operator is
    \begin{equation}
        \langle \mathcal{C}^{(\ell)}\rangle_\Qrot = \Pi^{(\mathcal{L})}_\ell \langle \Pi^{(H)}_{\ell,\Qrot}\rangle_\Qrot = \frac{1}{(2\ell+1)}\Pi^{(\mathcal{L})}_{\ell} \operatorname{tr}(\Pi^{(H)}_\ell),
    \end{equation}
    effectively disentangling the crystal and internal molecular symmetries. This is particularly evident in the angular average of the trace of the combined survival operator:
    \begin{equation}
        \left\langle \operatorname{tr}\left[\mathcal{C}^{(\ell)}\right]\right\rangle_\Qrot = \operatorname{tr}(\Pi^{(\mathcal{L})}_{\ell} )\operatorname{tr}(\Pi^{(H)}_\ell).
    \end{equation}
    As a result, our symmetry measures derived from the the combined survival operator factorise after taking the angular average
    \begin{equation}
        \langle \kappa_\ell\rangle_\Qrot =  \kappa_\ell^{(\mathcal{L})} \kappa_\ell^{(H)}, \qquad \langle \xi_\infty \rangle_\Qrot = \xi_\infty^{(\mathcal{L})} \xi_\infty^{(H)},
    \end{equation}
    supporting the statement that at large $\ell$ and small angular scales, the combined effect of the crystal and molecular symmetries is to maximise the loss of anisotropy in the molecular crystal system.

\section{Degenerate energy eigenstates}\label{app:degenerateLevels}

Crystals and single molecules with nontrivial symmetries often have degenerate excited states. When the degeneracy is exact, the form factor $|f_s(\vec q)|^2$ or $|f_\text{uc}(\vec q)|^2$ of \eqref{eq:fsFromTDM} and \eqref{eq:incoherentSum} should include a sum over all final states $a,b \ldots$ with the same excitation energy, $E_a = E_b = \ldots = E_s$, i.e.
\begin{align}
    |f_s(\vec q)|^2 = \sum_{i = a, b, \ldots} |f_i(\vec q)|^2 .
\end{align}
This is the physically observable material response that appears in \eqref{eq:dynamicS}. For molecules with nontrivial point group $H$, it is the combined form factor $|f_s(\vec q)|^2$ that is invariant under $H$; the individual final states $|f_a(\vec q)|^2$ and $|f_b(\vec q)|^2$ do not necessarily respect the symmetry. 

If states $\ket{\Phi_a}$ and $\ket{\Phi_b}$ are degenerate energy eigenstates, then linear combinations of $a$ and $b$ are equally valid descriptions of the system. We demonstrate here that the observable $|f_s(\vec q)|^2$ is invariant under relabellings of this type. 
Let us define a new basis for states with energy $E_s$:
\begin{align}
\ket{\Psi_\pm} = \frac{1}{\sqrt{2}} \left( \ket{\Psi_a } \pm \ket{\Psi_b} \right) ,
\end{align}
with transition densities 
\begin{align}
\Phi^{g \rightarrow \pm}(\vec x) &= \frac{1}{\sqrt{2}} \left( \Phi^{g \to a}(\vec x) \pm \Phi^{g \to b}(\vec x) \right) .
\end{align}
From the linearity of the Fourier transform, the momentum form factors $f_\pm(\vec q)$ are simply
\begin{align}
f_\pm(\vec q) &= \frac{1}{\sqrt{2}} \left( f_a(\vec q) \pm f_b(\vec q) \right), 
\\
|f_\pm(\vec q)|^2 &= \frac{1}{2} \left( |f_a(\vec q)|^2 \pm 2 \text{Re}\left[ f_a^\star(\vec{q}) f_b(\vec{q}) \right] + |f_b(\vec q)|^2 \right) . 
\end{align}
Although the interference terms $\pm f_a^\star f_b$ may add nontrivial momentum features to $|f_+|^2$ and $|f_-|^2$, they cancel in the sum over degenerate final states:
\begin{align}
|f_+(\vec q)|^2 + |f_-(\vec q)|^2 &= |f_a(\vec q)|^2 + |f_b(\vec q)|^2. 
\end{align}
In other cases, $\ket{\Psi_\pm}$ may represent the true energy eigenstates, with some small splitting $E_\pm = E_s \pm \delta E$, for some $\delta E \ll E_s$. Although the cancellation of the $f_a^\star f_b$ cross terms is no longer exact, the correction to the scattering rate is proportional to $\delta E$. This can be seen in the expression for the partial rate matrix, \eqref{eq:partialRateSingle}: a small change to $E_s$ merely changes $v_\text{min}(q)$ by an amount equal to $\delta E / q$.

This basis-independence of the labelling of energy eigenstates is particularly important when considering molecular crystals. In the absence of nearest-neighbour interactions between molecules, the spectrum of excited states includes a $g \rightarrow s$ transition from every molecule, $i = 1, 2, \ldots N$.
Our approach in \eqref{eq:incoherentSum} is to calculate the transition density $\Phi^{g \to s}(\vec x)$ at each individual molecule, find the corresponding $|f_s(\vec q)|^2$, and add the results. Here we demonstrate that any linear combination of exactly degenerate excited states from multiple molecules will produce the same crystal form factor, 
\begin{align}
|f_\text{uc}^{g \to s}(\vec q) |^2 = \sum_{i = 1}^N |f_{i}^{g \to s}(\vec q)|^2.
\end{align}
Labelling the excited state $s$ at the $i$th molecule as $s_i$, the basis of states can be redefined by rotation by a unitary matrix $A \in U(N)$: 
\begin{align}
\ket{\Psi_j} = A_{ji} \ket{ \Psi_{s_i} } , 
&&
\Phi^{g \rightarrow j} = A_{ji} \Phi^{g \to s_i} , 
&&
f_j(\vec q) = A_{ji} f_{s_i}(\vec q). 
\end{align}
Summing the squared form factors, we find again that 
\begin{align}
\sum_j f_j^\star(\vec q) f_j(\vec q) 
&= \left( A_{jk} f_{s_k}(\vec q) \right)^\star \left( A_{ji} f_{s_i}(\vec q) \right) = f_{s_k}^\star(\vec q) A_{jk}^\star A_{ji} f_{s_i} 
= \sum_i |f_{s_i}(\vec q)|^2, 
\end{align}
where by definition $A^\dagger A = \mathbb{I}$ for unitary matrix $A \in U(N)$. 

In conclusion, even though the multiple-molecule transition density $\Phi^{g \to j}(\vec x)$ may have a much wider spatial extent than the single-molecule $\Phi^{g \to s_i}$ functions, this by itself does not introduce new lower-momentum modes into the combined form factor $|f_s(\vec q)|^2$. As long as the true energy eigenstates of the system can be written as linear combinations of single-molecule states $\ket{\Psi_{s_i}}$, our treatment of the crystal form factor as an incoherent sum of molecular form factors in \eqref{eq:incoherentSum} is valid. 

If neighbouring molecules interact strongly with each other, on the other hand, the energy eigenstates of the crystal are not necessarily linear combinations of single-molecule states,  and $|f_\text{uc}(\vec q)|^2$ is not well described by \eqref{eq:incoherentSum}. 
Strong interactions between molecules will also tend to break the degeneracy between the excited states $E_j$, leading to some splitting $E_j = E_s + \Delta E_j$. This is the Davydov splitting~\cite{Davydov1948}. If the splitting is large, then likewise we cannot rely on \eqref{eq:incoherentSum} to determine $|f_\text{uc}(\vec q)|^2$. 
The Davydov splitting in molecular crystals can be estimated from the dipole--dipole nearest neighbour interactions, once the electronic structure of a single molecule is known.

\bibliographystyle{JHEP}
\bibliography{references}

\end{document}